\documentclass[]{aastex63}

\usepackage{amsmath}
\usepackage{url}
\usepackage{ulem}

\newcommand{\be}{\begin{equation}}
\newcommand{\ee}{\end{equation}}

\shorttitle{Parameter-Estimation Biases Caused by pulsar terms in PTAs}
\shortauthors{Chen \&. Wang}

\begin{document}

\title{Parameter-Estimation Biases for Eccentric Supermassive Binary Black Holes in Pulsar Timing Arrays: Biases Caused by Ignored Pulsar Terms
\footnote{Released on \today}}

\correspondingauthor{Jie-Wen Chen \&. Yan Wang}
\email{chjw@hust.edu.cn, ywang12@hust.edu.cn}

\author[0000-0002-8037-543X]{Jie-Wen Chen}
\author[0000-0001-8990-5700]{Yan Wang}
\affiliation{MOE Key Laboratory of Fundamental Physical Quantities Measurements,
Hubei Key Laboratory of Gravitation and Quantum Physics,
\&. Department of Astronomy, School of Physics,
Huazhong University of Science and Technology, Wuhan 430074, China}

\begin{abstract}

The continuous nanohertz gravitational waves (GWs) from individual supermassive binary black holes (SMBBHs)
can be encoded in the timing residuals of pulsar timing arrays (PTAs).
For each pulsar,
the residuals actually contain an Earth term and a pulsar term,
but usually only the Earth term is considered as signal and the pulsar term is dropped,
which leads to parameter-estimation biases (PEBs) for the SMBBHs,
and currently there are no convenient evaluations of the PEBs.
In this article, we formulate the PEBs for a SMBBH with an eccentric orbit.
In our analyses, the unknown phases of pulsar terms are treated as random variables obeying the uniform distribution $U[0,2\pi)$,
due to the fact that pulsar distances are generally poorly measured.
Our analytical results are in accordance with the numerical work by \citet{Zhu_et_al_2016} at $1.5\sigma$ level,
implying that our formulae are effective in estimating magnitudes of the PEBs.
Additionally, we find that for two parameters--- Earth term phase $\varphi^E$ and orbital eccentricity $e$,
their biases $\Delta \varphi^E$ and $\Delta e/e$ monotonically decrease as $e$ increases,
which partly confirms a hypothesis in our previous work \citep{Chen&Zhang2018}.
Furthermore, we also calculate the PEBs caused by the recently observed common-spectrum process (CSP),
finding that if the strain amplitude of the continuous GW is significantly stronger ($3$ times larger, in our cases) than the stochastic GW background,
the PEBs from pulsar terms are larger than those from the CSP.
Our formulae of the PEBs can be conveniently applied in the future PTA data analyses.
\end{abstract}




\section{Introduction} \label{sec: Intro}

Pulsar timing arrays (PTAs) provide a promising tool to search for the nanohertz band ($10^{-9}-10^{-6}$Hz)  gravitational waves (GWs) from inspiralling supermassive binary black holes (SMBBHs),
via precise measurements of the time of arrival (TOA) of radio pulses from arrays of millisecond pulsars (MSPs) \citep{Sazhin1978, Detweiler1979, Foster&Backer1990}.
Currently, the PTA observations are mainly carried out by three regional collaborations ---
the Parkes PTA (PPTA) \citep{manchester_et_al._2013, Hobbs_2013},
the North American Nanohertz Observatory for GWs (NANOGrav) \citep{McLaughlin2013, Ransom_et_al_2019},
and the European PTA (EPTA) \citep{Kramer&Champion2013},
and they have combined as the International PTA (IPTA) collaboration \citep{Hobbs_et_al_2010, Manchester_et_al_2013b}.
The IPTA has accumulated timing residual data for more than 10 years, offering nearly $100$ stable MSPs with white noises below $300$ ns \citep{Perera2019, Alam_et_al_2021, Alam_et_al_2021b,  Kerr_et_al_2020, Desvignes_et_al_2016}.
Furthermore, all the aforementioned collaborations have detected a common-spectrum process (CSP) recently, whose origin is still unclear,
and a potential explanation is the signal from stochastic GW background (SGWB) \citep{Arzoumanian_2020, Goncharov_2021, EPTA_GWB, IPTA_2022}.
Additionally, the Indian PTA (InPTA) \citep{Joshi_et_al_2018} has also joined the IPTA recently,
and the Chinese PTA (CPTA) \citep{Lee2016} is preparing to join the IPTA.
The next-generation radio telescopes,
such as
the Five-hundred-meter Aperture Spherical Telescope (FAST) \citep{Nan_et_al_2011} and
the Square Kilometre Array (SKA) \citep{Smits_et_al_2009},
are expected to enlarge the number of the well-timed MSPs (noise rms $\lesssim 100$~ns) to $O(10^3)$ \citep{2020PASA...37....2W, 2018JPhCS.957a2003W, Wang&Mohanty2017, Hobbs_et_al_2019, Feng_et_al_2020}.

The timing residuals for each MSP in a PTA are actually caused by both the GWs at Earth (hereafter, Earth term) and at the MSP (pulsar term).
The pulsar term has a phase earlier than the Earth term,
which is proportional to the distance of the MSP from Earth \citep{Jenet_et_al_2004, Burke-Spolaor_et_al_2019}.
Hence the Earth terms for all MSPs in the PTA are the same and can be coherently added to enlarge the total signal-to-noise ratio (S/N),
while the pulsar terms have individual phases and tend to cancel each other \citep{Corbin&Cornish2010, Lee_et_al_2011, Burke-Spolaor_et_al_2019}.
In addition, the pulsar distances are generally poorly measured,
with their uncertainties much larger than the wavelengths of nanohertz-GWs ($10^{-1}-10$ ly) \citep{Verbiest_et_al_2012, Sesana_and_Vecchio_2010},
so it is difficult to fix those phases of pulsar terms.
Furthermore, if we include both Earth term and pulsar terms in the signal templates,
and take all the pulsar phases as free parameters,
namely adopting a full-signal search \citep{Ellis_et_al_2012, Ellis_et_al_2012b, Zhu_et_al_2016},
the number of parameters will be larger than the number of MSPs in the data analyses,
and this will make the parameter estimations complicated and computationally expensive \citep{Sesana_and_Vecchio_2010, WMJ1, WMJ2}.
Therefore, in most of the current PTA data analyses, only the Earth term is deemed as a coherent GW signal,
and the pulsar terms are usually treated as a kind of self-noise and ignored.
This scheme is termed as the Earth-term-only search \citep{Sesana_and_Vecchio_2010, Ellis_et_al_2012, Ellis_et_al_2012b, Zhu_et_al_2016}.

However, the Earth-term-only search will lead to biases on parameter estimations,
since it drops all pulsar terms in the timing residual templates.
For instance, the works by \cite{Ellis_et_al_2012} and \cite{Zhu_et_al_2016}, which use synthetic data sets, have demonstrated that
the sky localizations of GW sources are biased in the Earth-term-only search,
and the biases in \citet{Zhu_et_al_2016} are not smaller than the localization uncertainties due to the injected white noise of the MSPs even
for a strong signal case (the network S/N $=100$).
Besides the sky localizations, other parameters can also be biased in the Earth-term-only search \citep{Corbin&Cornish2010},
and the parameter-estimation biases (PEBs) from the ignored pulsar terms cannot be systematically estimated.
Hence, in this article, we will investigate the issue and propose a technique to calculate the PEBs.

Due to the recent detection of the aforementioned CSP
and the studies in \citet{Rosado2015, Mingarelli:2017fbe, Taylor:2020zpk, Pol_2021, Ali-Haimoud_2021},
one can expect that the SGWB may be detected earlier than the continuous GWs from individual sources.
Even if the continuous GWs can also be detected in the future, the CSP will have considerable impacts on them  \citep{B_csy_2020}
and lead to PEBs for the individual SMBBHs, if the CSP is not properly considered in the noise model.
Hence it is worthwhile to calculate the PEBs arising from the CSP and compare them with the PEBs from pulsar terms.
In this article, we focus on the data analyses of a single SMBBH, and
assume that the SMBBH can be detected,
even in the case that its continuous GW is weaker than the SGWB
\footnote{Actually, if the continuous GW is weaker than the SGWB, it will be absorbed by the background, as indicated by \citet{B_csy_2020}.}.
The current research can serve as a foundation for further studies on the PEBs for multiple sources \citep{Songsheng:2021tri, Qian:2021bfp}
and even for SGWB, since SGWB in the astrophysical scenario is generally deemed as a composition of multiple unresolvable SMBBHs \citep{Phinney:2001di}.

Furthermore, the SMBBHs in PTA observations are allowed to have large orbital eccentricities \citep{Taylor_et_al_2016},
and some candidate sources are reported to contain non-negligible eccentricities \citep{Feng_et_al_2019},
e.g. OJ 287 has an eccentricity $e \simeq 0.7$ \citep{Lehto&Valtonen_1996, Dey_et_al_2019}
and NGC 4151 has $e \simeq 0.4$ \citep{Bon_et_al_2012}.
In addition, in our previous work \citep{Chen&Zhang2018},
a hypothesis is proposed that,
for the SMBBH with a large eccentricity (i.e. $e \gtrsim 0.5$),
the PEBs due to pulsar terms should be smaller than those for a quasi-circular SMBBH,
and we will test it in this work.
So our analyses in this article will be based on an eccentric SMBBH,
but we will also give the results for a circular SMBBH as a specific case.


The organization of this paper is as follows.
In Section \ref{Sec:Templates}, we revisit the template of timing residual from an eccentric SMBBH.
In Section \ref{Sec: Parameter Estimations}, we derive the PEBs from the ignored pulsar terms by using the noise-projection technique \citep{Cutler&Flanagan1994, Cutler&Harms2006, Harms_et_al_2008}.
In Section \ref{Sec: Computation of the Errors}, the formulae of the PEBs are derived.
In Section \ref{Sec:Applications}, we test the formulae of PEBs,
by comparing them with the numerical results given by \cite{Zhu_et_al_2016}.
In addition,
the hypothesis in \cite{Chen&Zhang2018} will be tested in Section \ref{Sec:Applications} as well.
Moreover, in Section \ref{Sec:CommonNoise}, the PEBs from CSP will be calculated and compared with the PEBs from pulsar terms.
Finally, we summarize in Section \ref{Sec: Summary}.

\section{Template of Timing Residual Signals}\label{Sec:Templates}

The GWs fluctuate the TOAs of the pulses emitted from the MSPs in a PTA,
and lead to timing residuals.
For the $I$-th MSP in a PTA ($I=1, 2, ..., N_{\rm P}$,
with $N_{\rm P}$ being the the number of MSPs),
its residual signal induced by the GWs at the time $t_i$ ($i=1, 2,..., N_I$, with $N_I$ being the number of data points for the $I$-th MSP) is
\be
s_I(t_i)= s_I^E(t_i)-s_I^{P}(t_i),
\ee
where $s^E$ denotes the Earth-term signal,
$s^P$ the pulsar-term signal,
and hereafter the superscripts ``$E$" and ``$P$" will denote ``Earth term" and ``pulsar term" respectively.
Both the Earth term and pulsar term can be written as \citep{Taylor_et_al_2016}
\be
\label{eq:sEP}
s^{E, P}_I(t_i)=(\cos 2\psi F_{I+}-\sin 2\psi F_{I\times}) s^{E, P}_{I+}(t_i)+(\sin 2\psi F_{I+}+\cos 2\psi F_{I\times}) s^{E, P}_{I\times}(t_i),
\ee
where the subscripts ``$+$" and ``$\times$" represent two GW polarization components respectively,
$\psi$ is the polarization angle,
and
\begin{align}
F_{I+}&=\frac{(1+\sin^2\delta)\cos^2\delta_I^P \cos(2\alpha-2\alpha_I^P)-\sin 2\delta \sin 2\delta_I^P\cos(\alpha-\alpha_I^P)+\cos^2\delta(2-3\cos^2\delta_I^P)  }{4[1-\cos \delta \cos \delta_I^P \cos(\alpha-\alpha_I^P)-\sin \delta \sin \delta_I^P]},\nonumber \\
F_{I\times}&=\frac{-\sin \delta \cos^2\delta_I^P \sin(2\alpha-2\alpha_I^P)+\cos \delta \sin 2\delta_I^P\sin(\alpha-\alpha_I^P)  }{2[1-\cos \delta \cos \delta_I^P \cos(\alpha-\alpha_I^P)-\sin \delta \sin \delta_I^P]},
\end{align}
are the antenna pattern functions,
with $\alpha$ and $\delta$ being the right ascension (RA) and declination (DEC) of the SMBBH respectively,
and $\alpha_I^P$ and $\delta_I^P$ being the RA and DEC of the MSP respectively \citep{Wahlquist1987, Zhu_et_al_2016}.

In general, the pulsar term corresponds to the GW at an earlier stage,
so it should have different amplitude and frequency (or period) from the Earth term, due to the orbital reduction of the SMBBH \citep{Lee_et_al_2011}.
However,
for simplicity, we adopt a non-evolving assumption here,
namely the Earth term and pulsar term have the same period and amplitude.
For an eccentric SMBBH,  two polarization components for both the Earth term and pulsar term in Eq. \eqref{eq:sEP} can be written as \citep{Taylor_et_al_2016}
\begin{align}
\label{eq:SEP-eccentric}
s^{E, P}_{I+}(t)  = & \mathcal A_m   {\mathop \sum_{n=1}^{\infty}} \Bigg[ (1+c_{i}^2)\bigg(X_n  \cos 2\phi_p  \mathcal S_n^{E, P} -Y_n   \sin 2\phi_p   \mathcal C_n^{E, P} \bigg)  + 2 (1-c_{i}^2)  Z_n \mathcal S_n^{E, P} \Bigg],    \nonumber \\
s^{E, P}_{I\times}(t) = & -2 \mathcal A_m   c_{i}  {\mathop \sum_{n=1}^{\infty}} \Bigg( X_n  \sin 2\phi_p \mathcal S_n^{E, P} +Y_n   \cos 2\phi_p  \mathcal C_n^{E, P} \Bigg),
\end{align}
where $\mathcal A_m$ is the characteristic amplitude of the timing residual signal, 
$c_i$ stands for the cosine of the inclination angle, $\phi_p$ denotes the orbital phase of the periastron,
and
\begin{align}
\label{eq:Xn-Yn-Zn}
X_n &=J_{n-2}(e n)-2 e J_{n-1}(e n)+\frac{2}{n} J_{n}(e n)+2 e J_{n+1}(e n)-J_{n+2}(e n), \nonumber \\
Y_n &=\sqrt{1-e^2} \left[ -J_{n-2}(e n)+2 J_{n}(e n)-J_{n+2}(e n) \right],  \\
Z_n &=\frac{1}{n} J_n(en), \nonumber
\end{align}
are time independent coefficients,
with $e$ being the orbital eccentricity,
$J_n$ being the first-kind Bessel function of order $n$;
$\mathcal S_n^{E, P}$ and $\mathcal C_n^{E, P}$ are time dependent functions
\begin{align}
\label{eq:Sn-Cn}
\mathcal S_n^E=\sin \left(\frac{2n\pi t}{T}+n \varphi^E \right), ~~~~~~~~~~~~~~~~~~
\mathcal C_n^E=\cos \left(\frac{2n\pi t}{T}+n \varphi^E \right),   \nonumber \\
\mathcal S_n^P=\sin \left(\frac{2n\pi t}{T}+n \varphi^P_I \right), ~~~~~~~~~~~~~~~~~~
\mathcal C_n^P=\cos \left(\frac{2n\pi t}{T}+n \varphi^P_I \right),
\end{align}
where $t=0$ denotes the initial observational time,
$T$ is the orbital period of the SMBBH,
$\varphi^E$ denotes the phase of Earth term,
and $\varphi_I^P$ represents the phase of pulsar term (pulsar phase, hereafter).

Given above, if we only detect the Earth-term signal, namely taking an Earth-term-only search,
$9$ parameters should be considered, and they can be included in a $9$-dimensional vector
\begin{align}
\label{eq:parameter-vector-eccentric}
\{ \lambda^a \}=\left\{ T, \mathcal A_m,  \alpha, \delta, \psi, c_i, \varphi^E, \phi_p, e \right\},
\end{align}
with $a=1, 2, ...,9$.
Moreover, if we detect pulsar terms for all MSPs, i.e. taking a full-signal search,
all the phases $\varphi^P_I$ (for $I=1, 2, ..., N_{\rm P}$) should also be considered,
and the total number of parameters in this case is $9+N_{\rm P}$.
The complexity grows as the number of unknown parameters enlarges,
so the full-signal search is not applied in most of the current PTA data analyses \citep{Sesana_and_Vecchio_2010, Ellis_et_al_2012, Ellis_et_al_2012b}.

Specifically, in the circular limit $e \rightarrow 0$, the coefficients $X_n, Y_n, Z_n$ in Eq. \eqref{eq:Xn-Yn-Zn}
approach  $\delta_{n, 2}, -\delta_{n, 2}$ and $0$ respectively,
with $\delta_{n, 2}$ being the Kronecker delta function,
so only the mode $n=2$ leaves,
and the GW in this case is monochromatic, with the GW frequency being twice the orbital frequency $f_{\rm gw}=2/T$.
In addition, in the circular case, the eccentricity $e$ vanishes in the template and the orbital phase of the periastron $\phi_p$ can be absorbed into the phase parameters $\varphi^E$ and $\varphi^P_I$.
As a result, for a circular SMBBH, one needs to consider 7 parameters in the Earth-term-only search (the vector $\{ \lambda^a \}$ in Eq. \eqref{eq:parameter-vector-eccentric} with $a=1, 2, ..., 7$) \citep{Lee_et_al_2011, WMJ1, WMJ2}, and $7+N_{\rm P}$ parameters in the full-signal search \citep{Sesana_and_Vecchio_2010}.

As is mentioned that the template \eqref{eq:SEP-eccentric} is based on a non-evolving assumption,
and if we relax the assumption and consider the evolution of the SMBBH,
the number of parameters will become larger in the full-signal search.
Specifically,
if the periastron precession is considered, the parameter $\phi_p$ in each pulsar term should be different,
so the total number of parameters is $9+2N_{\rm P}$ in this case;
if the orbital reduction due to gravitational radiation is also taken,
the amplitude $\mathcal A_m$, orbital period $T$ and eccentricity $e$ should vary from different pulsar terms,
so $9+5N_{\rm P}$ parameters should be included in this case;
if the spins of the SMBBH are contained as well,
the direction of the orbital plane generally changes \citep{Kidder1995},
and the parameters $c_i$ and $\psi$ in each pulsar term are different,
so $9+7N_{\rm P}$ parameter should be taken into account in this general case.
In conclusion, the full-signal search will become extremely complicated in these generalized cases.

\section{Parameter-Estimation Biases from pulsar terms} \label{Sec: Parameter Estimations}

\subsection{Generalized Likelihood Ratio Test} \label{Sec: GLRT}

Before deriving the PEBs, let us firstly introduce the  method of parameter estimation.
In this article, we adopt the method named Generalized Likelihood Ratio Test (GLRT) (see \cite{WMJ1, WMJ2} for details),
in which we estimate parameters by the detection statistic as follows
\begin{align}
\label{eq:GLRT}
{\rm \bf GLRT} (\{d \})={\mathop {\rm Max} \limits_{\{ \lambda^a \} }} \frac{\mathcal L(\{d \}; H_{ \{ \lambda^a \} })}{\mathcal L(\{d \}; H_0)}
=\frac{\mathcal L(\{d \}; H_{ \{ \hat  \lambda^a \}})}{\mathcal L(\{d \}; H_0)},
\end{align}
where $\mathcal L$ denotes likelihood,
$\{d \}=d_I (t_i)$ (with $I=1, 2, ..., N_{\rm P}$ and $i=1, 2, ..., N_I$) is the observational data set of the PTA,
$t_i$ is the TOA of the $i$-th data point,
$\{\lambda^a \} $ is the parameter vector given by Eq. \eqref{eq:parameter-vector-eccentric},
$\{ \hat  \lambda^a \}$ represents the best fitted parameters from the data,
$H_{\{ \lambda^a \}}$ stands for the hypothesis that the data contains signal with the parameter set $\{ \lambda^a \}$,
and  $H_0$ is the hypothesis that the data only contains noise.

The log-likelihood ratio can be expressed as \citep{Ellis_et_al_2012b, WMJ1, WMJ2}
\begin{align}
\label{eq:GLRT-2}
\log \frac{\mathcal L(\{d \}; H_{ \{ \lambda^a \} })}{\mathcal L(\{d \}; H_0)}
\equiv \Lambda \left(\{ d \}; \{ \lambda^a \} \right)= {\mathop \sum \limits_{I=1}^{N_{\rm P}}} \Lambda_I \left(d_I; \{ \lambda^a \}\right),
\end{align}
with
\begin{align}
\Lambda_I(d_I; \{ \lambda^a \})=\left< d_I \big| \mathcal T_I(\{ \lambda^a \}) \right>_I-\frac{1}{2}\left< \mathcal T_I(\{ \lambda^a \}) \big| \mathcal T_I(\{ \lambda^a \}) \right>_I,
\end{align}
where $\mathcal T_I$ is the template of timing residual induced by the GW for the $I$-th MSP,
and $\left<...\big|...\right>_I$ denotes the noise weighted inner product, which is defined as
$\left<x \big|y \right>_I \equiv {\mathop \sum \limits_{i, j}} x(t_i) (\Sigma_{I}^{-1})_{i j} y(t_j)$,
with $\Sigma_{I}^{ij}$ being the auto-covariance matrix of the noise for the $I$-th MSP \citep{WMJ1}.
In this article, we assume that
the noises are dominated by stationary white Gaussian noises with standard deviation $\sigma_I$,
and in this case the inner product reduces to
\be
\label{eq:inner-product-weak}
\left<x \big|y\right>_I=\left<y \big|x\right>_I=  \frac{1}{\sigma_I^2}{\mathop \sum \limits_{i=1}^{N_I}} x(t_i) y(t_i) \simeq \frac{\int x(t) y(t) dt}{\sigma_I^2 \delta t_I},
\ee
where $\delta t_I$ is the cadence between two data points for the $I$-th MSP.
Taking the partial differentiation of Eq. \eqref{eq:GLRT}, one obtains
\begin{align}
\label{eq:GLRT-3}
\frac{\partial \Lambda}{\partial \lambda^a}={\mathop \sum \limits_{I=1}^{N_{\rm P}}}  \left< d_I-\mathcal T_I ~ \bigg| ~ \frac{\partial \mathcal T_I}{\partial \lambda^a} \right>_I  =0 , ~~~~~~~~~~~~~~ \{ \lambda^a \}= \{ \hat \lambda^a \},
\end{align}
with $\lambda^a$ being the $a$-th element in the parameter vector $\{ \lambda^a \}$ given by Eq. \eqref{eq:parameter-vector-eccentric}.

In the following, we will study the full-signal and Earth-term-only searches in the frame of GLRT respectively.
In the full-signal search, both the Earth term and the pulsar terms are included in the template,
i.e. $\mathcal T_I(t_i)=s^E_I(t_i)-s^P_I(t_i)$,
and from \eqref{eq:GLRT-3} one obtains
\begin{align}
\label{eq:M1}
\frac{\partial \Lambda^{\rm fs}}{\partial \lambda^a}={\mathop \sum \limits_{I=1}^{N_{\rm P}}}  \left< d_I-s^E_I+s^P_I ~ \bigg| ~ \frac{\partial s^E_I}{\partial \lambda^a}-\frac{\partial s^P_I}{\partial \lambda^a} \right>_I  =0 , ~~~~~~~~~~~~~~ \{ \lambda^a \} = \{ \hat \lambda^a_{\rm fs} \},
\end{align}
where the super-/sub-script ``fs" denotes the full-signal search.
Similarly, in the Earth-term-only search, the template only includes
the Earth term $\mathcal T_I (t_i) =s^E_I (t_i)$,
and one yields
\begin{align}
\label{eq:M2}
\frac{\partial \Lambda^{\rm eto}}{\partial \lambda^a}={\mathop \sum \limits_{I=1}^{N_{\rm P}}}  \left< d_I-s_I^E ~ \bigg| ~ \frac{\partial s_I^E}{\partial \lambda^a} \right>_I  =0 , ~~~~~~~~~~~~~~ \{ \lambda^a \} = \{ \hat \lambda^a_{\rm eto} \},
\end{align}
where the super-/sub-script ``eto" denotes the Earth-term-only search.
It is clear that the difference between these two best fitted values $\delta \lambda^a=\hat \lambda^a_{\rm fs}-\hat \lambda^a_{\rm eto}$
is the PEBs caused by the ignored pulsar terms,
which will be derived in the following.

\subsection{Derivations of the Parameter-Estimation Biases}

Combining Eqs. \eqref{eq:M1} and \eqref{eq:M2}, one yields
\be
\label{eq:difference}
\frac{\partial \Lambda^{\rm fs}}{\partial\lambda^a}-\frac{\partial \Lambda^{\rm eto}}{\partial\lambda^a}=\xi_{a},
\ee
with
\[
\xi_{a}={\mathop \sum \limits_{I=1}^{N_{\rm P}}}\left( -\left< d_I-s^E_I+s^P_I \bigg| \frac{\partial s_I^P}{\partial \lambda^a} \right>_I+\left< s^P_I \bigg| \frac{\partial s_I^E}{\partial \lambda^a} \right>_I \right).
\]
Note that the observational data $d_I$ actually consists of Earth term, pulsar term and white noise $n_I$,
i.e. $ d_I(t_i)=s^E_I(t_i)-s^P_I(t_i)+n_I(t_i)$,
and we can assume that the white noise is independent from the pulsar terms $\left<n_I \big| \partial s_I^P/ \partial \lambda^k\right>_I=0$ \footnote{This assumption should be valid at least at the level of astrophysical origins,  since Pulsar terms are GW components from the SMBBH,
while white noises, such as jitter noise and radiometer noise, only depend on the properties of MSPs and telescopes, and are uncorrelated to the SMBBH.}.
Therefore, one obtains
\be
\label{eq:xi-sim}
\xi_{a}= {\mathop \sum \limits_{I=1}^{N_{\rm P}}} \left< s^P_I \bigg| \frac{\partial s_I^E}{\partial \lambda^a} \right>_I.
\ee

In the following, we will derive the PEBs and the analyses will be based on the Earth-term-only search.
At the best fitted value given by the full-signal search, Eq. \eqref{eq:difference} reduces to
\be
\label{eq:difference-2}
\xi_{a}=-\frac{\partial \Lambda^{\rm eto}}{\partial\lambda^a}, ~~~~~~~~~~~~~~~~~~ \{ \lambda^a \}= \{ \hat \lambda^a_{\rm fs} \}.
\ee
Then we expand Eq. \eqref{eq:difference-2} as $ \hat \lambda^a_{\rm fs} =  \hat \lambda^a_{\rm eto} + \delta \lambda^a$,
and assume that the PEBs are small $ \hat \lambda^a_{\rm fs} \simeq \hat \lambda^a_{\rm eto}$.
As a result, the leading order of the left hand side of \eqref{eq:difference-2} is
\[
\xi_a \big|_{ \{ \lambda^a \}= \{ \hat \lambda^a_{\rm fs} \} }=\xi_a \big|_{\{ \lambda^a \}= \{ \hat \lambda^a_{\rm eto} \}}+ \mathcal O (\delta \lambda^a),
\]
and the right hand side is
\[
\frac{\partial \Lambda^{\rm eto}}{\partial\lambda^a} \bigg|_{ \{ \lambda^a \}=\{\hat \lambda^a_{\rm fs} \} }
=\frac{\partial^2 \Lambda^{\rm eto}}{\partial\lambda^a \partial \lambda^b} \bigg|_{ \{ \lambda^a \}= \{ \hat \lambda^a_{\rm eto} \} }  \cdot \delta \lambda^b+ \mathcal O \left[(\delta \lambda^b)^2 \right].
\]
Therefore, we obtain
\be
\label{eq:difference-lead}
\frac{\partial^2 \Lambda^{\rm eto}}{\partial\lambda^a \partial \lambda^b} \delta \lambda^b =-\xi_a, ~~~~~~~~~~~~~~~~~~~~~~~~ \{ \lambda^a \}=\{ \hat \lambda^a_{\rm eto} \}.
\ee
Furthermore, from Eq. \eqref{eq:M2}, we have
\be
\label{eq:M2-2}
\frac{\partial^2 \Lambda^{\rm eto}}{\partial\lambda^a \partial \lambda^b}={\mathop \sum \limits_{I=1}^{N_{\rm P}}}  \left< d_I-s_I^E  \bigg|  \frac{\partial^2 s_I^E}{\partial \lambda^a \partial \lambda^b} \right>_I-{\mathop \sum \limits_{I=1}^{N_{\rm P}}}  \left< \frac{\partial s_I^E}{\partial \lambda^a}  \bigg|  \frac{\partial s_I^E}{\partial \lambda^b} \right>_I.
\ee
It is worth noting that in the analyses based on the Earth-term-only search,
the pulsar term $s_I^P$ is ignored, i.e. $d_I-s_I^E=n_I$,
so the first term on the right hand side of \eqref{eq:M2-2} should vanish, and
one obtains
\be
\label{eq:Fisher}
\frac{\partial^2 \Lambda^{\rm eto}}{\partial\lambda^a \partial \lambda^b}=-{\mathop \sum \limits_{I=1}^{N_{\rm P}}}  \left< \frac{\partial s_I^E}{\partial \lambda^a}  \bigg|  \frac{\partial s_I^E}{\partial \lambda^b} \right>_I \equiv -\Gamma_{a b},
\ee
where $\Gamma_{a b}$ is the so-called Fisher information matrix \citep{Sesana_and_Vecchio_2010}.
As a result, we obtain the PEBs due to the ignored pulsar terms
\be
\label{eq:parameter-estimation-error}
\delta \lambda^a = \Gamma^{a b} \xi_{b}= \Gamma^{a b} {\mathop \sum \limits_{I=1}^{N_{\rm P}}} \left< s^P_I \bigg| \frac{\partial s_I^E}{\partial \lambda^b} \right>_I, ~~~~~~~~~~~~~~~\{ \lambda^a \} = \{ \hat \lambda^a_{\rm eto} \},
\ee
where $\Gamma^{a b}$ is the inversed Fisher information matrix.
Note that the above analytical method is the so called noise-projection technique \citep{Cutler&Flanagan1994, Cutler&Harms2006, Harms_et_al_2008}, and $\xi_a$ in Eq. \eqref{eq:xi-sim} is termed as the noise-projection operator with the noises taken as $s_I^P$.

\subsection{Statistical Properties of the PEBs}

From Eq. \eqref{eq:parameter-estimation-error},
one sees that the PEBs $\delta \lambda^a$ are functions of pulsar terms $s_I^P$,
which contain unknown parameters $\varphi_I^P$, with $I= 1, 2,..., N_{\rm P}$.
In the following, we will take these unknown pulsar phases $\varphi_I^P$ as random variables,
and analyze the statistical signatures of $\delta \lambda^a$.

For the $I$-th MSP, its pulsar phase $\varphi_I^P$ satisfies the relation \citep{WMJ1, WMJ2}
\[
\Delta \varphi_I \equiv \varphi^P_I-\varphi^E=2\pi \frac{(1-\cos \mu_I) L_I}{c T} ,
\]
where $\mu_I$ is the opening angle between the SMBBH and the MSP, $L_I$ the distance from the MSP to Earth,
and $c$ the light speed.
For most MSPs, their distances $L_I$ are poorly measured, with uncertainties much larger than $c T$ (about $10^{-1}-10$ ly) \citep{Verbiest_et_al_2012},
so the phase difference $\Delta \varphi^P_I$ generally has uncertainty much larger than $2\pi$.
For convenience in calculation,
we will restrict the value of $\Delta \varphi^P_I$ within the span $[0, 2\pi)$,
i.e. taking the transformation
\be
\label{eq:transformation}
\Delta \varphi_I \rightarrow 2\pi \left[ \frac{\Delta \varphi_I}{2\pi}-\mathcal N\left( \frac{\Delta \varphi_I}{2\pi}\right) \right],
\ee
where $\mathcal N(x)$ denotes the integer part of $x$.
After this transformation, $\Delta \varphi_I$  can be considered as a random variable 
with a Uniform distribution function $\Delta \varphi_I \in U[0, 2\pi)$.
Equivalently, the pulsar-term phase $\varphi^P_I \equiv \Delta \varphi_I+\varphi^E$,
if taken the same transformation as \eqref{eq:transformation}, should also follow the Uniform distribution $\varphi^P_I \in U[0, 2\pi)$.

The value of each PEB $\delta \lambda^a$,
as a summation of $N_{\rm P}$ random terms according to Eq. \eqref{eq:parameter-estimation-error}, should also be a random variable,
and  it is reasonable to assume that $\delta \lambda^a$ follows a Gaussian distribution function when $N_{\rm P} \gg 1$, due to the central limit theorem.
Hence if we calculate the expected value $\left< \delta \lambda^a \right>$ and the covariance matrix $\Pi^{a b} \equiv \left< \delta \lambda^a \delta \lambda^b \right>$, we can completely fix its probability density function (PDF)
\be
\label{eq: Gaussian}
\mathcal L \left(\{\delta \lambda^a \} \right) \propto \exp \left( -\frac{1}{2} \big( \delta \lambda^b - \left< \delta \lambda^b \right> \big) \Pi^{-1}_{ b c}  \big( \delta \lambda^c - \left< \delta \lambda^c \right> \big) \right),
\ee
with $\Pi^{-1}_{a b}$ being the inverse of $\Pi^{a b}$.

It is not difficult to show that $\delta \lambda^a$ has a vanished expected value, under the assumption $\Delta \varphi_I^P \in U[0,2\pi)$.
From Eq. \eqref{eq:parameter-estimation-error}, one obtains that
\[
\left< \delta \lambda^a \right> =\Gamma^{a b} {\mathop \sum \limits_{I=1}^{N_{\rm P}}} \left< \left< s^P_I \right> \bigg| \frac{\partial s_I^E}{\partial \lambda^b} \right>_I,
\]
where $\left< x \right>=\int^{}_{} d\varphi_I^P \cdot x \cdot U[0,2\pi)=\int d(\Delta \varphi_I) \cdot x \cdot U[0,2\pi) $ denotes the average of $x$ over $\varphi_I^P$ or $\Delta \varphi_I$.
According to Eqs. \eqref{eq:SEP-eccentric} and \eqref{eq:Sn-Cn},
the pulsar term $s^P_I$ consists of a linear combination of $\mathcal S_n^P$ and  $\mathcal C_n^P$,
and they satisfy $\left< \mathcal S_n^P \right>=\left< \mathcal C_n^P \right>=0$.
As a result, one has $\left< \delta \lambda^a \right>=0$.

Therefore, in the following, we only need to express the covariance matrix
\be
\label{eq:parameter-uncertainty}
\Pi^{a b} \equiv \left< \delta \lambda^a \delta \lambda^b \right>=\Gamma^{a c}\Gamma^{b d} \left< \xi_c \xi_d  \right>.
\ee
It is seen that to obtain the covariance matrix $\Pi^{a b}$,
we firstly need to calculate all elements of the Fisher information matrix $\Gamma_{a b}$
and the mean-squared noise-projection matrix $\left< \xi_a \xi_b  \right>$.
The computation details are given in Section \ref{Sec: Computation of the Errors}.

\section{Computation of the Parameter-Estimation Biases} \label{Sec: Computation of the Errors}


From Eqs. \eqref{eq:xi-sim} and \eqref{eq:Fisher},
one can see that the expressions of $\Gamma_{a b}$ and $\left< \xi_{a}\xi_{b} \right>$ should contain terms like $s_I^P$ and $\partial s_I^E/\partial \lambda^a$.
Hence, for convenience, we firstly rewrite the template \eqref{eq:sEP} as
\be
\label{eq:SEP-eccentric-(2)}
s_I^{E, P}={\mathop \sum \limits_{n=1}^{N_I}} A_n \mathcal S_n^{E, P}+B_n \mathcal C_n^{E, P},
\ee
with
\begin{align}
\label{eq:An-Bn}
A_n=&\mathcal A_m \big\{ (\cos 2\psi F_{I +}-\sin 2\psi F_{I \times}) \left[ (1+c_i^2)X_n\cos 2\phi_p+2(1-c_i^2)Z_n\right] -2 (\sin 2\psi F_{I +}+\cos 2\psi F_{I \times})c_i X_n\sin 2\phi_p \big\},       \nonumber \\
B_n=&\mathcal A_m \big[ -(\cos 2\psi F_{I +}-\sin 2\psi F_{I \times})(1+c_i^2) Y_n\sin 2\phi_p -2 (\sin 2\psi F_{I +}+\cos 2\psi F_{I \times})c_i Y_n\cos 2\phi_p \big].
\end{align}
Note that the coefficients $A_n$ and $B_n$ actually vary from MSPs,
and should have the subscript ``$I$", but we ignore it for clarity without confusion.
Furthermore, it is seen that, unlike in the template \eqref{eq:SEP-eccentric}, the summation in \eqref{eq:SEP-eccentric-(2)}
has a maximum mode $N_I=\mathcal N(T/2\delta t_I)$,
which corresponds to the Nyquist frequency \citep{Bracewell2020},
and the higher-frequency modes $n>N_I$ cannot be detected by PTA with observational cadence $\delta t_I$.

The partial derivative of the template can be expressed as
\begin{align}
\label{eq:s_I^E-coefficient-eccentric}
\frac{\partial s_I^E}{\partial \lambda^a}={\mathop \sum \limits_{n=1}^{N_I}} D_{a, n} \mathcal S_n^E+E_{a, n} \mathcal C_n^E,
\end{align}
with the parameter $\lambda^a$ given in Eq. \eqref{eq:parameter-vector-eccentric},
and the coefficients $D_{a, n}$ and $E_{a, n}$ (for $a=1, 2, ...,9$) listed in Appendix \ref{Sec: D_a-E_a-eccentric}.

In the following, we shall derive the matrices  $\Gamma_{a b}$ and  $\left< \xi_a \xi_b \right>$ respectively.

\subsection{Fisher Information Matrix $\Gamma_{a b}$}\label{sec:Fisher-matrix}

From Eq. \eqref{eq:s_I^E-coefficient-eccentric}, one can expand the Fisher matrix as
\[
\Gamma_{a b} ={\mathop \sum \limits_{I=1}^{N_{\rm P}}} \left< \frac{\partial s_I^E}{\partial \lambda^a} \bigg| \frac{\partial s_I^E}{\partial \lambda^{b}} \right>_I ={\mathop \sum \limits_{I=1}^{N_{\rm P}}}{\mathop \sum \limits_{n_1=1}^{N_I}}{\mathop \sum \limits_{n_2=1}^{N_I}} \left< D_{a, n_1} \mathcal S^E_{n_1}+E_{a, n_1} \mathcal C^E_{n_1}  \big| D_{b, n_2} \mathcal S^E_{n_2}+E_{b, n_2} \mathcal C^E_{n_2} \right>_I.
\]
From the formulae of $D_{a, n}$ and $E_{a, n}$ given in Appendix \ref{Sec: D_a-E_a-eccentric},
it is seen that for the case $a \neq 1$,
the coefficients $D_{a, n}$ and $E_{a, n}$ are time independent,
and can be moved out of $\left< ... | ... \right>_I$.
However, for the case $a=1$, both coefficients $D_{1, n}$ and $E_{1, n}$ are proportional to the time $t$, and it should be left inside of $\left< ... | ... \right>_I$.
As a result, the Fisher matrix should be divided into three parts ---
the ``$11$"-component, ``$1a$"(or ``$a1$")-components (with $a \neq 1$) and ``$a b$"-components (with both $a \neq 1$ and $b \neq 1$).
The resulting formulae are
\begin{align}
\label{eq:Gamma_11-eccentric}
\Gamma_{1 1} =\frac{4\pi^2}{T^4}{\mathop \sum \limits_{I=1}^{N_{\rm P}}}{\mathop \sum \limits_{n_1=1}^{N_I}}{\mathop \sum \limits_{n_2=1}^{N_I}}  &
n_1 n_2  \bigg\{  B_{n_1}B_{n_2} \left<t \mathcal S^E_{n_1} \big| t \mathcal S^E_{n_2} \right>_I
-B_{n_1}A_{n_2}  \left<t \mathcal S^E_{n_1} \big| t \mathcal C^E_{n_2} \right>_I                            \\
&-A_{n_1}B_{n_2} \left<t\mathcal C^E_{n_1} \big| t\mathcal S^E_{n_2} \right>_I
+A_{n_1}A_{n_2} \left<t \mathcal C^E_{n_1} \big| t \mathcal C^E_{n_2} \right>_I \bigg\},  \nonumber
\end{align}
\begin{align}
\label{eq:Gamma_1a-eccentric}
\Gamma_{1 a}=\Gamma_{a 1} =\frac{2\pi}{T^2}{\mathop \sum \limits_{I=1}^{N_{\rm P}}}{\mathop \sum \limits_{n_1=1}^{N_I}}{\mathop \sum \limits_{n_2=1}^{N_I}}
n_2 \bigg\{
D_{a, n_1} B_{n_2} \left<\mathcal S^E_{n_1} \big| t \mathcal S^E_{n_2} \right>_I
-D_{a, n_1}A_{n_2} \left<\mathcal S^E_{n_1} \big| t \mathcal C^E_{n_2} \right>_I    \nonumber \\
+E_{a, n_1}B_{n_2} \left<\mathcal C^E_{n_1} \big| t \mathcal S^E_{n_2} \right>_I
-E_{a, n_1}A_{n_2} \left<\mathcal C^E_{n_1} \big| t \mathcal C^E_{n_2} \right>_I \bigg\},
\end{align}
for $a \neq 1$, and
\begin{align}
\label{eq:Gamma_ab-eccentric}
\Gamma_{a b} ={\mathop \sum \limits_{I=1}^{N_{\rm P}}}{\mathop \sum \limits_{n_1=1}^{N_I}}{\mathop \sum \limits_{n_2=1}^{N_I}} \bigg\{
 D_{a, n_1}D_{b, n_2} \left<\mathcal S^E_{n_1} \big| \mathcal S^E_{n_2} \right>_I
+D_{a, n_1}E_{b, n_2} \left<\mathcal S^E_{n_1} \big| \mathcal C^E_{n_2} \right>_I    \nonumber \\
+E_{a, n_1}D_{b, n_2} \left<\mathcal C^E_{n_1} \big| \mathcal S^E_{n_2} \right>_I
+E_{a, n_1}E_{b, n_2} \left<\mathcal C^E_{n_1} \big| \mathcal C^E_{n_2} \right>_I \bigg\},
\end{align}
for both $a \neq 1$ and $b \neq 1$.

Since $\mathcal S^E_{n}$ and $\mathcal C^E_{n}$ are trigonometric functions of time, their inner products can be analytically calculated from Eq. \eqref{eq:inner-product-weak}, if the white noise $\sigma_I$ and cadence $\delta t_I$ are time independent.
The formulae of all the inner-product terms in Eqs. \eqref{eq:Gamma_11-eccentric}-\eqref{eq:Gamma_ab-eccentric} are listed in Appendix \ref{Sec: eccentric-time-evolving}.


\subsection{Mean-Squared Noise-Projection Matrix $\left< \xi_a \xi_b  \right>$} \label{Sec:Noise-Projection Matrix}

From Eq. \eqref{eq:xi-sim}, we can express the matrix as
\begin{align}
\label{eq:xi_k^2(1)}
\left< \xi_{a}\xi_{b} \right> &= {\mathop \sum \limits_{I=1}^{N_{\rm P}}}{\mathop \sum \limits_{I'=1}^{N_{\rm P}}} \left< \left< s^P_I \bigg| \frac{\partial s_I^E}{\partial \lambda^a} \right>_I  \left< s^P_{I'} \bigg| \frac{\partial s_{I'}^E}{\partial \lambda^b} \right>_{I'}  \right>        \\
&={\mathop \sum \limits_{I=1}^{N_{\rm P}}} \left<  \left< s^P_I \bigg| \frac{\partial s_I^E}{\partial \lambda^a} \right>_I  \left< s^P_{I} \bigg| \frac{\partial s_{I}^E}{\partial \lambda^b} \right>_{I} \right>
+ {\mathop \sum \limits_{I=1}^{N_{\rm P}}}{\mathop \sum \limits_{I' \neq I}^{N_{\rm P}}} \left< \left< s^P_I \bigg| \frac{\partial s_I^E}{\partial \lambda^a} \right>_I  \left< s^P_{I'} \bigg| \frac{\partial s_{I'}^E}{\partial \lambda^b} \right>_{I'}  \right>.   \nonumber
\end{align}
For the case $I' \neq I$, the pulsar terms $s^P_I$ and $s^P_{I'}$ should be independent from each other,
so the second term on the right hand side of \eqref{eq:xi_k^2(1)} (in the 2nd row) should vanish,
and one obtains a simplified formula
\be
\label{eq:xi_k^2(2)}
\left< \xi_{a}\xi_{b} \right> = {\mathop \sum \limits_{I=1}^{N_{\rm P}}} \left<  \left< s^P_I \bigg| \frac{\partial s_I^E}{\partial \lambda^a} \right>_I  \left< s^P_{I} \bigg| \frac{\partial s_{I}^E}{\partial \lambda^b} \right>_{I} \right>.
\ee
Furthermore, taking Eqs. \eqref{eq:SEP-eccentric-(2)} and \eqref{eq:s_I^E-coefficient-eccentric} into \eqref{eq:xi_k^2(2)}, we obtain
\begin{align}
\label{eq:xi_k^2(3)}
 \left< \xi_{a}\xi_{b} \right>  = & {\mathop \sum \limits_{I=1}^{N_{\rm P}}}    {\mathop \sum \limits_{n_1=1}^{N_I}}{\mathop \sum \limits_{n_2=1}^{N_I}} {\mathop \sum \limits_{n_3=1}^{N_I}}{\mathop \sum \limits_{n_4=1}^{N_I}} \nonumber \\
& \left<  \left< A_{n_1} \mathcal S_{n_1}^{P}+B_{n_1} \mathcal C_{n_1}^{P} \bigg| D_{a, n_2} \mathcal S_{n_2}^E+E_{a, n_2} \mathcal C_{n_2}^E \right>_I  \left< A_{n_3} \mathcal S_{n_3}^{P}+B_{n_3} \mathcal C_{n_3}^{P} \bigg| D_{a, n_4} \mathcal S_{n_4}^E+E_{a, n_4} \mathcal C_{n_4}^E \right>_{I} \right>.
\end{align}
For the same reason as we calculate $\Gamma_{a b}$ in Section \ref{sec:Fisher-matrix}, the matrix $\left< \xi_a \xi_b  \right>$ should also be divided into three cases,
and the resulting formulae are shown in Eqs. \eqref{eq:xi_1-xi_1-eccentric}-\eqref{eq:xi_a-xi_b-eccentric}.

In the following, we will calculate the averaged terms $\left<...\right>$ in Eqs. \eqref{eq:xi_1-xi_1-eccentric}-\eqref{eq:xi_a-xi_b-eccentric}.
Recall that $\left< x \right>=\int d(\Delta \varphi_I^P) \cdot x \cdot U[0, 2\pi)$,
so one should firstly express $\mathcal S_n^P$ and $\mathcal C_n^P$ as functions of $\mathcal S_n^E$, $\mathcal C_n^E$ and $\Delta \varphi_I^P$.
From Eq. \eqref{eq:Sn-Cn}, one has
\begin{align}
\label{eq:SCP-SCE-eccentric}
\mathcal S_{n}^P= \cos(n \Delta \varphi_I) \mathcal S_{n}^E+\sin(n \Delta \varphi_I)\mathcal C_{n}^E , ~~~~~~~~~~~~
\mathcal C_{n}^P= \cos(n \Delta \varphi_I) \mathcal C_{n}^E -\sin(n \Delta \varphi_I)\mathcal S_{n}^E,
\end{align}
so the parameter $\Delta \varphi_I^P$  in Eqs. \eqref{eq:xi_1-xi_1-eccentric}-\eqref{eq:xi_a-xi_b-eccentric} can be
moved out of $\left< ... | ... \right>_I$, for example
\begin{align}
\left<\mathcal S_{n_1}^P \big|\mathcal  S_{n_2}^E \right>_I = \cos(n_1 \Delta \varphi_I)\left<\mathcal S_{n_1}^E\big| \mathcal S_{n_2}^E \right>_I+\sin(n_1 \Delta \varphi_I)\left<\mathcal C_{n_1}^E\big| \mathcal S_{n_2}^E \right>_I,   \nonumber
\end{align}
and all the other inner-product terms in \eqref{eq:xi_1-xi_1-eccentric}-\eqref{eq:xi_a-xi_b-eccentric} are rewritten in Eq. \eqref{eq:SnP-SnE}.
One can see that the matrix $\left< \xi_{a}\xi_b \right>$ has a quadratic form of the trigonometric functions of $\Delta \varphi_I$,
and their expected values can be easily calculated
\begin{align}
\label{eq:trigonometric-eccentric}
\left<\sin(n_1 \Delta \varphi_I)\sin(n_3  \Delta \varphi_I) \right> =\left<\cos(n_1  \Delta \varphi_I)\cos(n_3  \Delta \varphi_I) \right>=\frac{1}{2}\delta_{n_1, n_3},  ~~~~~~~~~~~
\left<\sin(n_1  \Delta \varphi_I)\cos(n_3  \Delta \varphi_I) \right> =0.
\end{align}
After calculating all the averaged terms in Eqs. \eqref{eq:xi_1-xi_1-eccentric}-\eqref{eq:xi_a-xi_b-eccentric} (see Appendix \ref{Sec: eccentric-expected} for details),
we finally obtain the elements of the matrix listed in \eqref{eq:xi_1-xi_1-eccentric-(2)}-\eqref{eq:xi_a-xi_b-eccentric-(2)}.

In conclusion of this part,
we can calculate the mean-squared noise-projection matrix $\left< \xi_a \xi_b \right>$ from Eqs. \eqref{eq:xi_1-xi_1-eccentric-(2)}-\eqref{eq:xi_a-xi_b-eccentric-(2)},
where the coefficients $A_n$, $B_n$ are given by \eqref{eq:An-Bn},
$D_{a, n}$ and $E_{a, n}$ listed in Appendix \ref{Sec: D_a-E_a-eccentric},
and all the inner-product terms are shown in Appendix \ref{Sec: eccentric-time-evolving}.

\subsection{Simplification: Leading-Order Expressions}

With the Fisher matrix $\Gamma_{a b}$ given by Eqs. \eqref{eq:Gamma_11-eccentric}-\eqref{eq:Gamma_ab-eccentric}
and the mean-squared noise-projection matrix by \eqref{eq:xi_1-xi_1-eccentric-(2)}-\eqref{eq:xi_a-xi_b-eccentric-(2)},
the covariance matrix of the PEBs $\Pi^{a b}$ can be computed from Eq. \eqref{eq:parameter-uncertainty}.
However,  one can see that the formulae of the inner-product terms in Appendix \ref{Sec: eccentric-time-evolving} are very complicated,
so in the following we will simplify these results for the efficient evaluation of  $\Pi^{a b}$,
by keeping them up to leading orders, in the limit $t \gg T$.

For example, from Eq. \eqref{eq:inner-product-eccentric-(1)}, we can see that $\left<\mathcal S_{n_1}^E\big| \mathcal S_{n_2}^E \right>_I $ contains a linearly growing term ($\propto t$, when $n_1=n_2$), oscillating terms (e.g. terms $\propto \mathcal S_{n_1 \pm n_2}^E$ ) and time-independent terms (e.g. terms $\propto \sin[(n_1\pm n_2)\varphi^E]$).
In the limit $t \gg T$, the growing term dominates the results, so all the other terms can be dropped,
and one obtains $\left<\mathcal S_{n_1}^E\big| \mathcal S_{n_2}^E \right>_I^{\rm lead} \propto t $.
As a result, we can express all the simplified formulae of the inner products in Appendix \ref{Sec: eccentric-time-evolving} as follows
\begin{align}
\label{eq:leading}
\left<\mathcal S_{n_1}^E\big| \mathcal S_{n_2}^E \right>_I^{\rm lead}=\left<\mathcal C_{n_1}^E\big| \mathcal C_{n_2}^E \right>_I^{\rm lead} =\frac{t}{2\sigma_I^2 \delta t_I} \delta_{n_1,n_2},  ~~~~~~~~~
\left<\mathcal S_{n_1}^E\big| \mathcal C_{n_2}^E \right>_I^{\rm lead}=\left<\mathcal C_{n_1}^E\big| \mathcal S_{n_2}^E \right>_I^{\rm lead}=0,  \nonumber \\
\left<\mathcal S_{n_1}^E\big| t \mathcal S_{n_2}^E \right>_I^{\rm lead}=\left<\mathcal C_{n_1}^E\big| t \mathcal C_{n_2}^E \right>_I^{\rm lead} =\frac{t^2}{4\sigma_I^2 \delta t_I} \delta_{n_1,n_2},   ~~~~~~~~
\left<\mathcal S_{n_1}^E\big| t \mathcal C_{n_2}^E \right>_I^{\rm lead}=\left<\mathcal C_{n_1}^E\big| t \mathcal S_{n_2}^E \right>_I^{\rm lead}=0,   \\
\left<t \mathcal S_{n_1}^E\big| t \mathcal S_{n_2}^E \right>_I^{\rm lead}=\left<t \mathcal C_{n_1}^E\big| t \mathcal C_{n_2}^E \right>_I^{\rm lead} =\frac{t^3}{6\sigma_I^2 \delta t_I} \delta_{n_1,n_2},   ~~~~~
\left<t \mathcal S_{n_1}^E\big| t \mathcal C_{n_2}^E \right>_I^{\rm lead}=\left<t \mathcal C_{n_1}^E\big| t \mathcal S_{n_2}^E \right>_I^{\rm lead}=0.  \nonumber
\end{align}
It is worth noting that, the vanished terms in \eqref{eq:leading} actually have non-zero results: $\left< \mathcal S_{n_1}^E\big| \mathcal C_{n_2}^E \right>_I^{\rm lead} \rightarrow {\rm const}$, $\left< \mathcal S_{n_1}^E\big| t \mathcal C_{n_2}^E \right>_I^{\rm lead} \propto t$, and $\left< \mathcal S_{n_1}^E\big| \mathcal C_{n_2}^E \right>_I^{\rm lead} \propto t^2$,
and we ignore them because they grow slower than their congeners ($\left< \mathcal S_{n_1}^E\big| \mathcal S_{n_2}^E \right>_I^{\rm lead}$, $\left< \mathcal S_{n_1}^E\big| t \mathcal S_{n_2}^E \right>_I^{\rm lead}$ and $\left< t \mathcal S_{n_1}^E\big| t \mathcal S_{n_2}^E \right>_I^{\rm lead}$ respectively).

Taking Eq. \eqref{eq:leading} into Eqs. \eqref{eq:Gamma_11-eccentric}-\eqref{eq:Gamma_ab-eccentric}, we obtain the simplified expressions for the Fisher matrix
\begin{align}
\label{eq:leading-Gamma-eccentric}
\begin{cases}
\Gamma_{1 1}^{\rm lead} = {\mathop \sum \limits_{I=1}^{N_{\rm P}}} {\mathop \sum \limits_{n=1}^{N_I}}
\frac{8 \pi^2 t^3}{3 T^4 \sigma_I^2\delta t_I}  (A_{n}^2+B_{n}^2),  & ~\\
\Gamma_{1 a}^{\rm lead}=    {\mathop \sum \limits_{I=1}^{N_{\rm P}}}  {\mathop \sum \limits_{n=1}^{N_I}}
 \frac{\pi t^2}{T^2 \sigma_I^2\delta t_I}  (B_n D_{a, n} -A_n E_{a, n}),  & a \neq 1\\
 \Gamma_{a b}^{\rm lead} = {\mathop \sum \limits_{I=1}^{N_{\rm P}}} {\mathop \sum \limits_{n=1}^{N_I}}
\frac{t}{2\sigma_I^2\delta t_I}
( D_{a, n}D_{b, n}+E_{a, n}E_{b, n} ),  &  a \neq 1 ~\&~ b \neq 1
\end{cases},
\end{align}
and taking \eqref{eq:leading} into Eqs. \eqref{eq:xi_1-xi_1-eccentric-(2)}-\eqref{eq:xi_a-xi_b-eccentric-(2)}, we yield the simplified mean-squared noise-projection matrix
\begin{align}
\label{eq:leading-xixi-eccentric}
\begin{cases}
\left< \xi_1 \xi_1 \right>^{\rm lead} = \frac{1}{2}  {\mathop \sum \limits_{I=1}^{N_{\rm P}}} {\mathop \sum \limits_{n=1}^{N_I}}
\left(\frac{\pi t^2}{ T^2 \sigma_I^2\delta t_I}\right)^2 (A_{n}^2+B_{n}^2)^2,          & ~ \\
\left< \xi_1 \xi_a \right>^{\rm lead} = \frac{1}{2} {\mathop \sum \limits_{I=1}^{N_{\rm P}}} {\mathop \sum \limits_{n=1}^{N_I}}
\frac{\pi t^3}{2 T^2 \sigma_I^4\delta t_I^2}  (A_{n}^2+B_{n}^2)(B_n D_{a, n}- A_n E_{a, n}),   & a \neq 1    \\
\left< \xi_a \xi_b \right>^{\rm lead} = \frac{1}{2} {\mathop \sum \limits_{I=1}^{N_{\rm P}}} {\mathop \sum \limits_{n=1}^{N_I}}
 \left(\frac{t}{2\sigma_I^2\delta t_I}\right)^2 (A_{n}^2+B_{n}^2)(D_{a, n}D_{b, n} +E_{a, n}E_{b, n} ),    & a \neq 1 ~\&~ b \neq 1
\end{cases}.
\end{align}

Finally, the leading-order formula of covariance matrix of the PEBs $\Pi^{a b}$ is
\be
\label{eq:parameter-uncertainty-lead}
\left(\Pi^{a b} \right)^{\rm lead}  = (\Gamma^{a c})^{\rm lead}(\Gamma^{b d})^{\rm lead} \left< \xi_c \xi_d  \right>^{\rm lead},
\ee
where $(\Gamma^{a b})^{\rm lead}$ is the inverse of $\Gamma_{a b}^{\rm lead}$.

\subsection{Circular Case}\label{Sec: circular}

In this part, we will formulate the PEBs for the circular case $e=0$.

As is mentioned in Section \ref{Sec:Templates},
the orbital phase of the periastron $\phi_P$ can be absorbed by the phases $\varphi^E$ or $\varphi^P_I$ in this case,
so we can take $\phi_P=0$ in the following analyses.
Therefore, one needs to consider only 7 parameters in the Earth-term-only search,
i.e. the parameter vector $\{\lambda^a\}$ given by \eqref{eq:parameter-vector-eccentric} should be considered only for $a=1, 2, ...,7$.

Furthermore, only the $n=2$ mode survives in the formulae when $e=0$, so the template in Eqs. \eqref{eq:SEP-eccentric-(2)} and  \eqref{eq:s_I^E-coefficient-eccentric} reduces to
\be
\label{eq:template-rewritten-circular}
s_I^{E, P}=  A   \mathcal S^{E, P}+B  \mathcal C^{E, P},
\ee
and
\begin{align}
\label{eq:s_I^E-coefficient-circular}
\frac{\partial s_I^E}{\partial \lambda^a}=D_{a} \mathcal S^E+E_{a} \mathcal C^E,
\end{align}
where $A$, $B$, $D_a$, $E_a$, $\mathcal S^{E, P}$ and $\mathcal C^{E, P}$ denote $A_2$, $B_2$, $D_{a, 2}$, $E_{a, 2}$, $\mathcal S^{E, P}_2$ and $\mathcal C^{E, P}_2$ respectively, and we default the subscript ``$2$" in the analyses of the circular case.
The coefficients $A$, $B$, $D_a$ and $E_a$ are formulated in Appendix \ref{Sec: D_a-E_a}.

As a result, the Fisher information matrix $\Gamma_{a b}$ in Eqs. \eqref{eq:Gamma_11-eccentric}-\eqref{eq:Gamma_ab-eccentric} reduces to
\begin{align}
\label{eq:Gamma_11-circular}
\Gamma_{1 1} = \frac{16 \pi^2}{T^4}{\mathop \sum \limits_{I=1}^{N_{\rm P}}}  B^2 \left<t \mathcal S^E \big| t \mathcal S^E \right>_I
-AB \left< t \mathcal S^E  \big| t \mathcal C^E \right>_I
-AB \left< t \mathcal C^E \big| t \mathcal S^E \right>_I
+A^2 \left< t \mathcal C^E \big| t \mathcal C^E \right>_I,
\end{align}
\begin{align}
\label{eq:Gamma_1a-circular}
\Gamma_{a 1} =\Gamma_{1 a}=\frac{4\pi}{T^2}{\mathop \sum \limits_{I=1}^{N_{\rm P}}}  B D_{a} \left<\mathcal S^E \big| t \mathcal S^E \right>_I
-A D_{a} \left<\mathcal S^E  \big| t \mathcal C^E \right>_I
+B E_{a} \left<\mathcal C^E \big| t \mathcal S^E \right>_I
-A E_{a} \left<\mathcal C^E \big| t \mathcal C^E \right>_I,
\end{align}
for $a \neq 1$,
\begin{align}
\label{eq:Gamma_ab-circular}
\Gamma_{a b} ={\mathop \sum \limits_{I=1}^{N_{\rm P}}}  D_{a}D_{b} \left<\mathcal S^E \big| \mathcal S^E \right>_I
+D_{a}E_{b} \left<\mathcal S^E  \big| \mathcal C^E \right>_I
+E_{a}D_{b} \left<\mathcal C^E \big| \mathcal S^E \right>_I
+E_{a}E_{b} \left<\mathcal C^E \big| \mathcal C^E \right>_I,
\end{align}
for both $a \neq 1$ and $b \neq 1$,
and the mean-squared noise-projection matrix $\left< \xi_a \xi_b \right>$ given by \eqref{eq:xi_1-xi_1-eccentric-(2)}-\eqref{eq:xi_a-xi_b-eccentric-(2)} reduces to Eqs. \eqref{eq:xi_1-xi_1-circular-(2)}-\eqref{eq:xi_a-xi_b-circular-(2)}.
Given above, we can calculate the matrices $\Gamma_{a b}$ and $\left< \xi_a \xi_b \right>$ for the circular case, from Eqs. \eqref{eq:Gamma_11-circular}-\eqref{eq:Gamma_ab-circular} and \eqref{eq:xi_1-xi_1-circular-(2)}-\eqref{eq:xi_a-xi_b-circular-(2)},
with all the inner-product terms therein given in Appendix \ref{Sec: circular-time-evolving}.

\begin{figure}[h]
\begin{center}
\includegraphics[width=0.8\textwidth]{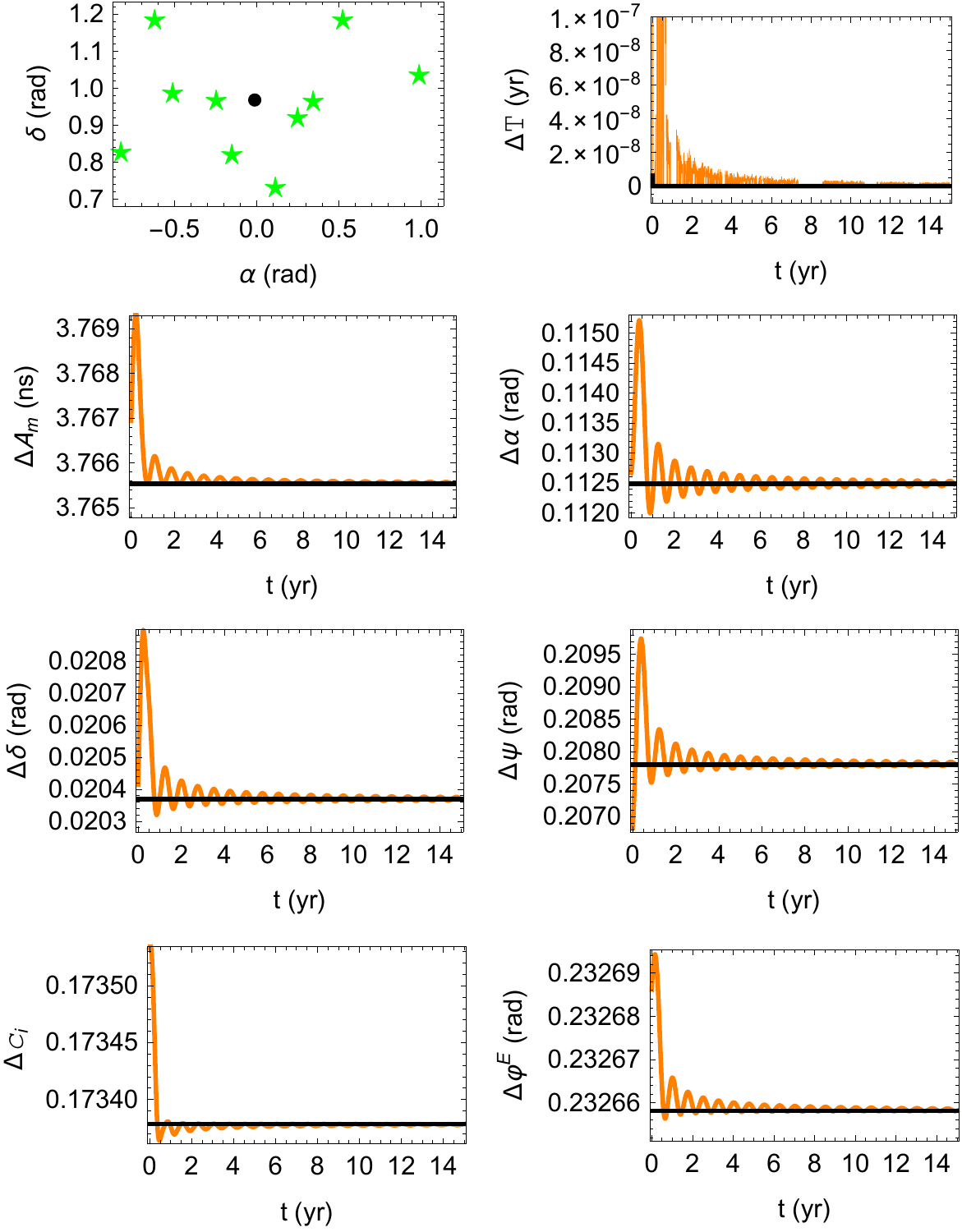}
\caption{The top-left panel shows the sky locations of the injected GW source (black dot) and 10 MSPs (green stars),
and the remaining panels illustrate the evolutionary behaviors of $\Delta \lambda^a$ (for $a=1, 2, ..., 7$ respectively),
with the orange curves representing the results given by the full formulae \eqref{eq:Gamma_11-circular}-\eqref{eq:Gamma_ab-circular} and \eqref{eq:xi_1-xi_1-circular-(2)}-\eqref{eq:xi_a-xi_b-circular-(2)},
and the black-thick lines standing for the results from leading-order approximation \eqref{eq:Gamma_ab-circular-lead}-\eqref{eq:xi_a-xi_b-circular-lead}.
The white noise for the 10 MSPs is $100$ ns,
and the injected GW parameters are: $ T=3 {~\rm yr}, \mathcal A_m=10 {~\rm ns}, \psi=-45^\circ, c_i=0.3, ~{\rm and}~ \varphi^E=1.0 {~\rm rad} $.
Note that, the $\Delta T$ curve on the right-top panel actually represents the numerical error, and one can effectively yield $\Delta T=0$.}
\label{fig:PEE-circular}
\end{center}
\end{figure}

Furthermore, the leading-order formulae of $\Gamma_{a b}$ and $\left< \xi_a \xi_b \right>$ in \eqref{eq:leading-Gamma-eccentric} and  \eqref{eq:leading-xixi-eccentric} reduce to
\begin{align}
\label{eq:Gamma_ab-circular-lead}
\begin{cases}
\Gamma_{1 1}^{\rm lead} = {\mathop \sum \limits_{I=1}^{N_{\rm P}}} \frac{8 \pi^2 t^3}{3 T^4 \sigma_I^2\delta t_I}  (A^2+B^2),  & ~\\
\Gamma_{1 a}^{\rm lead} = {\mathop \sum \limits_{I=1}^{N_{\rm P}}} \frac{\pi t^2}{T^2 \sigma_I^2\delta t_I}  (B D_{a} -A E_{a}),     &   a \neq 1\\
\Gamma_{a b}^{\rm lead} = {\mathop \sum \limits_{I=1}^{N_{\rm P}}}\frac{t}{2\sigma_I^2\delta t_I}( D_{a}D_{b}+E_{a}E_{b} ),        & a \neq 1 ~ \& ~ b \neq 1
\end{cases}
\end{align}
and
\begin{align}
\label{eq:xi_a-xi_b-circular-lead}
\begin{cases}
\left< \xi_1 \xi_1 \right>^{\rm lead} = \frac{1}{2}  {\mathop \sum \limits_{I=1}^{N_{\rm P}}}
\left(\frac{\pi t^2}{ T^2 \sigma_I^2\delta t_I}\right)^2 (A^2+B^2)^2,    & ~   \\
\left< \xi_1 \xi_a \right>^{\rm lead} = \frac{1}{2}  {\mathop \sum \limits_{I=1}^{N_{\rm P}}}
\frac{\pi t^3}{2 T^2 \sigma_I^4\delta t_I^2} (A^2+B^2)(B D_a- A E_a),       & a \neq 1     \\
\left< \xi_a \xi_b \right>^{\rm lead} = \frac{1}{2} {\mathop \sum \limits_{I=1}^{N_{\rm P}}}
 \left(\frac{t}{2\sigma_I^2\delta t_I}\right)^2 (A^2+B^2)(D_a D_b +E_a E_b),   & a \neq 1 ~ \& ~ b \neq 1
\end{cases}.
\end{align}

The time evolutionary behaviors of standard deviations of the PEBs $\Delta \lambda^a \equiv \sqrt{\Pi^{aa}}$ (for $a=1, 2, ..., 7$) are illustrated in Figure \ref{fig:PEE-circular},
and both the full results (from Eqs. \eqref{eq:Gamma_11-circular}-\eqref{eq:Gamma_ab-circular} and \eqref{eq:xi_1-xi_1-circular-(2)}-\eqref{eq:xi_a-xi_b-circular-(2)}) and the leading-order ones (from Eqs. \eqref{eq:Gamma_ab-circular-lead}-\eqref{eq:xi_a-xi_b-circular-lead}) are plotted as a comparison.
It is seen that, after an orbital period $t>T$ ($3$ yr in this case),
the full and leading-order results almost overlap,
which implies that the leading-order approximation is quite accurate.
Furthermore, we can conclude that $\Delta T=0$,
and the other PEBs $\Delta \lambda^a$ ($a \neq 1$) approach constants when $t \gg T$.
This result indicates that pulsar terms do not practically affect the measurement of $T$ when the observational time is long enough,
but may have strong impacts on the measurements of the other parameters.
For example, the PEB of the amplitude in Figure \ref{fig:PEE-circular} is $\Delta \mathcal A_m \simeq 3.766$ ns,
with the true amplitude being $\mathcal A_m=10$ ns,
so the corresponding relative deviation is as large as $\Delta \mathcal A_m/\mathcal A_m \simeq 37.66\%$.
For the result $\Delta T =0$, we infer that this is because we have taken a non-evolving assumption in our analyses,
namely the pulsar terms have the same orbital period as the Earth term, so they have no bias in the measurement of $T$.
We predict that if the non-evolving assumption is eliminated, one will have $\Delta T \neq 0$ (as will be shown in Section \ref{Sec:CommonNoise}).
In addition, we argue that even if $\Delta T  \neq 0$, its value should decay soon.
This is because from \eqref{eq:parameter-uncertainty}, one yields
\[
\Delta T =\sqrt{\Gamma^{1 1}\Gamma^{1 1} \left< \xi_1 \xi_1  \right>+\Gamma^{1 1}\Gamma^{a 1} \left< \xi_1 \xi_a  \right>+\Gamma^{1 a}\Gamma^{1 b} \left< \xi_a \xi_b  \right> },
\]
for both $a \neq 1$ and $b \neq 1$,
where $\left( \Gamma^{1 1} \right)^{\rm lead} \propto t^{-3}$, $\left( \Gamma^{a 1} \right)^{\rm lead} \propto t^{-2}$, $\left( \Gamma^{a b} \right)^{\rm lead} \propto t^{-1}$, $\left< \xi_1 \xi_1  \right>^{\rm lead} \propto t^4$, $\left< \xi_1 \xi_a  \right>^{\rm lead} \propto t^3$ and $\left< \xi_a \xi_b  \right>^{\rm lead} \propto t^2$ according to \eqref{eq:Gamma_ab-circular-lead}-\eqref{eq:xi_a-xi_b-circular-lead},
so finally we obtain a decaying result $(\Delta T)^{\rm lead} \propto t^{-1}$.
Similarly,
we have
\[
\Delta \lambda^a =\sqrt{\Gamma^{a 1}\Gamma^{a 1} \left< \xi_1 \xi_1  \right>+\Gamma^{a 1}\Gamma^{a b} \left< \xi_1 \xi_b  \right>+\Gamma^{a b}\Gamma^{a c} \left< \xi_b \xi_c  \right> },
\]
for $a \neq 1$, $b \neq 1$ and $c \neq 1$,
and from \eqref{eq:Gamma_ab-circular-lead}-\eqref{eq:xi_a-xi_b-circular-lead}, we can obtain $\Delta \lambda^a \rightarrow {\rm const}$ (for $a \neq 1$), as is shown above.

\section{Tests and Applications}\label{Sec:Applications}
\subsection{Sky Localization Biases of a Circular SMBBH}\label{sec: Zhu-test}

\begin{figure}[h]
\begin{center}
\includegraphics[width=0.5\textwidth]{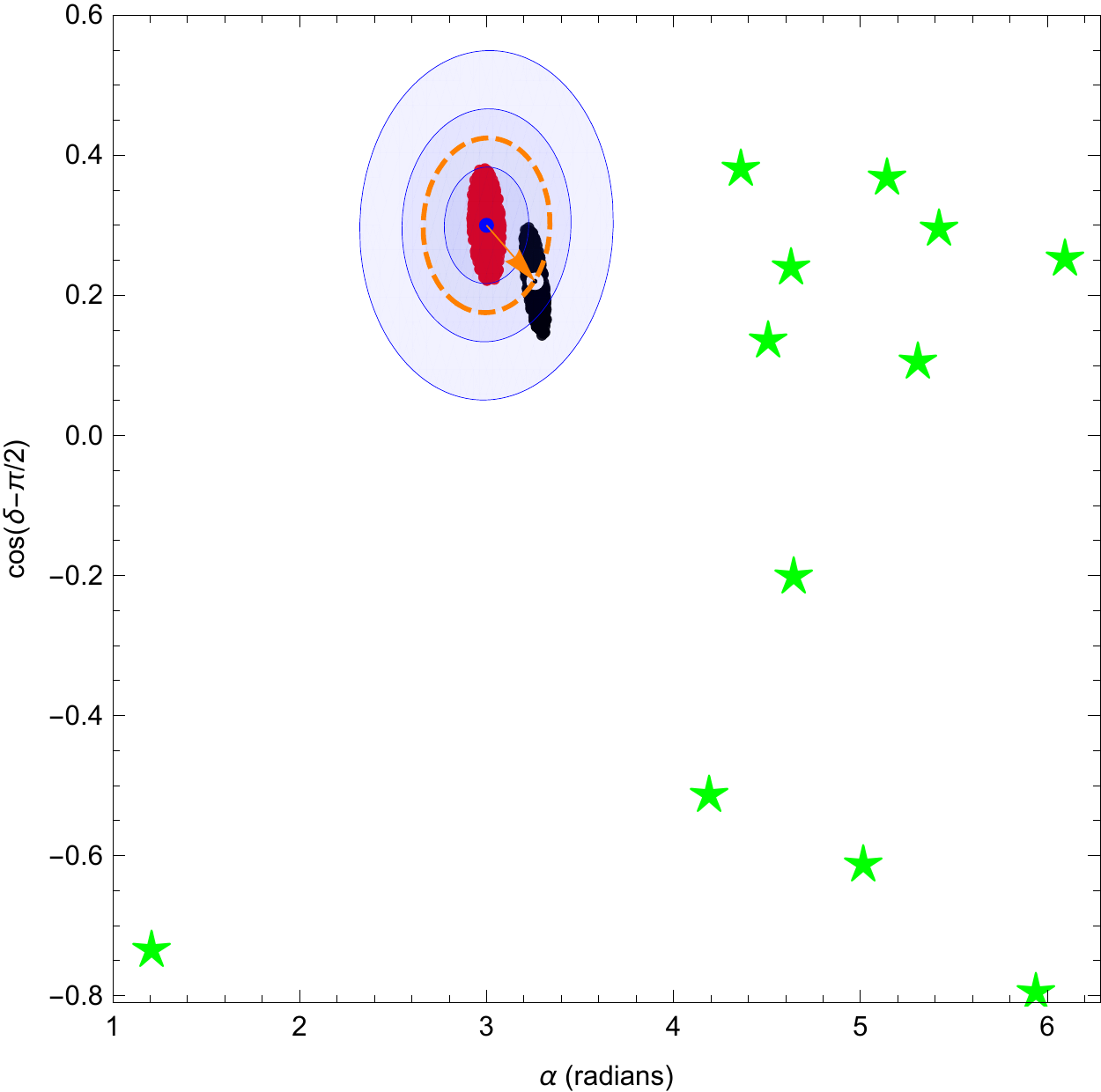}
\caption{Sky localization and its PEBs of a circular SMBBH,
where the blue dot denotes the best fitted location [$\hat \alpha_{\rm eto}=3$, $\cos(\hat \delta_{\rm eto}-\pi/2)=0.3$],
the blue contours represent the $1, 2, 3-\sigma$ regions of the PEBs from Eq. \eqref{eq:2-dim-likelihood},
the orange arrow indicates the true PEBs given by \cite{Zhu_et_al_2016},
the orange-dashed curve stands for the ``$1.5\sigma$" region,
the  filled white circle marks the true SMBBH location (as well as the best fitted location in the full-signal search),
and the red and black scatter regions sketch the parameter uncertainties due to the MSP white noise in Earth-term-only and full-signal searches respectively,
and the sky locations of the 12 MSPs are marked by green ``$\star$".
The estimated source locations for both earth-term only search and full signal search are reproduced from \cite{Zhu_et_al_2016}.}
\label{fig:Zhu-strong}
\end{center}
\end{figure}

In Section \ref{Sec: Computation of the Errors}, we have presented the formulae of the PEBs caused by pulsar terms,
and in this part we will test the validity of the formulae.
As mentioned earlier, \cite{Zhu_et_al_2016} has used synthetic data
to demonstrate that the ignored pulsar terms can lead to biased localization of the SMBBH.
In Figure \ref{fig:Zhu-strong},
we show the estimated sky locations by \cite{Zhu_et_al_2016} for both Earth-term-only and full-signal searches,
and it is seen that the estimated locations in two schemes (red and black regions respectively) do not overlap,
implying that the PEBs due to pulsar terms are no smaller than the measurement errors due to the white noise (the sizes of the black and red regions).
Hereafter we want to apply our formulae to calculate sky localization biases due to pulsar terms,
and compare our results with those given by \cite{Zhu_et_al_2016}.
Note that, to convincingly test the formulae,
more results given by other authors \citep{Corbin&Cornish2010, Ellis_et_al_2012} should be considered,
but here we only illustrate the test using the case in \citet{Zhu_et_al_2016}, due to the data availability.

We adopt the same parameters used in \cite{Zhu_et_al_2016},
that the SMBBH has a circular orbit with (injected values): $T=6.34$ yr (the GW frequency $f_0=10$ nHz in the paper), $\mathcal A_m=213.3$ ns ($\mathcal A_m=h_0 T/4\pi$, with GW amplitude $h_0=1.34\times 10^{-14}$ for the strong signal case S/N=$100$ in that paper), $\alpha=3.2594$ rad, $\delta=0.2219$ rad, $c_i=0.88$, $\psi=0.5$,
$\varphi^E=2.89$.
Furthermore, we assume that all parameters except $\alpha$ and $\delta$ have been accurately
 measured in both Earth-term-only and full-signal searches ($\hat \lambda^a_{\rm eto} = \hat \lambda^a_{\rm fs}$, for $a \neq 3, 4$),
and the best fitted location given by the Earth-term-only search are $\hat \alpha_{\rm eto}=3$ and $\hat \delta_{\rm eto}=0.305$
(i.e. $\cos( \hat \delta_{\rm eto}-\pi/2)=0.3$ in that paper).
Note that, here we ignore the PEBs of the other parameters because we intend to recover the case in \cite{Zhu_et_al_2016},
and if they are considered, the results may change, as will be shown in Section \ref{Sec:CommonNoise}.
The 12 MSPs chosen in \cite{Zhu_et_al_2016} are: J0437-1475, J1600-3053, J1640+2224, J1713+0747, J1741+1351, J1744-1134, J1909-3744, J1939+2134, J2017+0603, J2043+1711, J2241-5236, and J2317+1439 respectively, with their sky locations ($\{ \alpha_I, \delta_I \}$ with $I=1,...,12$) given by the ATNF pulsar Catalogue \footnote{http://www.atnf.csiro.au/research/pulsar/psrcat/} \citep{Manchester_et_al_2005}, and the RMS of their white noises ($\sigma_I$) are: $58$, $202$, $158$, $116$, $233$, $203$, $102$, $104$, $238$, $170$, $300$ and $267$ ns respectively.
In addition, the total observational time is $t=15$ yr and the cadence is taken as $\delta t_I=2$ weeks for all MSPs.

We will calculate the marginalized likelihood with respect to the sky locations.
Since we have assumed that all parameters follow Gaussian distributions in Eq. \eqref{eq: Gaussian},
the marginalized likelihood is obtained simply by dropping the irrelevant variables
\be
\label{eq:2-dim-likelihood}
\mathcal L \propto \exp \left(-\frac{1}{2} \delta \lambda^a \Pi^{-1}_{a b} \delta \lambda^b \right)=\exp \left(-\frac{1}{2} (\hat \lambda^a_{\rm fs}-\hat \lambda^a_{\rm eto}) \Pi^{-1}_{a b} (\hat \lambda^b_{\rm fs}-\hat \lambda^b_{\rm eto}) \right),
\ee
with both $a= 3, 4$ and $b= 3, 4$, and the covariance matrix being
\[
\begin{cases}
\Pi^{33} = \Gamma^{33}\Gamma^{33}\left<\xi_3 \xi_3\right>+2\Gamma^{33}\Gamma^{34}\left<\xi_3 \xi_4\right>+\Gamma^{34}\Gamma^{34}\left<\xi_4 \xi_4\right> & ~ \\
\Pi^{34} = \Gamma^{33}\Gamma^{43}\left<\xi_3 \xi_3\right>+(\Gamma^{33}\Gamma^{44}+\Gamma^{34}\Gamma^{43})\left<\xi_3 \xi_4\right>+\Gamma^{34}\Gamma^{44}\left<\xi_4 \xi_4\right> & ~ \\
\Pi^{44} = \Gamma^{43}\Gamma^{43}\left<\xi_3 \xi_3\right>+2\Gamma^{43}\Gamma^{44}\left<\xi_3 \xi_4\right>+\Gamma^{44}\Gamma^{44}\left<\xi_4 \xi_4\right> & ~
\end{cases}.
\]
Note that, since $t>T$ is satisfied in this example, we can use the leading-order approximations in Eqs.
\eqref{eq:Gamma_ab-circular-lead}-\eqref{eq:xi_a-xi_b-circular-lead} to calculate  the matrices $\Gamma_{a b}$ and $\left<\xi_a \xi_b\right>$ (for $a, b= 3, 4$) above.
The results are shown in Figure \ref{fig:Zhu-strong},
and we can see that the sky localization biases given by \cite{Zhu_et_al_2016} are in accordance with our formulae at $1.5\sigma$ level.
It implies that our formulae are effective, at least as an estimation on the order of magnitude.

\subsection{PEBs as Functions of Eccentricity}

In our previous work \cite{Chen&Zhang2018}, we propose a hypothesis as follows.
For a highly-eccentric SMBBH (say, $e \gtrsim 0.5$),
its GW in each orbital period is strong only in a short duration around the periastron,
so the waveforms (for both Earth term and pulsar terms) should have periodic-burst-like profiles.
As a result, the ``bursts" of Earth term and pulsar terms have small probabilities to overlap.
In this case, the Earth term and pulsar terms can be distinguished clearly,
so pulsar terms will have small impact on the detection of Earth-term signals,
namely, the PEBs due to pulsar terms for a highly-eccentric SMBBH should be smaller than those for a quasi-circular SMBBH.

\begin{figure}[h]
\begin{center}
\includegraphics[width=0.9\textwidth]{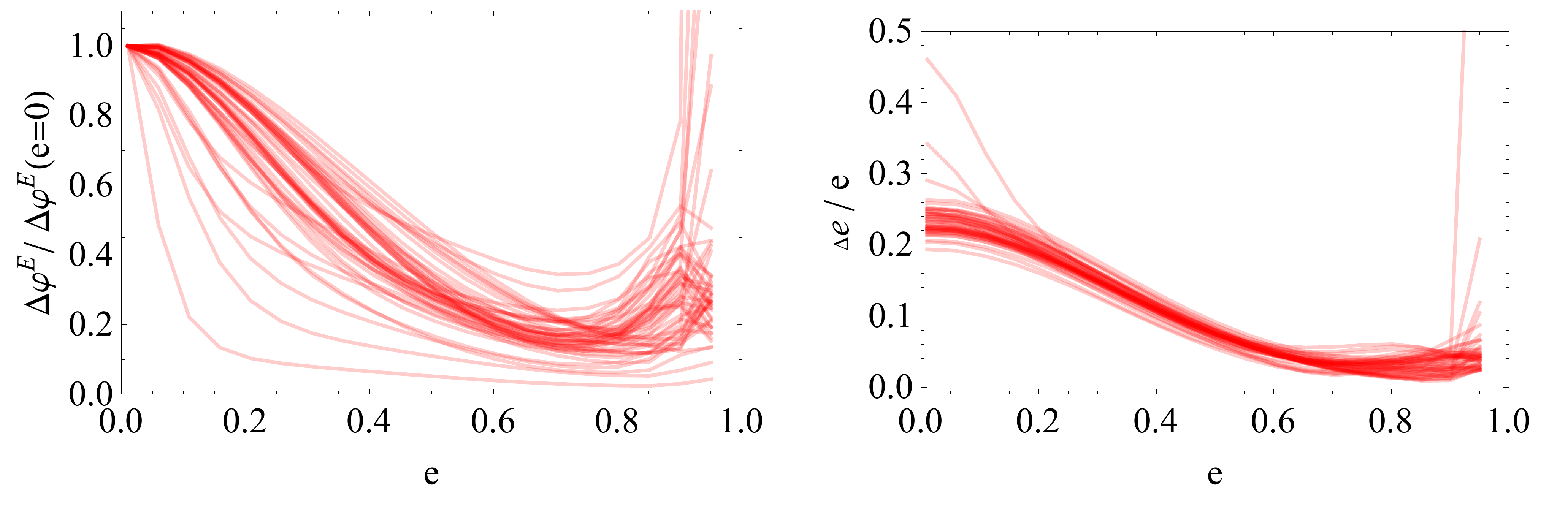}
\caption{Left: the normalized PEB $\Delta \varphi^E/\Delta \varphi^E(e=0)$ as a function of eccentricity $e$, for various parameters (various curves).
Right: the relative PEB $\Delta e/e$ as a function of eccentricity $e$.
The results in each panel are illustrated by a superposition of 50 curves.}
\label{fig:PEB-e}
\end{center}
\end{figure}

To test this hypothesis, here we calculate the standard deviations  $\Delta \lambda^a$ (for $a= 2, 3, ..., 9$)\footnote{Since the non-evolving assumption is taken, one has $\Delta \lambda^1=\Delta T=0$, as is stated in Section \ref{Sec: circular}.},
to see whether their values decrease as the eccentricity $e$ increases, for various parameters.
The parameters of the GW/SMBBH in our calculation are chosen as: $T=1$ yr, $\mathcal A_m=10$ ns, $\alpha=0$, $\delta=0$, $\psi$, $c_i$, $\varphi^E $ and $\phi_P$ are randomly taken by $\psi \in U[0, \pi]$, $c_i \in U[0,1]$, $\varphi^E \in U[0, 2\pi)$, $\phi_p \in U[0, 2\pi)$.
For each case of the parameters above, we consider $e=\{0, 0.05, 0.1, ..., 0.95\}$.
In addition, we simulate 20 MSPs to detect the GW,
with their sky locations randomly distributed in a region around the SMBBH
$\{ \alpha_I \in U[-0.1, 0.1] ~\&~ \sin \delta_I \in U[-0.1, 0.1] \}$,
their white noise taken as $\sigma_I=100$ ns,
the sampling cadence $\delta t_I=2$ weeks,
and the total observational time $t=15$ yr for all MSPs ($I=1, 2, ..., 20$).
As the observational time is much larger than the orbital period $t \gg T$, we can use the simplified formulae \eqref{eq:leading-Gamma-eccentric} and  \eqref{eq:leading-xixi-eccentric} to calculate the PEBs.

By repeating this calculation, we find that for only two parameters $\varphi^E$ and $e$,
their PEBs (or functions of PEBs)--- $\Delta \varphi^E$ and $\Delta e/e$,
present monotonically decreasing relations with the eccentricity $e$ for various parameters,
as is shown in Figure \ref{fig:PEB-e}.
We argue that the two parameters $\varphi^E$ and $e$ have more direct relations with the profile of the waveform than other parameters,
in details, $\varphi^E$ determines the moments of ``bursts" and $e$ determines the widths of the ``bursts".
That is why their PEBs (or functions of PEBs) present obvious correlations with $e$, while other parameters do not.
Furthermore,
from Figure \ref{fig:PEB-e}, one can see that the decreasing relations are valid only when $e \leq 0.7$.
The anomaly for  $e \geq 0.7$ can be explained that,
for the GWs with $T=1$ yr and $e \geq 0.7$,
the widths of ``bursts" are smaller than the cadence $\delta t_I=2$ weeks, namely,
the super-Nyquist modes $n>N_I=\mathcal N(T/2\delta t_I)$ have important contributions in this case,
and they cannot be detected by the PTA.
This implies that, to detect highly-eccentric GWs by PTA,
high-cadence observations or staggered samplings \citep{Wang_et_al_2020} are required in PTA.
Note that there have been a few MSPs with high-cadence (daily or higher) observations, such as J1939+2143 (B1937+21) \citep{Yi2014} and J1713+0747 \citep{Dolch_2016, Perera2018} etc.,
which may help the searches for highly-eccentric SMBBHs in the future.

As a conclusion, we find that two PEBs--- $\Delta \varphi^E$ and $\Delta e/e$, monotonically decrease as the eccentricity $e$ increases,
which partly confirms our hypothesis in \cite{Chen&Zhang2018}.

\section{ Comparison between PEBs from Pulsar Terms and from CSP}\label{Sec:CommonNoise}

As mentioned in Section \ref{sec: Intro},
multiple collaborations have detected a CSP recently,
which also leads to PEBs for the individual SMBBH,
if the CSP is not properly considered in the noise model.
In this part, we will calculate the PEBs arising from the CSP,
and compare them with the PEBs caused by pulsar terms.

\begin{figure}[h]
\begin{center}
\includegraphics[width=0.8\textwidth]{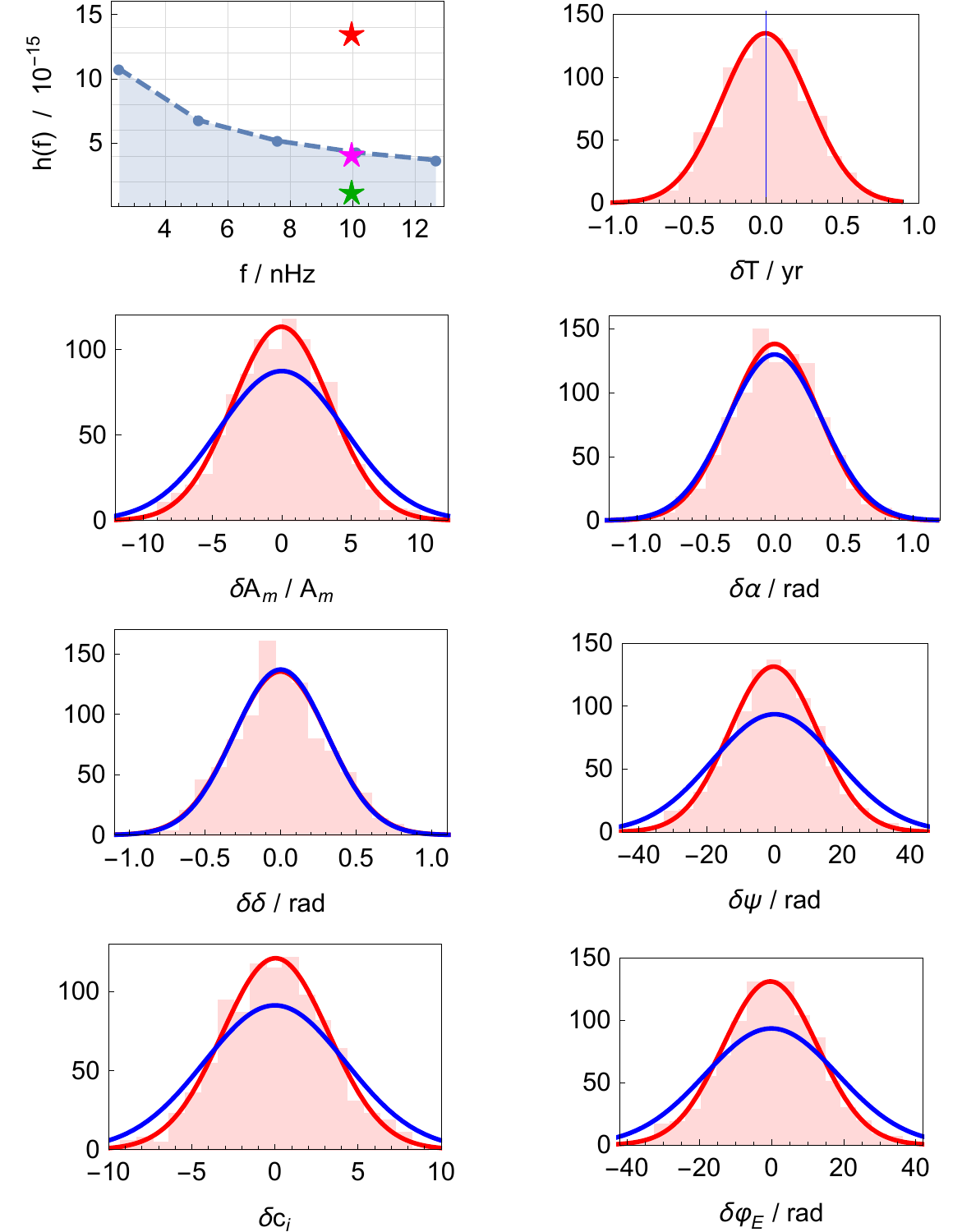}
\caption{The top-left panel shows the strain of the SGWB (blue-dashed curve) given by \eqref{eq:SGWB}
 and the continuous GWs discussed in \citet{Zhu_et_al_2016},
with the red, magenta and green stars representing the strong, moderate and weak cases, respectively.
The remaining panels illustrate the PDFs of the PEBs$^{\rm (CSP)}$ and PEBs$^{\rm (PT)}$ for the strong case,
where the light-pink histograms illustrate the PDFs of $\lambda^{a {\rm (CSP)}}$ from 1000 realizations,
the red curves are the fitted Gaussian PDFs of $\lambda^{a {\rm (CSP)}}$, and the blue curves represent the PDFs of $\lambda^{a {\rm (PT)}}$.}
\label{fig:PGWB}
\end{center}
\end{figure}

Although the origin of the CSP is still unclear,
it is usually considered as an inkling of the SGWB with a characteristic strain
\be
\label{eq:SGWB}
h^{\rm SGWB}(f)=A^{\rm SGWB} \left( \frac{f}{f_{\rm yr}} \right)^\beta,
\ee
with $A^{\rm SGWB} \simeq 2\times 10^{-15}$, $\beta=-2/3$ and $f_{\rm yr}=31.7$ nHz \citep{Phinney:2001di, Arzoumanian_2020, Goncharov_2021, EPTA_GWB, IPTA_2022}.
Hence, the time domain gravitational waveform of the SGWB for the $I$-th MSP in the PTA can be written as
\be
h_I(t)={\mathop \sum \limits_{f}} h_{I, f} \cos (2\pi f t+\phi_{I, f}),
\ee
where the amplitude $h_{I, f}$ is a random variable with a standard deviation $h^{\rm SGWB}(f)$ given by \eqref{eq:SGWB},
and the phase $\varphi_{I, f} \in  U[0, 2\pi)$.
According to the NANOGrav 12.5-yr results \citep{Arzoumanian_2020}, the five lowest frequencies ($f \in \{1, 2, 3, 4, 5\} \times f_{\rm yr}/12.5$)
contribute $99.98\%$ of the total S/N,
hence we will only consider these five frequencies in the following computations.
Furthermore, if the CSP truly arises from the SGWB,
the amplitudes $h_{I, f}$ among different MSPs should be correlated
to ensure that the overlap reduction function of the timing residuals follows the Hellings-Downs (HD) correlation \citep{1983ApJ-Hellings-Downs}.
However, the reported overlap reduction functions by various collaborations \citep{Arzoumanian_2020, Goncharov_2021, EPTA_GWB, IPTA_2022} tend to be uncorrelated, compared with the HD correlation \footnote{For simplicity, we do not consider the complicated cases which include multiple components of uncorrelated, monopolar, dipolar and quadrupolar (HD) correlations.}.
Therefore, in our approach, the amplitudes can be taken as random variables independently from MSPs $h_{I, f} \in N[0, h^{\rm SGWB}(f)^2]$.
As a result, the timing residual corresponding to the CSP can be expressed as
\be
s^{\rm (CSP)}_I(t)=\int h_I(t') dt'={\mathop \sum \limits_{f}} \frac{h_{I, f}}{2\pi f} \sin (2\pi f t+\phi_{I, f}),
\ee
and hereafter the superscript ``${\rm (CSP)}$" denotes signals or PEBs arising from the CSP.

In the following, we still take the circular case used in \citet{Zhu_et_al_2016} and in Section \ref{sec: Zhu-test}.
In this case, the PEBs$^{\rm (CSP)}$ can also be evaluated by the noise-projection technique \citep{Cutler&Flanagan1994, Cutler&Harms2006, Harms_et_al_2008}
\be
\label{eq:PEB-CN}
\delta \lambda^{a {\rm (CSP)}}=\Gamma^{a b} {\mathop \sum \limits_{I=1}^{N_P}} \left< s_I^{\rm (CSP)} \bigg| \frac{\partial s_I^E}{\partial \lambda^b} \right>_I
=\Gamma^{a b} {\mathop \sum \limits_{I=1}^{N_P}} \left< s_I^{\rm (CSP)} \big| D_b \mathcal S^E+ E_b \mathcal C^E\right>_I.
\ee
Combining Eqs. \eqref{eq:SGWB}-\eqref{eq:PEB-CN}, it is clear that the PEBs$^{\rm (CSP)}$ are proportional to $A^{\rm SGWB}$.
Additionally, the ratios between the PEBs$^{\rm (CSP)}$ and the PEBs from pulsar terms (hereafter, noted by the superscript ``${\rm (PT)}$")
depends on the relative strength between the SGWB and the continuous GW $h^{\rm SGWB}(f_0)/h_0$, with $h_0$ and $f_0$ being the amplitude and frequency of the continuous GW respectively.
To illustrate it, we will allow a variation of $h_0$ in the following, considering all the strong ($h_0=1.34\times 10^{-14}$ or S/N=100), moderate ($h_0=4.03\times 10^{-15}$ or S/N=30) and weak ($h_0=1.07\times 10^{-15}$ or S/N=8) cases in \citet{Zhu_et_al_2016} (see the first panel in Figure \ref{fig:PGWB}).
Note that, the strong case in \citet{Zhu_et_al_2016} has actually been ruled out by \citet{Zhu_2014, Babak_2016, Aggarwal_2020, Taylor:2020zpk}, and here we consider it only for the purpose of illustrating the analyses.
Furthermore,
we expect that the formulae of the covariance matrix of PEB$^{\rm (CSP)}$ would be more complicated than Eqs. \eqref{eq:Gamma_ab-circular-lead} and \eqref{eq:xi_a-xi_b-circular-lead} for the PEB$^{\rm (PT)}$, so numerical simulations will be applied to calculate $\delta \lambda^{a {\rm (CSP)}}$ (for $a=1, 2, ..., 7$) and obtain their PDFs and standard deviations $\Delta \lambda^{a {\rm (CSP)}}$.

\begin{table}[h]
\caption{A comparison between the standard deviations of the PEBs$^{\rm (CSP)}$ and PEBs$^{\rm (PT)}$ given by 1000 simulations,
 for the strong, moderate and week cases in Figure \ref{fig:PGWB} respctively. }\label{tab:1}
\begin{tabular}{|c|c|c|c|c|c|c|c|}
  \hline
  case  & $\Delta T^{\rm (CSP)}$/T & $\frac{\Delta \mathcal A_m^{\rm (CSP)}}{\Delta \mathcal A_m^{\rm (PT)}}$ & $\frac{\Delta \alpha^{\rm (CSP)}}{\Delta \alpha^{\rm (PT)} }$ & $\frac{\Delta \delta^{\rm (CSP)}}{\Delta \delta^{\rm (PT)}}$ & $\frac{\Delta \psi^{\rm (CSP)}}{\Delta \psi^{\rm (PT)}}$ & $\frac{\Delta c_i^{\rm (CSP)}}{\Delta c_i^{\rm (PT)}}$  & $\frac{\Delta \varphi^{E {\rm (CSP)}}}{\Delta \varphi^{E {\rm (PT)}}}$ \\
  \hline
  strong & $0.09$ & $0.77$ & $0.94$ & $1.01$ & $0.71$ & $0.75$ & $0.71$ \\
  \hline
  moderate & $0.32$ & $2.53$ & $3.16$ & $3.28$ & $2.43$ & $2.46$ & $2.44$ \\
  \hline
  weak & $1.14$ & $9.74$ & $13.0$ & $12.8$ & $9.55$ & $9.48$ & $9.57$ \\
  \hline
\end{tabular}
\end{table}

The resulting PDFs of the PEBs$^{\rm (CSP)}$ for the strong case are shown in Figure \ref{fig:PGWB}.
Firstly, it is seen that the CSP affects the estimation of the period, i.e. $\delta T^{\rm (CSP)} \neq 0$,
unlike pulsar terms.
This is because the pulsar terms are assumed to have the same frequency as the Earth term,
while the CSP/SGWB generally contains components with different frequencies.
Hence this result confirms our inference in Section \ref{Sec: circular}.
Furthermore, we note that the localization biases from pulsar terms ($\Delta \alpha^{\rm (PT)}=0.34$ and $\Delta \delta^{\rm (PT)}=0.31$) are different from the results in Figure \ref{fig:Zhu-strong} ($\Delta \alpha^{\rm (PT)}=0.23$ and $\Delta \delta^{\rm (PT)}=0.33$).
This is because the results in Section \ref{sec: Zhu-test} are based on the assumption that all the other parameters are precisely measured,
which is removed in this case: $\Delta \mathcal A_m^{\rm (PT)}/ \mathcal A_m\simeq 4.57$, $\Delta \psi^{\rm (PT)} \simeq 18.3$, $\Delta c_i^{\rm (PT)} \simeq 4.26$ and $\Delta \varphi^{E {\rm (PT)}} \simeq 18.4$.
Additionally, in this strong case $h_0/h^{\rm SGWB}(f_0) \simeq 3$,
the localization biases ($\delta \alpha$ and $\delta \delta$) given by pulsar terms and by the CSP are nearly the same,
while the PEBs$^{\rm (CSP)}$ for the other parameters (except the period $T$) are considerably smaller than PEBs$^{\rm (PT)}$.
In this sense, the CSP will have larger impacts on the localization of the SMBBH than the other parameters (except $T$).
To better compare PEBs$^{\rm (CSP)}$ and PEBs$^{\rm (PT)}$,
we present the ratios between their standard deviations in Table \ref{tab:1},
which indicate that $\Delta \lambda^{a {\rm (CSP)}}/\Delta \lambda^{a {\rm (PT)}}$ (for $a=2, 3, ..., 7$) are inversely proportional to $h_0/h^{\rm SGWB}(f_0)$ approximately.
Finally, we can conclude that, to ensure PEBs$^{\rm (CSP)}$ are smaller than PEBs$^{\rm (PT)}$,
the continuous GW is required to be significantly stronger than the SGWB, $h_0/h^{\rm SGWB}(f_0)\gtrsim 3$ at least for the cases investigated in this work.

\section{Summary and Further Discussions}\label{Sec: Summary}

In this article, we have presented an analytical approach to estimate the PEBs caused by the ignored pulsar terms in PTA data analyses.
Our formulae can be applied conveniently,
as long as a GW event is announced and the measured parameters based on an Earth-term-only search are released by a PTA collaboration in the future.
The analyses based on a single SMBBH in this work will be extended to the cases of multiple SMBBHs or even the SGWB in the future.

Our formulae of sky localization biases are in accordance with the numerical results given by \cite{Zhu_et_al_2016} at $1.5\sigma$ level,
which implies that our results are effective at least as an estimation on the order of magnitude for this case.
Testing the formulae through more numerical simulations will be one of the subjects of our future works.

We also investigate the PEBs in eccentric cases,
and find that $\Delta \varphi^E$ and $\Delta e/e$  monotonically decrease as the eccentricity $e$ increases.
The decreasing relations are helpful to project future PTA observations.
For example, the PTA observations are usually planned with averaged cadence $\delta t_I={\rm const}$,
but this is not a very good strategy for detecting the GWs from an eccentric SMBBH.
One can expect that if the data points are centralized in the ``bursts" of the Earth-term waveform (with the total number of data points $N_I$ fixed),
the PEBs  from pulsar terms (or other GW sources) will decrease.
In addition, more frequent observations around the ``bursts" ($\delta t_I<2$ weeks) will increase the Nyquist Frequency, so higher-frequency modes can be well detected.
Therefore, if the residuals from a highly-eccentric SMBBH are observed, i.e. measure $\varphi^E$ and $e$ with small biases,
we will know the time and durations of the following ``bursts" accurately,
and then we can re-arrange the subsequent observational time to ensure the data points are around the ``bursts" to decrease the PEBs.

Moreover, we also numerically calculate the PEBs arising from the CSP,
and find that the PEBs$^{\rm (PT)}$ are larger when the continuous GW is significantly stronger than the SGWB,
and otherwise the PEBs$^{\rm (CSP)}$ are larger.
To better understand the properties of the PEBs$^{\rm (CSP)}$,
we plan to derive their formulae in the same manner as we treat the PEBs$^{\rm (PT)}$ in the future.
Additionally, we intend to extend the analyses to the case considering a joint search for both the continuous GWs and SGWB.
Note that \citet{B_csy_2020} have considered the case using Bayesian methods,
finding that the SGWB decreases the significance/Bayes factors for low-frequency continuous GWs $f_0 \leq 20$ nHz.
Hence, we want to compare our analytical results with \citet{B_csy_2020} in the future.

Currently, the distances of MSPs are generally poorly measured,
so we have treated the pulsar-term phases $\varphi^P_I$ as  random variables obeying $U[0, 2\pi)$ for all the MSPs.
However, note that there are still a few MSPs with precisely measured distances, e.g.
PSR J0437-4715 has a distance $156.79\pm0.25$ pc \citep{Reardon_et_al_2015},
with the uncertainty $0.25$ pc comparable or even smaller than the typical GW wavelengthes of PTA ($0.1-10$ ly).
As a result, when these MSPs are contained in the PTA, the PDF $\varphi^P_{I} \in U[0, 2\pi)$ no longer holds,
and the statistical results of $\delta \lambda^a$ should change,
e.g. the expected value $\left< \delta \lambda^a \right>=0$ is not satisfied any more.
Therefore, it is worthy studying the PEBs for PTA with these MSPs.

Furthermore, from Eqs.  \eqref{eq:Gamma_11-eccentric}-\eqref{eq:Gamma_ab-eccentric}  and \eqref{eq:xi_1-xi_1-eccentric}-\eqref{eq:xi_a-xi_b-eccentric},
we see that $\Gamma^{a b} \propto N_{\rm P}^{-1}$ and $\left< \xi_a \xi_b \right> \propto N_{\rm P}$,
implying $\Delta \lambda^a \propto N_{\rm P}^{-1/2}$,
namely the PEBs decay as the MSP number increases.
It implies that in the future PTA based on SKA or FAST,
as more MSPs are expected to be detected,
it is possible that the PEBs are small enough, and the Earth term can be detected accurately.
For example, as we estimated,
for SMBBHs with various parameters, the typical relative PEB of the amplitude is $\Delta \mathcal A_m/\mathcal A_m \simeq 15\%$ at the current level ($N_P \sim \mathcal O(10^2)$),
and it can achieve the accuracy $\simeq 5\%$ in the SKA or FAST era ($N_P \sim \mathcal O(10^3)$).
If the Earth term has been accurately measured
and a few MSPs have high individual S/Ns $\rho_I=\left<s_I^E \big| s_I^E \right>_I \simeq \left<s_I^P \big| s_I^P \right>_I \gg 1$,
we can further measure their pulsar phases $\varphi_I^P$ precisely,
i.e. detect their pulsar terms.
This scheme seems simpler than the full-signal research, requiring less parameters in the data analyses,
and merits further studies by our follow-up works.
Precisely measured pulsar terms can serve as a powerful tool for astrophysical and cosmological researches,
e.g. improving the pulsar distance measurements to sub-parsec precision \citep{Lee_et_al_2011},
 probing the long-time ($\gtrsim 10^3$ yr) evolutionary histories of GW sources \citep{Mingarelli_et_al_2012, Chen&Zhang2018},
 and yielding a standard siren with purely GW measurement \citep{DOrazio&Loeb2020} etc.

\acknowledgments

We are grateful to A. Gopakumar, S. D. Mohanty, X. J. Zhu, Y. Q. Qian, and Y. Zhang for valuable discussions, comments and guidance.
We thank the anonymous referee for the careful review and thoughtful comments that improve the manuscript significantly.
J. W. Chen acknowledges the support from China Postdoctoral Science Foundation under Grant No. 2021M691146.
Y. Wang acknowledges the support from National Natural Science Foundation of China (NSFC) under Grants No. 11973024 and No. 11690021,
 Guangdong Major Project of Basic and Applied Basic Research (Grant No. 2019B030302001),
and the National Key Research and Development Program of China (No. 2020YFC2201400).

\appendix
\section{Coefficients $D_{a, n}$ and $E_{a, n}$ } \label{Sec: D_a-E_a-eccentric}

The coefficients $D_{a, n}$ and $E_{a, n}$ for $a=1, 2, ..., 9$ in Eq. \eqref{eq:s_I^E-coefficient-eccentric} are as follows
\begin{align}
\label{eq:D1n-E1n}
D_{1, n}=\frac{2\pi n B_n t}{T^2},~~~~~~~~~~~~~~~~~~~~~~
E_{1, n}=-\frac{2\pi n A_n t}{T^2},
\end{align}
\begin{align}
D_{2, n}=\frac{A_n}{\mathcal A_m},~~~~~~~~~~~~~~~~~~~~~~~~~~~~~~~~~
E_{2, n}=\frac{B_n}{\mathcal A_m},
\end{align}
\begin{align}
D_{3, n}=&\mathcal A_m \big\{ (\cos 2\psi F_{I +}^{(3)}-\sin 2\psi F_{I \times}^{(3)}) \left[ (1+c_i^2)X_n\cos 2\phi_p+2(1-c_i^2)Z_n\right] -2(\sin 2\psi F_{I +}^{(3)}+\cos 2\psi F_{I \times}^{(3)})c_i X_n\sin 2\phi_p \big\}, \nonumber \\
E_{3, n}=&\mathcal A_m \big[ -(\cos 2\psi F_{I +}^{(3)}-\sin 2\psi F_{I \times}^{(3)})(1+c_i^2) Y_n\sin 2\phi_p  -2(\sin 2\psi F_{I +}^{(3)}+\cos 2\psi F_{I \times}^{(3)})c_i Y_n\cos 2\phi_p \big],
\end{align}
with
\begin{align}
\label{eq:F^(3)}
F_{I+}^{(3)}=&\frac{1}{8\big[1-\cos \delta \cos \delta_I^P \cos(\alpha-\alpha_I^P)-\sin \delta \sin \delta_I^P \big]^2}
\bigg\{  -4\cos ^3\delta \cos \delta_I^P \sin (\alpha -\alpha_I^P)  \nonumber \\
&+ \cos \delta \cos ^3 \delta_I^P \sin (\alpha -\alpha_I^P) \big[-\cos 2 \delta \cos (2 \alpha -2 \alpha_I^P)+3 \cos (\alpha -\alpha_I^P)+\cos 2 \delta +9 \big]  \nonumber \\
&+2 (1-\sin \delta  \sin \delta_I^P ) \big[ \sin 2 \delta \sin 2\delta_I^P \sin (\alpha -\alpha_I^P)+(\cos 2 \delta -3) \cos ^2\delta_I^P \sin (2 \alpha -2 \alpha_I^P)   \big]                   \bigg\},     \\
F_{I\times}^{(3)}=&\frac{1}{2\big[1-\cos \delta \cos \delta_I^P \cos(\alpha-\alpha_I^P)-\sin \delta \sin \delta_I^P \big]^2} \bigg\{
2 \sin \delta  \cos ^2\delta_I^P \cos (2\alpha -2\alpha_I^P)  \big[ \sin \delta \sin \delta_I^P    \nonumber \\
&~~~~ +\cos \delta  \cos \delta_I^P \cos (\alpha -\alpha_I^P) -1 \big] +\cos \delta \big[-2 \cos \delta \sin \delta_I^P \cos ^2\delta_I^P   \nonumber \\
&~~~~ + \cos (\alpha -\alpha_I^P) \big(2 \sin (\delta ) \cos ^3\delta_I^P \sin ^2(\alpha -\alpha_I^P)-2 \sin \delta  \sin ^2\delta_I^P \cos \delta_I^P +\sin 2\delta_I^P \big) \big]
 \bigg\},                    \nonumber
\end{align}
\begin{align}
D_{4, n}=&\mathcal A_m \big\{ (\cos 2\psi F_{I +}^{(4)}-\sin 2\psi F_{I \times}^{(4)}) \left[ (1+c_i^2)X_n\cos 2\phi_p+2(1-c_i^2)Z_n\right] -2(\sin 2\psi F_{I +}^{(4)}+\cos 2\psi F_{I \times}^{(4)})c_i X_n\sin 2\phi_p \big\}, \nonumber \\
E_{4, n}=&\mathcal A_m \big[ -(\cos 2\psi F_{I +}^{(4)}-\sin 2\psi F_{I \times}^{(4)})(1+c_i^2) Y_n\sin 2\phi_p -2(\sin 2\psi F_{I +}^{(4)}+\cos 2\psi F_{I \times}^{(4)})c_i Y_n\cos 2\phi_p \big],
\end{align}
with
\begin{align}
\label{eq:F^(4)}
F_{I+}^{(4)}=&\frac{1}{4\big[1-\cos \delta \cos \delta_I^P \cos(\alpha-\alpha_I^P)-\sin \delta \sin \delta_I^P \big]^2}
\bigg\{ 2 \big[ \cos \delta \cos \delta_I^P \cos (\alpha -\alpha_I^P)+\sin \delta  \sin \delta_I^P  \nonumber \\
&~~~~ -1 \big] \big[ -\sin \delta  \cos \delta  \cos ^2\delta_I^P
\big(\cos (2\alpha -2\alpha_I^P)+3 \big)
+\cos 2 \delta  \sin 2 \delta_I^P \cos (\alpha -\alpha_I^P)+\sin 2 \delta \big]      \nonumber \\
&~~~~ -\big( \sin \delta  \cos\delta_I^P \cos (\alpha -\alpha_I^P)-\cos \delta  \sin \delta_I^P\big)
\big[ \big( \sin ^2\delta +1\big) \cos ^2\delta_I^P \cos (2\alpha -2\alpha_I^P)    \nonumber \\
&~~~~ -\sin 2 \delta  \sin 2\delta_I^P \cos (\alpha -\alpha_I^P)+\cos ^2\delta  (2-3 \cos ^2\delta_I^P)\big]
 \bigg\},  \\
F_{I\times}^{(4)}=&\frac{1}{2\big[1-\cos \delta \cos \delta_I^P \cos(\alpha-\alpha_I^P)-\sin \delta \sin \delta_I^P \big]^2} \bigg\{-\cos \delta \cos ^2\delta_I^P \sin (2\alpha -2\alpha_I^P)  \nonumber\\
&~~~~ +\sin \delta \big[ \sin 2\delta_I^P  \sin (\alpha -\alpha_I^P) (\sin \delta  \sin \delta_I^P-1)+\sin \delta  \cos ^3\delta_I^P \sin (2 \alpha -2\alpha_I^P) \cos (\alpha -\alpha_I^P)\big]  \nonumber \\
&~~~~ +\cos ^2\delta  \big[\sin \delta_I^P \sin 2\delta_I^P \sin (\alpha -\alpha_I^P)+\cos ^3\delta_I^P \sin (2  \alpha -2\alpha_I^P) \cos ( \alpha -\alpha_I^P)\big]
 \bigg\},  \nonumber
\end{align}
\begin{align}
D_{5, n}=&2\mathcal A_m \big\{ (-\sin 2\psi F_{I +}-\cos 2\psi F_{I \times}) \left[ (1+c_i^2)X_n\cos 2\phi_p+2(1-c_i^2)Z_n\right]  -2(\cos 2\psi F_{I +}-\sin 2\psi F_{I \times})c_i X_n\sin 2\phi_p \big\}, \nonumber \\
E_{5, n}=&2\mathcal A_m \big[ (\sin 2\psi F_{I +}+\cos 2\psi F_{I \times})(1+c_i^2) Y_n\sin 2\phi_p  -2(\cos 2\psi F_{I +}-\sin 2\psi F_{I \times})c_i Y_n\cos 2\phi_p \big],
\end{align}
\begin{align}
D_{6, n}=&\mathcal A_m \big\{ 2c_i(\cos 2\psi F_{I +}-\sin 2\psi F_{I \times}) \big( X_n\cos 2\phi_p -2 Z_n \big)   -2(\sin 2\psi F_{I +}+\cos 2\psi F_{I \times}) X_n\sin 2\phi_p \big\}, \nonumber \\
E_{6, n}=&\mathcal A_m \big[ -2(\cos 2\psi F_{I +}-\sin 2\psi F_{I \times})c_i Y_n\sin 2\phi_p  -2(\sin 2\psi F_{I +}+\cos 2\psi F_{I \times}) Y_n\cos 2\phi_p \big];
\end{align}
\begin{align}
D_{7, n}=n B_n,~~~~~~~~~~~~~~~~~~~~~~~~~~~~~~~~~
E_{7, n}=-n A_n.
\end{align}
\begin{align}
D_{8, n}=&2 \mathcal A_m \big\{ -(\cos 2\psi F_{I +}-\sin 2\psi F_{I \times})  (1+c_i^2)X_n\sin 2\phi_p -2 (\sin 2\psi F_{I +}+\cos 2\psi F_{I \times})c_i X_n\cos 2\phi_p \big\}, \nonumber \\
E_{8, n}=&2 \mathcal A_m \big[ -(\cos 2\psi F_{I +}-\sin 2\psi F_{I \times})(1+c_i^2) Y_n\cos 2\phi_p  +2(\sin 2\psi F_{I +}+\cos 2\psi F_{I \times})c_i Y_n\sin 2\phi_p \big],
\end{align}
and
\begin{align}
\label{eq:D9n-E9n}
D_{9, n}=&\mathcal A_m \big\{ (\cos 2\psi F_{I +}-\sin 2\psi F_{I \times}) \left[ (1+c_i^2)X_n^{(e)}\cos 2\phi_p+2(1-c_i^2)Z_n^{(e)}\right] -2 (\sin 2\psi F_{I +}+\cos 2\psi F_{I \times})c_i X_n^{(e)}\sin 2\phi_p \big\}, \nonumber \\
E_{9, n}=&\mathcal A_m \big[ -(\cos 2\psi F_{I +}-\sin 2\psi F_{I \times})(1+c_i^2) Y_n^{(e)}\sin 2\phi_p -2(\sin 2\psi F_{I +}+\cos 2\psi F_{I \times})c_i Y_n^{(e)}\cos 2\phi_p \big],
\end{align}
with
\begin{align}
X_n^{(e)} =&\frac{n}{2}J_{n-3}(e n)-e n J_{n-2}(e n)-\frac{2+n}{2} J_{n-1}(e n)+2 e n J_{n}(e n)+\frac{2-n}{2} J_{n+1}(e n) -e n J_{n+2}(e n)+\frac{n}{2}J_{n+3}(e n), \nonumber \\
Y_n^{(e)}  =&-\frac{\sqrt{1-e^2} n}{2}  J_{n-3}(e n)+\frac{e}{\sqrt{1-e^2}} J_{n-2}(e n)+\frac{3\sqrt{1-e^2} n}{2}  J_{n-1}(e n)-\frac{2 e }{\sqrt{1-e^2}}J_n(e n)-\frac{3\sqrt{1-e^2} n}{2}  J_{n+1}(e n) \nonumber \\
&+\frac{e }{\sqrt{1-e^2}}J_{n+2}(e n) +\frac{\sqrt{1-e^2} n}{2}  J_{n+3}(e n), \nonumber \\
Z_n^{(e)} =& J_{n-1}(e n)-J_{n+1}(e n). \nonumber
\end{align}

\section{Analytical Results of the Inner-Product terms}\label{Sec: eccentric-time-evolving}

From Eq. \eqref{eq:inner-product-weak}, we can analytically solve the inner-product terms in \eqref{eq:xi_1-xi_1-eccentric}-\eqref{eq:xi_a-xi_b-eccentric} that
\begin{align}
\label{eq:inner-product-eccentric-(1)}
\left<\mathcal S_{n_1}^E\big| \mathcal S_{n_2}^E \right>_I =
\begin{cases}
\frac{1}{\sigma_I^2 \delta t_I} \left[\frac{t}{2}-\frac{T \mathcal S_{2n_1}^E}{8 \pi n_1}+\frac{T \sin(2n_1 \varphi^E)}{8 \pi n_1} \right], & n_1=n_2, \\
\frac{T}{4\pi \sigma_I^2 \delta t_I} \left[\frac{\mathcal S_{n_1-n_2}^E}{n_1-n_2}-\frac{\sin[(n_1-n_2)\varphi^E]}{n_1-n_2}-\frac{\mathcal S_{n_1+n_2}^E}{n_1+n_2}+\frac{\sin[(n_1+n_2)\varphi^E]}{n_1+n_2} \right], & n_1\neq n_2,
\end{cases}
\end{align}
\begin{align}
\left<\mathcal C_{n_1}^E\big| \mathcal C_{n_2}^E \right>_I =
\begin{cases}
\frac{1}{\sigma_I^2 \delta t_I} \left[\frac{t}{2}+\frac{T \mathcal S_{2n_1}^E}{8 \pi n_1}-\frac{T \sin(2n_1 \varphi^E)}{8 \pi n_1}  \right], & n_1=n_2, \\
\frac{T}{4\pi \sigma_I^2 \delta t_I} \left[\frac{\mathcal S_{n_1-n_2}^E}{n_1-n_2}-\frac{\sin[(n_1-n_2)\varphi^E]}{n_1-n_2}+\frac{\mathcal S_{n_1+n_2}^E}{n_1+n_2}-\frac{\sin[(n_1+n_2)\varphi^E]}{n_1+n_2} \right], & n_1\neq n_2,
\end{cases}
\end{align}
\begin{align}
\left<\mathcal S_{n_1}^E\big| \mathcal C_{n_2}^E \right>_I =
\begin{cases}
-\frac{T \mathcal C_{2n_1}^E}{8 \pi n_1 \sigma_I^2 \delta t_I}+\frac{T \mathcal \cos(2n_1 \varphi^E)}{8 \pi n_1 \sigma_I^2 \delta t_I} , & n_1=n_2, \\
\frac{T}{4\pi \sigma_I^2 \delta t_I} \left[-\frac{\mathcal C_{n_1-n_2}^E}{n_1-n_2}+\frac{\cos[(n_1-n_2)\varphi^E]}{n_1-n_2}-\frac{\mathcal C_{n_1+n_2}^E}{n_1+n_2}+\frac{\cos[(n_1+n_2)\varphi^E]}{n_1+n_2} \right], & n_1\neq n_2,
\end{cases}
\end{align}
\begin{align}
\left<\mathcal S_{n_1}^E\big| t \mathcal S_{n_2}^E \right>_I =
\begin{cases}
\frac{1}{\sigma_I^2 \delta t_I} \left[\frac{t^2}{4}-\frac{t T \mathcal S_{2n_1}^E}{8\pi n_1}-\frac{T^2 \mathcal C_{2n_1}^E}{32 \pi^2 n_1^2}+\frac{T^2 \cos(2n_1 \varphi^E)}{32 \pi^2 n_1^2}  \right], & n_1=n_2, \\
\frac{T^2}{8\pi^2 \sigma_I^2 \delta t_I} \bigg[  \frac{2\pi t \mathcal S_{n_1-n_2}^E}{(n_1-n_2)T}-\frac{2\pi t \mathcal S_{n_1+n_2}^E}{(n_1+n_2)T}+\frac{\mathcal C_{n_1-n_2}^E}{(n_1-n_2)^2}   \\
 ~~~~~~~~~~~~
 -\frac{\cos[(n_1-n_2) \varphi^E]}{(n_1-n_2)^2} -\frac{\mathcal C_{n_1+n_2}^E}{(n_1+n_2)^2}+\frac{\cos[(n_1+n_2) \varphi^E]}{(n_1+n_2)^2} \bigg], & n_1\neq n_2,
\end{cases}
\end{align}
\begin{align}
\left<\mathcal C_{n_1}^E\big| t \mathcal C_{n_2}^E \right>_I =
\begin{cases}
\frac{1}{\sigma_I^2 \delta t_I} \left[\frac{t^2}{4}+\frac{t T \mathcal S_{2n_1}^E}{8\pi n_1}+\frac{T^2 \mathcal C_{2n_1}^E}{32 \pi^2 n_1^2}-\frac{T^2 \cos(2n_1 \varphi^E)}{32 \pi^2 n_1^2}  \right], & n_1=n_2, \\
\frac{T^2}{8\pi^2 \sigma_I^2 \delta t_I} \bigg[  \frac{2\pi t \mathcal S_{n_1-n_2}^E}{(n_1-n_2)T}+\frac{2\pi t \mathcal S_{n_1+n_2}^E}{(n_1+n_2)T}+\frac{\mathcal C_{n_1-n_2}^E}{(n_1-n_2)^2}  \\
 ~~~~~~~~~~~~
 -\frac{\cos[(n_1-n_2) \varphi^E]}{(n_1-n_2)^2} +\frac{\mathcal C_{n_1+n_2}^E}{(n_1+n_2)^2}-\frac{\cos[(n_1+n_2) \varphi^E]}{(n_1+n_2)^2} \bigg], & n_1\neq n_2,
\end{cases}
\end{align}
and
\begin{align}
\label{eq:inner-product-eccentric-(6)}
\left<\mathcal S_{n_1}^E\big| t \mathcal C_{n_2}^E \right>_I =\left<\mathcal C_{n_2}^E\big| t \mathcal S_{n_1}^E \right>_I
=
\begin{cases}
\frac{1}{\sigma_I^2 \delta t_I} \left[\frac{t T \mathcal C_{2n_1}^E}{8n_1 \pi}+\frac{T^2 \mathcal S_{2n_1}^E}{32n_1 \pi} -\frac{T^2 \sin(2n_1 \varphi^E)}{32n_1 \pi}                  \right], & n_1=n_2,           \\
\frac{T^2}{8\pi^2 \sigma_I^2 \delta t_I} \bigg[  -\frac{2\pi t \mathcal C_{n_1-n_2}^E}{(n_1-n_2)T}-\frac{2\pi t \mathcal C_{n_1+n_2}^E}{(n_1+n_2)T}+\frac{\mathcal S_{n_1-n_2}^E}{(n_1-n_2)^2}    \\
 ~~~~~~~~~~~~
 -\frac{\sin[(n_1-n_2) \varphi^E]}{(n_1-n_2)^2} +\frac{\mathcal S_{n_1+n_2}^E}{(n_1+n_2)^2}-\frac{\sin[(n_1+n_2) \varphi^E]}{(n_1+n_2)^2} \bigg], & n_1\neq n_2
\end{cases}
.
\end{align}

\begin{align}
\left<t \mathcal S_{n_1}^E\big| t \mathcal S_{n_2}^E \right>_I =
\begin{cases}
\frac{1}{\sigma_I^2 \delta t_I} \bigg[  \frac{t^3}{6}-\frac{t^2 T \mathcal S_{2n_1}^E}{8 \pi n_1} -\frac{t T^2 \mathcal C_{2n_1}^E}{16 \pi^2 n_1^2} +\frac{T^3  \mathcal S_{2n_1}^E }{64 \pi^3 n_1^3}-\frac{T^3 \sin (2n_1 \varphi^E)}{64 \pi^3 n_1^3}  \bigg], & n_1=n_2, \\
\frac{T}{8\pi^3 \sigma_I^2 \delta t_I} \bigg[\frac{2\pi^2 t^2  \mathcal S_{n_1-n_2}^E}{n_1-n_2}+\frac{2\pi t T \mathcal C_{n_1-n_2}^E}{(n_1-n_2)^2}-\frac{T^2 \mathcal S_{n_1-n_2}^E}{(n_1-n_2)^3}+\frac{T^2 \sin [(n_1-n_2)\varphi^E]}{(n_1-n_2)^3}   \\
~~~~~~-\frac{2\pi^2 t^2  \mathcal S_{n_1+n_2}^E}{n_1+n_2}-\frac{2\pi t T \mathcal C_{n_1+n_2}^E}{(n_1+n_2)^2}+\frac{T^2 \mathcal S_{n_1+n_2}^E}{(n_1+n_2)^3}-\frac{T^2 \sin [(n_1+n_2)\varphi^E]}{(n_1+n_2)^3}\bigg], & n_1\neq n_2,
\end{cases}
\end{align}
\begin{align}
\left<t \mathcal C_{n_1}^E\big| t \mathcal C_{n_2}^E \right>_I =
\begin{cases}
\frac{1}{\sigma_I^2 \delta t_I} \bigg[ \frac{t^3}{6}+\frac{t^2 T \mathcal S_{2n_1}^E}{8\pi n_1}+\frac{t T^2 \mathcal C_{2n_1}^E}{16\pi^2 n_1^2}-\frac{T^3 \mathcal S_{2n_1}^E}{64\pi^3 n_1^3}+\frac{T^3 \sin (2n_1 \varphi^E)}{64 \pi^3 n_1^3} \bigg], & n_1=n_2, \\
\frac{T}{8\pi^3 \sigma_I^2 \delta t_I} \bigg[\frac{2\pi^2 t^2  \mathcal S_{n_1-n_2}^E}{n_1-n_2}+\frac{2\pi t T \mathcal C_{n_1-n_2}^E}{(n_1-n_2)^2}-\frac{T^2 \mathcal S_{n_1-n_2}^E}{(n_1-n_2)^3}+\frac{T^2 \sin [(n_1-n_2)\varphi^E]}{(n_1-n_2)^3}    \\
~~~~~~-\frac{2\pi^2 t^2  \mathcal S_{n_1+n_2}^E}{n_1+n_2}-\frac{2\pi t T \mathcal C_{n_1+n_2}^E}{(n_1+n_2)^2}+\frac{T^2 \mathcal S_{n_1+n_2}^E}{(n_1+n_2)^3}-\frac{T^2 \sin [(n_1+n_2)\varphi^E]}{(n_1+n_2)^3}\bigg], & n_1\neq n_2,
\end{cases}
\end{align}
and
\begin{align}
\left<t \mathcal S_{n_1}^E\big| t \mathcal C_{n_2}^E \right>_I =
\begin{cases}
\frac{1}{\sigma_I^2 \delta t_I} \bigg[  -\frac{t^2 T \mathcal C_{2n_1}^E}{8 \pi n_1} +\frac{t T^2 \mathcal S_{2n_1}^E}{16 \pi^2 n_1^2} +\frac{T^3  \mathcal S_{2n_1}^E }{64 \pi^3 n_1^3}-\frac{T^3 \sin (2n_1 \varphi^E)}{64 \pi^3 n_1^3}  \bigg], & n_1=n_2, \\
\frac{T}{8\pi^3 \sigma_I^2 \delta t_I} \bigg[-\frac{2\pi^2 t^2  \mathcal C_{n_1-n_2}^E}{n_1-n_2}+\frac{2\pi t T \mathcal S_{n_1-n_2}^E}{(n_1-n_2)^2}+\frac{T^2 \mathcal C_{n_1-n_2}^E}{(n_1-n_2)^3}-\frac{T^2 \cos [(n_1-n_2)\varphi^E]}{(n_1-n_2)^3}    \\
~~~~~~-\frac{2\pi^2 t^2  \mathcal C_{n_1+n_2}^E}{n_1-n_2}+\frac{2\pi t T \mathcal S_{n_1+n_2}^E}{(n_1+n_2)^2}+\frac{T^2 \mathcal C_{n_1+n_2}^E}{(n_1+n_2)^3}-\frac{T^2 \cos [(n_1+n_2)\varphi^E]}{(n_1+n_2)^3}\bigg], & n_1\neq n_2.
\end{cases}
\end{align}

\section{Calculation of the Expected Values}\label{Sec: eccentric-expected}

Taking  Eq. \eqref{eq:SCP-SCE-eccentric} into Eqs. \eqref{eq:xi_1-xi_1-eccentric}-\eqref{eq:xi_a-xi_b-eccentric}, all the inner-product terms are expanded as
\begin{align}
\label{eq:SnP-SnE}
\left<\mathcal S_{n_1}^P \big|\mathcal  S_{n_2}^E \right>_I = \cos(n_1 \Delta \varphi_I)\left<\mathcal S_{n_1}^E\big| \mathcal S_{n_2}^E \right>_I+\sin(n_1 \Delta \varphi_I)\left<\mathcal C_{n_1}^E\big| \mathcal S_{n_2}^E \right>_I, \nonumber \\
\left<\mathcal S_{n_1}^P \big|\mathcal  C_{n_2}^E \right>_I = \cos(n_1 \Delta \varphi_I)\left<\mathcal S_{n_1}^E\big| \mathcal C_{n_2}^E \right>_I+\sin(n_1 \Delta \varphi_I)\left<\mathcal C_{n_1}^E\big| \mathcal C_{n_2}^E \right>_I, \nonumber \\
\left<\mathcal C_{n_1}^P \big|\mathcal  S_{n_2}^E \right>_I = \cos(n_1 \Delta \varphi_I)\left<\mathcal C_{n_1}^E\big| \mathcal S_{n_2}^E \right>_I-\sin(n_1 \Delta \varphi_I)\left<\mathcal S_{n_1}^E\big| \mathcal S_{n_2}^E \right>_I, \nonumber \\
\left<\mathcal C_{n_1}^P \big|\mathcal  C_{n_2}^E \right>_I = \cos(n_1 \Delta \varphi_I)\left<\mathcal C_{n_1}^E\big| \mathcal C_{n_2}^E \right>_I-\sin(n_1 \Delta \varphi_I)\left<\mathcal S_{n_1}^E\big| \mathcal C_{n_2}^E \right>_I,   \\
\left<\mathcal S_{n_1}^P \big| t\mathcal  S_{n_2}^E \right>_I = \cos(n_1 \Delta \varphi_I)\left<\mathcal S_{n_1}^E\big| t \mathcal S_{n_2}^E \right>_I+\sin(n_1 \Delta \varphi_I)\left<\mathcal C_{n_1}^E\big| t \mathcal S_{n_2}^E \right>_I, \nonumber \\
\left<\mathcal S_{n_1}^P \big| t\mathcal  C_{n_2}^E \right>_I = \cos(n_1 \Delta \varphi_I)\left<\mathcal S_{n_1}^E\big| t \mathcal C_{n_2}^E \right>_I+\sin(n_1 \Delta \varphi_I)\left<\mathcal C_{n_1}^E\big| t \mathcal C_{n_2}^E \right>_I, \nonumber \\
\left<\mathcal C_{n_1}^P \big| t\mathcal  S_{n_2}^E \right>_I = \cos(n_1 \Delta \varphi_I)\left<\mathcal C_{n_1}^E\big| t \mathcal S_{n_2}^E \right>_I-\sin(n_1 \Delta \varphi_I)\left<\mathcal S_{n_1}^E\big| t \mathcal S_{n_2}^E \right>_I, \nonumber \\
\left<\mathcal C_{n_1}^P \big| t\mathcal  C_{n_2}^E \right>_I = \cos(n_1 \Delta \varphi_I)\left<\mathcal C_{n_1}^E\big| t \mathcal C_{n_2}^E \right>_I-\sin(n_1 \Delta \varphi_I)\left<\mathcal S_{n_1}^E\big| t \mathcal C_{n_2}^E \right>_I.  \nonumber
\end{align}

Subsequently, from Eq. \eqref{eq:trigonometric-eccentric}, we can obtain all the expected terms in Eqs. \eqref{eq:xi_1-xi_1-eccentric}-\eqref{eq:xi_a-xi_b-eccentric} that
\begin{align}
\left<\left<\mathcal S_{n_1}^P \big|\mathcal  S_{n_2}^E \right>_I \cdot \left<\mathcal S_{n_3}^P \big|\mathcal  S_{n_4}^E \right>_I \right>= \frac{1}{2}\delta_{n_1,n_3} \left(\left<\mathcal S_{n_1}^E\big| \mathcal S_{n_2}^E \right>_I\cdot\left<\mathcal S_{n_3}^E\big| \mathcal S_{n_4}^E \right>_I+ \left<\mathcal C_{n_1}^E\big| \mathcal S_{n_2}^E  \right>_I\cdot \left<\mathcal C_{n_3}^E\big| \mathcal S_{n_4}^E \right>_I \right),         \nonumber \\
\left<\left<\mathcal S_{n_1}^P \big|\mathcal  S_{n_2}^E \right>_I \cdot \left<\mathcal S_{n_3}^P \big|\mathcal  C_{n_4}^E \right>_I \right>= \frac{1}{2}\delta_{n_1,n_3} \left(\left<\mathcal S_{n_1}^E\big| \mathcal S_{n_2}^E \right>_I\cdot\left<\mathcal S_{n_3}^E\big| \mathcal C_{n_4}^E \right>_I+ \left<\mathcal C_{n_1}^E\big| \mathcal S_{n_2}^E  \right>_I\cdot \left<\mathcal C_{n_3}^E\big| \mathcal C_{n_4}^E \right>_I \right),         \nonumber \\
\left<\left<\mathcal S_{n_1}^P \big|\mathcal  S_{n_2}^E \right>_I \cdot \left<\mathcal C_{n_3}^P \big|\mathcal  S_{n_4}^E \right>_I \right>= \frac{1}{2}\delta_{n_1,n_3} \left(\left<\mathcal S_{n_1}^E\big| \mathcal S_{n_2}^E \right>_I\cdot\left<\mathcal C_{n_3}^E\big| \mathcal S_{n_4}^E \right>_I -\left<\mathcal C_{n_1}^E\big| \mathcal S_{n_2}^E  \right>_I\cdot \left<\mathcal S_{n_3}^E\big| \mathcal S_{n_4}^E \right>_I \right),         \nonumber \\
\left<\left<\mathcal S_{n_1}^P \big|\mathcal  S_{n_2}^E \right>_I \cdot \left<\mathcal C_{n_3}^P \big|\mathcal  C_{n_4}^E \right>_I \right>= \frac{1}{2}\delta_{n_1,n_3} \left(\left<\mathcal S_{n_1}^E\big| \mathcal S_{n_2}^E \right>_I\cdot\left<\mathcal C_{n_3}^E\big| \mathcal C_{n_4}^E \right>_I -\left<\mathcal C_{n_1}^E\big| \mathcal S_{n_2}^E  \right>_I\cdot \left<\mathcal S_{n_3}^E\big| \mathcal C_{n_4}^E \right>_I \right),         \nonumber \\
\left<\left<\mathcal S_{n_1}^P \big|\mathcal  C_{n_2}^E \right>_I  \cdot \left<\mathcal S_{n_3}^P \big|\mathcal  S_{n_4}^E \right>_I \right>=\frac{1}{2}\delta_{n_1,n_3} \left(\left<\mathcal S_{n_1}^E\big| \mathcal C_{n_2}^E \right>_I \cdot\left<\mathcal S_{n_3}^E\big| \mathcal S_{n_4}^E \right>_I +\left<\mathcal C_{n_1}^E\big| \mathcal C_{n_2}^E  \right>_I\cdot \left<\mathcal C_{n_3}^E\big| \mathcal S_{n_4}^E \right>_I \right),      \nonumber \\
\left<\left<\mathcal S_{n_1}^P \big|\mathcal  C_{n_2}^E \right>_I  \cdot \left<\mathcal S_{n_3}^P \big|\mathcal C_{n_4}^E \right>_I \right>=\frac{1}{2}\delta_{n_1,n_3} \left(\left<\mathcal S_{n_1}^E\big| \mathcal C_{n_2}^E \right>_I \cdot\left<\mathcal S_{n_3}^E\big| \mathcal C_{n_4}^E \right>_I +\left<\mathcal C_{n_1}^E\big| \mathcal C_{n_2}^E  \right>_I\cdot \left<\mathcal C_{n_3}^E\big| \mathcal C_{n_4}^E \right>_I \right),      \nonumber \\
\left<\left<\mathcal S_{n_1}^P \big|\mathcal  C_{n_2}^E \right>_I  \cdot \left<\mathcal C_{n_3}^P \big|\mathcal S_{n_4}^E \right>_I \right>=\frac{1}{2}\delta_{n_1,n_3} \left(\left<\mathcal S_{n_1}^E\big| \mathcal C_{n_2}^E \right>_I \cdot\left<\mathcal C_{n_3}^E\big| \mathcal S_{n_4}^E \right>_I -\left<\mathcal C_{n_1}^E\big| \mathcal C_{n_2}^E  \right>_I\cdot \left<\mathcal S_{n_3}^E\big| \mathcal S_{n_4}^E \right>_I \right),      \nonumber \\
\left<\left<\mathcal S_{n_1}^P \big|\mathcal  C_{n_2}^E \right>_I  \cdot \left<\mathcal C_{n_3}^P \big|\mathcal C_{n_4}^E \right>_I \right>=\frac{1}{2}\delta_{n_1,n_3} \left(\left<\mathcal S_{n_1}^E\big| \mathcal C_{n_2}^E \right>_I\cdot\left<\mathcal C_{n_3}^E\big| \mathcal C_{n_4}^E \right>_I -\left<\mathcal C_{n_1}^E\big| \mathcal C_{n_2}^E  \right>_I\cdot \left<\mathcal S_{n_3}^E\big| \mathcal C_{n_4}^E \right>_I \right),     \nonumber \\
\left<\left<\mathcal C_{n_1}^P \big|\mathcal  S_{n_2}^E \right>_I\cdot \left<\mathcal S_{n_3}^P \big|\mathcal  S_{n_4}^E \right>_I \right>=\frac{1}{2}\delta_{n_1,n_3}\left(\left<\mathcal C_{n_1}^E\big| \mathcal S_{n_2}^E \right>_I\cdot\left<\mathcal S_{n_3}^E\big| \mathcal S_{n_4}^E \right>_I -\left<\mathcal S_{n_1}^E\big| \mathcal S_{n_2}^E  \right>_I\cdot \left<\mathcal C_{n_3}^E\big| \mathcal S_{n_4}^E \right>_I \right),      \nonumber \\
\left<\left<\mathcal C_{n_1}^P \big|\mathcal  S_{n_2}^E \right>_I\cdot \left<\mathcal S_{n_3}^P \big|\mathcal  C_{n_4}^E \right>_I \right>=\frac{1}{2}\delta_{n_1,n_3}\left(\left<\mathcal C_{n_1}^E\big| \mathcal S_{n_2}^E \right>_I\cdot\left<\mathcal S_{n_3}^E\big| \mathcal C_{n_4}^E \right>_I -\left<\mathcal S_{n_1}^E\big| \mathcal S_{n_2}^E  \right>_I \cdot \left<\mathcal C_{n_3}^E\big| \mathcal C_{n_4}^E \right>_I \right),      \nonumber \\
\left<\left<\mathcal C_{n_1}^P \big|\mathcal  S_{n_2}^E \right>_I\cdot \left<\mathcal C_{n_3}^P \big|\mathcal  S_{n_4}^E \right>_I \right>=\frac{1}{2}\delta_{n_1,n_3}\left(\left<\mathcal C_{n_1}^E\big| \mathcal S_{n_2}^E \right>_I\cdot\left<\mathcal C_{n_3}^E\big| \mathcal S_{n_4}^E \right>_I +\left<\mathcal S_{n_1}^E\big| \mathcal S_{n_2}^E  \right>_I\cdot \left<\mathcal S_{n_3}^E\big| \mathcal S_{n_4}^E \right>_I \right),      \nonumber \\
\left<\left<\mathcal C_{n_1}^P \big|\mathcal  S_{n_2}^E \right>_I\cdot \left<\mathcal C_{n_3}^P \big|\mathcal  C_{n_4}^E \right>_I \right>=\frac{1}{2}\delta_{n_1,n_3}\left(\left<\mathcal C_{n_1}^E\big| \mathcal S_{n_2}^E \right>_I\cdot\left<\mathcal C_{n_3}^E\big| \mathcal C_{n_4}^E \right>_I +\left<\mathcal S_{n_1}^E\big| \mathcal S_{n_2}^E  \right>_I\cdot \left<\mathcal S_{n_3}^E\big| \mathcal C_{n_4}^E \right>_I \right),      \nonumber \\
\left<\left<\mathcal C_{n_1}^P \big|\mathcal  C_{n_2}^E \right>_I\cdot \left<\mathcal S_{n_3}^P \big|\mathcal  S_{n_4}^E \right>_I \right>=\frac{1}{2}\delta_{n_1,n_3}\left(\left<\mathcal C_{n_1}^E\big| \mathcal C_{n_2}^E \right>_I\cdot\left<\mathcal S_{n_3}^E\big| \mathcal S_{n_4}^E \right>_I -\left<\mathcal S_{n_1}^E\big| \mathcal C_{n_2}^E  \right>_I \cdot \left<\mathcal C_{n_3}^E\big| \mathcal S_{n_4}^E \right>_I \right),      \nonumber \\
\left<\left<\mathcal C_{n_1}^P \big|\mathcal  C_{n_2}^E \right>_I\cdot \left<\mathcal S_{n_3}^P \big|\mathcal  C_{n_4}^E \right>_I \right>=\frac{1}{2}\delta_{n_1,n_3}\left(\left<\mathcal C_{n_1}^E\big| \mathcal C_{n_2}^E \right>_I\cdot\left<\mathcal S_{n_3}^E\big| \mathcal C_{n_4}^E \right>_I -\left<\mathcal S_{n_1}^E\big| \mathcal C_{n_2}^E  \right>_I \cdot \left<\mathcal C_{n_3}^E\big| \mathcal C_{n_4}^E \right>_I \right),      \nonumber \\
\left<\left<\mathcal C_{n_1}^P \big|\mathcal  C_{n_2}^E \right>_I\cdot \left<\mathcal C_{n_3}^P \big|\mathcal  S_{n_4}^E \right>_I \right>=\frac{1}{2}\delta_{n_1,n_3}\left(\left<\mathcal C_{n_1}^E\big| \mathcal C_{n_2}^E \right>_I \cdot\left<\mathcal C_{n_3}^E\big| \mathcal S_{n_4}^E \right>_I +\left<\mathcal S_{n_1}^E\big| \mathcal C_{n_2}^E  \right>_I \cdot \left<\mathcal S_{n_3}^E\big| \mathcal S_{n_4}^E \right>_I \right),      \nonumber \\
\left<\left<\mathcal C_{n_1}^P \big|\mathcal  C_{n_2}^E \right>_I\cdot \left<\mathcal C_{n_3}^P \big|\mathcal  C_{n_4}^E \right>_I \right>=\frac{1}{2}\delta_{n_1,n_3}\left(\left<\mathcal C_{n_1}^E\big| \mathcal C_{n_2}^E \right>_I \cdot\left<\mathcal C_{n_3}^E\big| \mathcal C_{n_4}^E \right>_I +\left<\mathcal S_{n_1}^E\big| \mathcal C_{n_2}^E  \right>_I \cdot \left<\mathcal S_{n_3}^E\big| \mathcal C_{n_4}^E \right>_I \right), \nonumber \\
\left<\left<\mathcal S_{n_1}^P \big| t \mathcal  S_{n_2}^E \right>_I \cdot \left<\mathcal S_{n_3}^P \big| t \mathcal  S_{n_4}^E \right>_I \right>= \frac{1}{2}\delta_{n_1,n_3} \left(\left<\mathcal S_{n_1}^E\big| t  \mathcal S_{n_2}^E \right>_I \cdot\left<\mathcal S_{n_3}^E\big| t  \mathcal S_{n_4}^E \right>_I+ \left<\mathcal C_{n_1}^E\big| t  \mathcal S_{n_2}^E  \right>_I \cdot \left<\mathcal C_{n_3}^E\big| t  \mathcal S_{n_4}^E \right>_I \right), \nonumber \\
\left<\left<\mathcal S_{n_1}^P \big| t \mathcal  S_{n_2}^E \right>_I \cdot \left<\mathcal S_{n_3}^P \big| t \mathcal  C_{n_4}^E \right>_I \right>= \frac{1}{2}\delta_{n_1,n_3} \left(\left<\mathcal S_{n_1}^E\big| t  \mathcal S_{n_2}^E \right>_I \cdot\left<\mathcal S_{n_3}^E\big| t  \mathcal C_{n_4}^E \right>_I+ \left<\mathcal C_{n_1}^E\big| t  \mathcal S_{n_2}^E  \right>_I \cdot \left<\mathcal C_{n_3}^E\big| t  \mathcal C_{n_4}^E \right>_I \right),         
\end{align}
and
\begin{align}
\left<\left<\mathcal S_{n_1}^P \big| t \mathcal  S_{n_2}^E \right>_I \cdot \left<\mathcal C_{n_3}^P \big| t \mathcal  S_{n_4}^E \right>_I \right>= \frac{1}{2}\delta_{n_1,n_3} \left(\left<\mathcal S_{n_1}^E\big| t  \mathcal S_{n_2}^E \right>_I \cdot\left<\mathcal C_{n_3}^E\big| t  \mathcal S_{n_4}^E \right>_I -\left<\mathcal C_{n_1}^E\big| t  \mathcal S_{n_2}^E  \right>_I \cdot \left<\mathcal S_{n_3}^E\big| t  \mathcal S_{n_4}^E \right>_I \right),         \nonumber \\
\left<\left<\mathcal S_{n_1}^P \big| t \mathcal  S_{n_2}^E \right>_I \cdot \left<\mathcal C_{n_3}^P \big| t \mathcal  C_{n_4}^E \right>_I \right>= \frac{1}{2}\delta_{n_1,n_3} \left(\left<\mathcal S_{n_1}^E\big| t  \mathcal S_{n_2}^E \right>_I \cdot\left<\mathcal C_{n_3}^E\big| t  \mathcal C_{n_4}^E \right>_I -\left<\mathcal C_{n_1}^E\big| t  \mathcal S_{n_2}^E  \right>_I \cdot \left<\mathcal S_{n_3}^E\big| t  \mathcal C_{n_4}^E \right>_I \right), \nonumber \\
\left<\left<\mathcal S_{n_1}^P \big| t \mathcal  C_{n_2}^E \right>_I \cdot \left<\mathcal S_{n_3}^P \big| t \mathcal  S_{n_4}^E \right>_I \right>=\frac{1}{2}\delta_{n_1,n_3} \left(\left<\mathcal S_{n_1}^E\big| t  \mathcal C_{n_2}^E \right>_I \cdot\left<\mathcal S_{n_3}^E\big| t  \mathcal S_{n_4}^E \right>_I +\left<\mathcal C_{n_1}^E\big| t  \mathcal C_{n_2}^E  \right>_I \cdot \left<\mathcal C_{n_3}^E\big| t  \mathcal S_{n_4}^E \right>_I \right), ~ \nonumber \\
\left<\left<\mathcal S_{n_1}^P \big| t \mathcal  C_{n_2}^E \right>_I \cdot \left<\mathcal S_{n_3}^P \big| t \mathcal C_{n_4}^E \right>_I \right>=\frac{1}{2}\delta_{n_1,n_3} \left(\left<\mathcal S_{n_1}^E\big| t  \mathcal C_{n_2}^E \right>_I \cdot\left<\mathcal S_{n_3}^E\big| t  \mathcal C_{n_4}^E \right>_I +\left<\mathcal C_{n_1}^E\big| t  \mathcal C_{n_2}^E  \right>_I \cdot \left<\mathcal C_{n_3}^E\big| t  \mathcal C_{n_4}^E \right>_I \right),     \nonumber \\
\left<\left<\mathcal S_{n_1}^P \big| t \mathcal  C_{n_2}^E \right>_I \cdot \left<\mathcal C_{n_3}^P \big| t \mathcal S_{n_4}^E \right>_I \right>=\frac{1}{2}\delta_{n_1,n_3} \left(\left<\mathcal S_{n_1}^E\big| t  \mathcal C_{n_2}^E \right>_I \cdot\left<\mathcal C_{n_3}^E\big| t  \mathcal S_{n_4}^E \right>_I -\left<\mathcal C_{n_1}^E\big| t  \mathcal C_{n_2}^E  \right>_I \cdot \left<\mathcal S_{n_3}^E\big| t  \mathcal S_{n_4}^E \right>_I \right),      \nonumber \\
\left<\left<\mathcal S_{n_1}^P \big| t \mathcal  C_{n_2}^E \right>_I \cdot \left<\mathcal C_{n_3}^P \big| t \mathcal C_{n_4}^E \right>_I \right>=\frac{1}{2}\delta_{n_1,n_3} \left(\left<\mathcal S_{n_1}^E\big| t  \mathcal C_{n_2}^E \right>_I \cdot\left<\mathcal C_{n_3}^E\big| t  \mathcal C_{n_4}^E \right>_I -\left<\mathcal C_{n_1}^E\big| t  \mathcal C_{n_2}^E  \right>_I \cdot \left<\mathcal S_{n_3}^E\big| t  \mathcal C_{n_4}^E \right>_I \right),   \nonumber \\
\left<\left<\mathcal C_{n_1}^P \big| t \mathcal  S_{n_2}^E \right>_I \cdot \left<\mathcal S_{n_3}^P \big| t \mathcal  S_{n_4}^E \right>_I \right>=\frac{1}{2}\delta_{n_1,n_3}\left(\left<\mathcal C_{n_1}^E\big| t  \mathcal S_{n_2}^E \right>_I \cdot\left<\mathcal S_{n_3}^E\big| t  \mathcal S_{n_4}^E \right>_I -\left<\mathcal S_{n_1}^E\big| t  \mathcal S_{n_2}^E  \right>_I \cdot \left<\mathcal C_{n_3}^E\big| t  \mathcal S_{n_4}^E \right>_I \right),    \nonumber \\
\left<\left<\mathcal C_{n_1}^P \big| t \mathcal  S_{n_2}^E \right>_I \cdot \left<\mathcal S_{n_3}^P \big| t \mathcal  C_{n_4}^E \right>_I \right>=\frac{1}{2}\delta_{n_1,n_3}\left(\left<\mathcal C_{n_1}^E\big| t  \mathcal S_{n_2}^E \right>_I \cdot\left<\mathcal S_{n_3}^E\big| t  \mathcal C_{n_4}^E \right>_I -\left<\mathcal S_{n_1}^E\big| t  \mathcal S_{n_2}^E  \right>_I \cdot \left<\mathcal C_{n_3}^E\big| t  \mathcal C_{n_4}^E \right>_I \right), \nonumber \\
\left<\left<\mathcal C_{n_1}^P \big| t \mathcal  S_{n_2}^E \right>_I \cdot \left<\mathcal C_{n_3}^P \big| t \mathcal  S_{n_4}^E \right>_I \right>=\frac{1}{2}\delta_{n_1,n_3}\left(\left<\mathcal C_{n_1}^E\big| t  \mathcal S_{n_2}^E \right>_I \cdot\left<\mathcal C_{n_3}^E\big| t  \mathcal S_{n_4}^E \right>_I +\left<\mathcal S_{n_1}^E\big| t  \mathcal S_{n_2}^E  \right>_I \cdot \left<\mathcal S_{n_3}^E\big| t  \mathcal S_{n_4}^E \right>_I \right),      \nonumber \\
\left<\left<\mathcal C_{n_1}^P \big| t \mathcal  S_{n_2}^E \right>_I \cdot \left<\mathcal C_{n_3}^P \big| t \mathcal  C_{n_4}^E \right>_I \right>=\frac{1}{2}\delta_{n_1,n_3}\left(\left<\mathcal C_{n_1}^E\big| t  \mathcal S_{n_2}^E \right>_I \cdot\left<\mathcal C_{n_3}^E\big| t  \mathcal C_{n_4}^E \right>_I +\left<\mathcal S_{n_1}^E\big| t  \mathcal S_{n_2}^E  \right>_I \cdot \left<\mathcal S_{n_3}^E\big| t  \mathcal C_{n_4}^E \right>_I \right),      \nonumber \\
\left<\left<\mathcal C_{n_1}^P \big| t \mathcal  C_{n_2}^E \right>_I \cdot \left<\mathcal S_{n_3}^P \big| t \mathcal  S_{n_4}^E \right>_I \right>=\frac{1}{2}\delta_{n_1,n_3}\left(\left<\mathcal C_{n_1}^E\big| t  \mathcal C_{n_2}^E \right>_I \cdot\left<\mathcal S_{n_3}^E\big| t  \mathcal S_{n_4}^E \right>_I -\left<\mathcal S_{n_1}^E\big| t  \mathcal C_{n_2}^E  \right>_I \cdot \left<\mathcal C_{n_3}^E\big| t  \mathcal S_{n_4}^E \right>_I \right),      \nonumber \\
\left<\left<\mathcal C_{n_1}^P \big| t \mathcal  C_{n_2}^E \right>_I \cdot \left<\mathcal S_{n_3}^P \big| t \mathcal  C_{n_4}^E \right>_I \right>=\frac{1}{2}\delta_{n_1,n_3}\left(\left<\mathcal C_{n_1}^E\big| t  \mathcal C_{n_2}^E \right>_I \cdot\left<\mathcal S_{n_3}^E\big| t  \mathcal C_{n_4}^E \right>_I -\left<\mathcal S_{n_1}^E\big| t  \mathcal C_{n_2}^E  \right>_I \cdot \left<\mathcal C_{n_3}^E\big| t  \mathcal C_{n_4}^E \right>_I \right),      \nonumber \\
\left<\left<\mathcal C_{n_1}^P \big| t \mathcal  C_{n_2}^E \right>_I \cdot \left<\mathcal C_{n_3}^P \big| t \mathcal  S_{n_4}^E \right>_I \right>=\frac{1}{2}\delta_{n_1,n_3}\left(\left<\mathcal C_{n_1}^E\big| t  \mathcal C_{n_2}^E \right>_I \cdot\left<\mathcal C_{n_3}^E\big| t  \mathcal S_{n_4}^E \right>_I +\left<\mathcal S_{n_1}^E\big| t  \mathcal C_{n_2}^E  \right>_I \cdot \left<\mathcal S_{n_3}^E\big| t  \mathcal S_{n_4}^E \right>_I \right),      \nonumber \\
\left<\left<\mathcal C_{n_1}^P \big| t \mathcal  C_{n_2}^E \right>_I \cdot \left<\mathcal C_{n_3}^P \big| t \mathcal  C_{n_4}^E \right>_I \right>=\frac{1}{2}\delta_{n_1,n_3}\left(\left<\mathcal C_{n_1}^E\big| t  \mathcal C_{n_2}^E \right>_I \cdot\left<\mathcal C_{n_3}^E\big| t  \mathcal C_{n_4}^E \right>_I +\left<\mathcal S_{n_1}^E\big| t  \mathcal C_{n_2}^E  \right>_I \cdot \left<\mathcal S_{n_3}^E\big| t  \mathcal C_{n_4}^E \right>_I \right).
\end{align}

Finally, inserting the above results into Eqs. \eqref{eq:xi_1-xi_1-eccentric}-\eqref{eq:xi_a-xi_b-eccentric},
one obtains the resulting Eqs. \eqref{eq:xi_1-xi_1-eccentric-(2)}-\eqref{eq:xi_a-xi_b-eccentric-(2)}.

\section{Components of the Matrix $\left< \xi_a \xi_b  \right>$} \label{Sec: results-eccentric}~

Expand Eq. \eqref{eq:xi_k^2(3)},
all the elements of the matrix $\left< \xi_a \xi_b  \right>$ are written as
\begin{align}
\label{eq:xi_1-xi_1-eccentric}
\left< \xi_1 \xi_1 \right>
&=\frac{4 \pi^2 }{T^4}{\mathop \sum \limits_{I=1}^{N_{\rm P}}}{\mathop \sum \limits_{n_1=1}^{N_I}}{\mathop \sum \limits_{n_2=1}^{N_I}} {\mathop \sum \limits_{n_3=1}^{N_I}}{\mathop \sum \limits_{n_4=1}^{N_I}}  n_2 n_4  \times \bigg[
A_{n_1} B_{n_2} A_{n_3} B_{n_4} \left<\left<\mathcal S_{n_1}^P \big|t \mathcal  S_{n_2}^E \right>_I   \left<\mathcal S_{n_3}^P \big|t \mathcal  S_{n_4}^E \right>_I \right>    \\
&-A_{n_1} B_{n_2} A_{n_3} A_{n_4} \left< \left<\mathcal S_{n_1}^P \big|t \mathcal  S_{n_2}^E \right>_I    \left<\mathcal S_{n_1}^P \big|t \mathcal  C_{n_2}^E\right>_I \right>
+A_{n_1} B_{n_2} B_{n_3}B_{n_4} \left< \left<\mathcal S_{n_1}^P \big|t \mathcal  S_{n_2}^E \right>_I   \left<\mathcal C_{n_3}^P \big|t \mathcal S_{n_4}^E\right>_I \right>    \nonumber \\
&-A_{n_1} B_{n_2} B_{n_3}A_{n_4} \left< \left<\mathcal S_{n_1}^P \big|t \mathcal  S_{n_2}^E \right>_I   \left<\mathcal C_{n_3}^P \big|t \mathcal C_{n_4}^E\right>_I \right>
-A_{n_1} A_{n_2} A_{n_3} B_{n_4} \left<\left<\mathcal S_{n_1}^P \big|t \mathcal  C_{n_2}^E \right>_I    \left<\mathcal S_{n_3}^P \big|t \mathcal  S_{n_4}^E \right>_I \right>   \nonumber \\
&+A_{n_1} A_{n_2} A_{n_3} A_{n_4} \left< \left<\mathcal S_{n_1}^P \big|t \mathcal  C_{n_2}^E \right>_I    \left<\mathcal S_{n_1}^P \big|t \mathcal  C_{n_2}^E\right>_I \right>
-A_{n_1} A_{n_2} B_{n_3}B_{n_4} \left< \left<\mathcal S_{n_1}^P \big|t \mathcal  C_{n_2}^E \right>_I   \left<\mathcal C_{n_3}^P \big|t \mathcal S_{n_4}^E\right>_I \right>  \nonumber \\
&+A_{n_1} A_{n_2} B_{n_3} A_{n_4} \left< \left<\mathcal S_{n_1}^P \big|t \mathcal  C_{n_2}^E \right>_I   \left<\mathcal C_{n_3}^P \big|t \mathcal  C_{n_4}^E\right>_I \right>
+B_{n_1} B_{n_2} A_{n_3} B_{n_4} \left<\left<\mathcal C_{n_1}^P \big|t \mathcal  S_{n_2}^E \right>_I   \left<\mathcal S_{n_3}^P \big|t \mathcal  S_{n_4}^E \right>_I \right>           \nonumber \\
&-B_{n_1} B_{n_2} A_{n_3} A_{n_4} \left< \left<\mathcal C_{n_1}^P \big|t \mathcal  S_{n_2}^E \right>_I    \left<\mathcal S_{n_1}^P \big|t \mathcal  C_{n_2}^E\right>_I \right>
+B_{n_1} B_{n_2} B_{n_3}B_{n_4} \left< \left<\mathcal C_{n_1}^P \big|t  \mathcal S_{n_2}^E \right>_I   \left<\mathcal C_{n_3}^P \big|t \mathcal  S_{n_4}^E\right>_I \right>   \nonumber \\
&-B_{n_1} B_{n_2} B_{n_3} A_{n_4} \left< \left<\mathcal C_{n_1}^P \big|t \mathcal  S_{n_2}^E \right>_I   \left<\mathcal C_{n_3}^P \big|t \mathcal  C_{n_4}^E\right>_I \right>
-B_{n_1} A_{n_2} A_{n_3} B_{n_4} \left<\left<\mathcal C_{n_1}^P \big|t \mathcal  C_{n_2}^E \right>_I   \left<\mathcal S_{n_3}^P \big|t \mathcal  S_{n_4}^E \right>_I \right>          \nonumber \\
&+B_{n_1} A_{n_2} A_{n_3} A_{n_4} \left< \left<\mathcal C_{n_1}^P \big|t \mathcal  C_{n_2}^E \right>_I    \left<\mathcal S_{n_1}^P \big|t \mathcal  C_{n_2}^E\right>_I \right>
-B_{n_1} A_{n_2} B_{n_3}B_{n_4} \left< \left<\mathcal C_{n_1}^P \big|t \mathcal  C_{n_2}^E \right>_I   \left<\mathcal C_{n_3}^P \big|t \mathcal  S_{n_4}^E\right>_I \right>   \nonumber \\
&+B_{n_1} A_{n_2} B_{n_3} A_{n_4} \left< \left<\mathcal C_{n_1}^P \big|t \mathcal  C_{n_2}^E \right>_I   \left<\mathcal C_{n_3}^P \big|t \mathcal  C_{n_4}^E\right>_I \right> \bigg], \nonumber
\end{align}

\begin{align}
\label{eq:xi_1-xi_a-eccentric}
\left< \xi_1 \xi_a \right>=   &   \left< \xi_a \xi_1 \right>   = \frac{2\pi}{T^2}   {\mathop \sum \limits_{I=1}^{N_{\rm P}}}    {\mathop \sum \limits_{n_1=1}^{N_I}}{\mathop \sum \limits_{n_2=1}^{N_I}} {\mathop \sum \limits_{n_3=1}^{N_I}}{\mathop \sum \limits_{n_4=1}^{N_I}} n_2 \times
\bigg[  A_{n_1} B_{n_2} A_{n_3} D_{a, n_4} \left<\left<\mathcal S_{n_1}^P \big| t \mathcal  S_{n_2}^E \right>_I    \left<\mathcal S_{n_3}^P \big| \mathcal  S_{n_4}^E \right>_I \right>  \\
& +A_{n_1}B _{n_2} A_{n_3} E_{a, n_4} \left<\left<\mathcal S_{n_1}^P \big| t \mathcal  S_{n_2}^E \right>_I    \left<\mathcal S_{n_3}^P \big| \mathcal  C_{n_4}^E \right>_I \right>
+A_{n_1} B_{n_2} B_{n_3} D_{a, n_4} \left<\left<\mathcal S_{n_1}^P \big| t \mathcal  S_{n_2}^E \right>_I    \left<\mathcal C_{n_3}^P \big| \mathcal  S_{n_4}^E \right>_I \right>            \nonumber \\
&+A_{n_1} B_{n_2} B_{n_3} E_{a, n_4} \left<\left<\mathcal S_{n_1}^P \big| t \mathcal  S_{n_2}^E \right>_I    \left<\mathcal C_{n_3}^P \big| \mathcal  C_{n_4}^E \right>_I \right>
 -A_{n_1} A_{n_2} A_{n_3} D_{a, n_4} \left<\left<\mathcal S_{n_1}^P \big| t \mathcal  C_{n_2}^E \right>_I    \left<\mathcal S_{n_3}^P \big| \mathcal  S_{n_4}^E \right>_I \right>        \nonumber \\
& -A_{n_1} A_{n_2} A_{n_3} E_{a, n_4} \left<\left<\mathcal S_{n_1}^P \big| t \mathcal  C_{n_2}^E \right>_I    \left<\mathcal S_{n_3}^P \big| \mathcal  C_{n_4}^E \right>_I \right>
-A_{n_1} A_{n_2} B_{n_3} D_{a, n_4} \left<\left<\mathcal S_{n_1}^P \big| t \mathcal  C_{n_2}^E \right>_I    \left<\mathcal C_{n_3}^P \big| \mathcal  S_{n_4}^E \right>_I \right>            \nonumber \\
&-A_{n_1} A_{n_2} B_{n_3} E_{a, n_4} \left<\left<\mathcal S_{n_1}^P \big| t \mathcal  C_{n_2}^E \right>_I    \left<\mathcal C_{n_3}^P \big| \mathcal  C_{n_4}^E \right>_I \right>
+ B_{n_1} B_{n_2} A_{n_3} D_{a, n_4} \left<\left<\mathcal C_{n_1}^P \big| t \mathcal  S_{n_2}^E \right>_I    \left<\mathcal S_{n_3}^P \big| \mathcal  S_{n_4}^E \right>_I \right>           \nonumber \\
& +B_{n_1} B_{n_2} A_{n_3} E_{a, n_4} \left<\left<\mathcal C_{n_1}^P \big| t \mathcal  S_{n_2}^E \right>_I    \left<\mathcal S_{n_3}^P \big| \mathcal  C_{n_4}^E \right>_I \right>
+B_{n_1} B_{n_2} B_{n_3} D_{a, n_4} \left<\left<\mathcal C_{n_1}^P \big| t \mathcal  S_{n_2}^E \right>_I    \left<\mathcal C_{n_3}^P \big| \mathcal  S_{n_4}^E \right>_I \right>            \nonumber \\
&+B_{n_1} B_{n_2} B_{n_3} E_{a, n_4} \left<\left<\mathcal C_{n_1}^P \big| t \mathcal  S_{n_2}^E \right>_I    \left<\mathcal C_{n_3}^P \big| \mathcal  C_{n_4}^E \right>_I \right>
-B_{n_1} A_{n_2} A_{n_3} D_{a, n_4} \left<\left<\mathcal C_{n_1}^P \big|t \mathcal  C_{n_2}^E \right>_I    \left<\mathcal S_{n_3}^P \big| \mathcal  S_{n_4}^E \right>_I \right>           \nonumber \\
& -B_{n_1} A_{n_2} A_{n_3} E_{a, n_4} \left<\left<\mathcal C_{n_1}^P \big| t \mathcal  C_{n_2}^E \right>_I    \left<\mathcal S_{n_3}^P \big| \mathcal  C_{n_4}^E \right>_I \right>
-B_{n_1} A_{n_2} B_{n_3} D_{a, n_4} \left<\left<\mathcal C_{n_1}^P \big| t \mathcal  C_{n_2}^E \right>_I    \left<\mathcal C_{n_3}^P \big| \mathcal  S_{n_4}^E \right>_I \right>            \nonumber \\
&-B_{n_1} A_{n_2} B_{n_3} E_{a, n_4} \left<\left<\mathcal C_{n_1}^P \big| t \mathcal  C_{n_2}^E \right>_I    \left<\mathcal C_{n_3}^P \big| \mathcal  C_{n_4}^E \right>_I \right>   \bigg],          \nonumber
\end{align}
for $a \neq 1$, and
\begin{align}
\label{eq:xi_a-xi_b-eccentric}
\left< \xi_a \xi_b \right> = {\mathop \sum \limits_{I=1}^{N_{\rm P}}}      &      {\mathop \sum \limits_{n_1=1}^{N_I}}{\mathop \sum \limits_{n_2=1}^{N_I}} {\mathop \sum \limits_{n_3=1}^{N_I}}{\mathop \sum \limits_{n_4=1}^{N_I}} \bigg[  A_{n_1} D_{a,n_2} A_{n_3} D_{b, n_4} \left<\left<\mathcal S_{n_1}^P \big| \mathcal  S_{n_2}^E \right>_I    \left<\mathcal S_{n_3}^P \big| \mathcal  S_{n_4}^E \right>_I \right>  \\
& +A_{n_1} D_{a,n_2} A_{n_3} E_{b, n_4} \left<\left<\mathcal S_{n_1}^P \big| \mathcal  S_{n_2}^E \right>_I    \left<\mathcal S_{n_3}^P \big| \mathcal  C_{n_4}^E \right>_I \right>
+A_{n_1} D_{a,n_2} B_{n_3} D_{b, n_4} \left<\left<\mathcal S_{n_1}^P \big| \mathcal  S_{n_2}^E \right>_I    \left<\mathcal C_{n_3}^P \big| \mathcal  S_{n_4}^E \right>_I \right>            \nonumber \\
&+A_{n_1} D_{a,n_2} B_{n_3} E_{b, n_4} \left<\left<\mathcal S_{n_1}^P \big| \mathcal  S_{n_2}^E \right>_I    \left<\mathcal C_{n_3}^P \big| \mathcal  C_{n_4}^E \right>_I \right>
+ A_{n_1} E_{a,n_2} A_{n_3} D_{b, n_4} \left<\left<\mathcal S_{n_1}^P \big| \mathcal  C_{n_2}^E \right>_I    \left<\mathcal S_{n_3}^P \big| \mathcal  S_{n_4}^E \right>_I \right>        \nonumber \\
& +A_{n_1} E_{a,n_2} A_{n_3} E_{b, n_4} \left<\left<\mathcal S_{n_1}^P \big| \mathcal  C_{n_2}^E \right>_I    \left<\mathcal S_{n_3}^P \big| \mathcal  C_{n_4}^E \right>_I \right>
+A_{n_1} E_{a,n_2} B_{n_3} D_{b, n_4} \left<\left<\mathcal S_{n_1}^P \big| \mathcal  C_{n_2}^E \right>_I    \left<\mathcal C_{n_3}^P \big| \mathcal  S_{n_4}^E \right>_I \right>            \nonumber \\
&+A_{n_1} E_{a,n_2} B_{n_3} E_{b, n_4} \left<\left<\mathcal S_{n_1}^P \big| \mathcal  C_{n_2}^E \right>_I    \left<\mathcal C_{n_3}^P \big| \mathcal  C_{n_4}^E \right>_I \right>
+ B_{n_1} D_{a,n_2} A_{n_3} D_{b, n_4} \left<\left<\mathcal C_{n_1}^P \big| \mathcal  S_{n_2}^E \right>_I    \left<\mathcal S_{n_3}^P \big| \mathcal  S_{n_4}^E \right>_I \right>           \nonumber \\
& +B_{n_1} D_{a,n_2} A_{n_3} E_{b, n_4} \left<\left<\mathcal C_{n_1}^P \big| \mathcal  S_{n_2}^E \right>_I    \left<\mathcal S_{n_3}^P \big| \mathcal  C_{n_4}^E \right>_I \right>
+B_{n_1} D_{a,n_2} B_{n_3} D_{b, n_4} \left<\left<\mathcal C_{n_1}^P \big| \mathcal  S_{n_2}^E \right>_I    \left<\mathcal C_{n_3}^P \big| \mathcal  S_{n_4}^E \right>_I \right>            \nonumber \\
&+B_{n_1} D_{a,n_2} B_{n_3} E_{b, n_4} \left<\left<\mathcal C_{n_1}^P \big| \mathcal  S_{n_2}^E \right>_I    \left<\mathcal C_{n_3}^P \big| \mathcal  C_{n_4}^E \right>_I \right>
+ B_{n_1} E_{a,n_2} A_{n_3} D_{b, n_4} \left<\left<\mathcal C_{n_1}^P \big| \mathcal  C_{n_2}^E \right>_I    \left<\mathcal S_{n_3}^P \big| \mathcal  S_{n_4}^E \right>_I \right>           \nonumber \\
& +B_{n_1} E_{a,n_2} A_{n_3} E_{b, n_4} \left<\left<\mathcal C_{n_1}^P \big| \mathcal  C_{n_2}^E \right>_I    \left<\mathcal S_{n_3}^P \big| \mathcal  C_{n_4}^E \right>_I \right>
+B_{n_1} E_{a,n_2} B_{n_3} D_{b, n_4} \left<\left<\mathcal C_{n_1}^P \big| \mathcal  C_{n_2}^E \right>_I    \left<\mathcal C_{n_3}^P \big| \mathcal  S_{n_4}^E \right>_I \right>            \nonumber \\
&+B_{n_1} E_{a,n_2} B_{n_3} E_{b, n_4} \left<\left<\mathcal C_{n_1}^P \big| \mathcal  C_{n_2}^E \right>_I    \left<\mathcal C_{n_3}^P \big| \mathcal  C_{n_4}^E \right>_I \right>   \bigg].          \nonumber
\end{align}
for both $a\neq 1$ and $b \neq 1$.

After calculating all the averaged terms in Eqs. \eqref{eq:xi_1-xi_1-eccentric}-\eqref{eq:xi_a-xi_b-eccentric} (see Appendix \ref{Sec: eccentric-expected} for details),
we finally obtain the elements of the matrix as follows
\begin{align}
\label{eq:xi_1-xi_1-eccentric-(2)}
\left< \xi_1 \xi_1 \right> = \frac{2\pi^2 }{T^4} {\mathop \sum \limits_{I=1}^{N_{\rm P}}}{\mathop \sum \limits_{n_1=1}^{N_I}}   &    {\mathop \sum \limits_{n_2=1}^{N_I}}{\mathop \sum \limits_{n_4=1}^{N_I}}    n_2 n_4  \bigg\{
\ A_{n_1}^2 B_{n_2} B_{n_4}  \left(\left<\mathcal S_{n_1}^E\big| t  \mathcal S_{n_2}^E \right>_I  \left<\mathcal S_{n_1}^E\big| t  \mathcal S_{n_4}^E \right>_I+ \left<\mathcal C_{n_1}^E\big| t  \mathcal S_{n_2}^E  \right>_I   \left<\mathcal C_{n_1}^E\big| t  \mathcal S_{n_4}^E \right>_I \right)  \\  
&-A_{n_1}^2 B_{n_2} A_{n_4} \left(\left<\mathcal S_{n_1}^E\big| t  \mathcal S_{n_2}^E \right>_I  \left<\mathcal S_{n_1}^E\big| t  \mathcal C_{n_4}^E \right>_I+ \left<\mathcal C_{n_1}^E\big| t  \mathcal S_{n_2}^E  \right>_I   \left<\mathcal C_{n_1}^E\big| t  \mathcal C_{n_4}^E \right>_I \right) \nonumber \\  
&+A_{n_1} B_{n_2} B_{n_1}B_{n_4}\left(\left<\mathcal S_{n_1}^E\big| t  \mathcal S_{n_2}^E \right>_I  \left<\mathcal C_{n_1}^E\big| t  \mathcal S_{n_4}^E \right>_I -\left<\mathcal C_{n_1}^E\big| t  \mathcal S_{n_2}^E  \right>_I   \left<\mathcal S_{n_1}^E\big| t  \mathcal S_{n_4}^E \right>_I \right)  \nonumber \\  
&-A_{n_1} B_{n_2} B_{n_1} A_{n_4} \left(\left<\mathcal S_{n_1}^E\big| t  \mathcal S_{n_2}^E \right>_I  \left<\mathcal C_{n_1}^E\big| t  \mathcal C_{n_4}^E \right>_I -\left<\mathcal C_{n_1}^E\big| t  \mathcal S_{n_2}^E  \right>_I   \left<\mathcal S_{n_1}^E\big| t  \mathcal C_{n_4}^E \right>_I \right)  \nonumber \\  
&-A_{n_1}^2 A_{n_2} B_{n_4} \left(\left<\mathcal S_{n_1}^E\big| t  \mathcal C_{n_2}^E \right>_I  \left<\mathcal S_{n_1}^E\big| t  \mathcal S_{n_4}^E \right>_I +\left<\mathcal C_{n_1}^E\big| t  \mathcal C_{n_2}^E  \right>_I   \left<\mathcal C_{n_1}^E\big| t  \mathcal S_{n_4}^E \right>_I \right) \nonumber \\             
&+A_{n_1}^2 A_{n_2}  A_{n_4} \left(\left<\mathcal S_{n_1}^E\big| t  \mathcal C_{n_2}^E \right>_I  \left<\mathcal S_{n_1}^E\big| t  \mathcal C_{n_4}^E \right>_I +\left<\mathcal C_{n_1}^E\big| t  \mathcal C_{n_2}^E  \right>_I   \left<\mathcal C_{n_1}^E\big| t  \mathcal C_{n_4}^E \right>_I \right) \nonumber \\    
&-A_{n_1} A_{n_2} B_{n_1}B_{n_4} \left(\left<\mathcal S_{n_1}^E\big| t  \mathcal C_{n_2}^E \right>_I  \left<\mathcal C_{n_1}^E\big| t  \mathcal S_{n_4}^E \right>_I -\left<\mathcal C_{n_1}^E\big| t  \mathcal C_{n_2}^E  \right>_I   \left<\mathcal S_{n_1}^E\big| t  \mathcal S_{n_4}^E \right>_I \right)  \nonumber \\   
&+A_{n_1} A_{n_2} B_{n_1} A_{n_4} \left(\left<\mathcal S_{n_1}^E\big| t  \mathcal C_{n_2}^E \right>_I  \left<\mathcal C_{n_1}^E\big| t  \mathcal C_{n_4}^E \right>_I -\left<\mathcal C_{n_1}^E\big| t  \mathcal C_{n_2}^E  \right>_I   \left<\mathcal S_{n_1}^E\big| t  \mathcal C_{n_4}^E \right>_I \right)  \nonumber \\   
&+B_{n_1} B_{n_2} A_{n_1} B_{n_4} \left(\left<\mathcal C_{n_1}^E\big| t  \mathcal S_{n_2}^E \right>_I  \left<\mathcal S_{n_1}^E\big| t  \mathcal S_{n_4}^E \right>_I -\left<\mathcal S_{n_1}^E\big| t  \mathcal S_{n_2}^E  \right>_I   \left<\mathcal C_{n_1}^E\big| t  \mathcal S_{n_4}^E \right>_I \right) \nonumber \\    
&-B_{n_1} B_{n_2} A_{n_1} A_{n_4} \left(\left<\mathcal C_{n_1}^E\big| t  \mathcal S_{n_2}^E \right>_I  \left<\mathcal S_{n_1}^E\big| t  \mathcal C_{n_4}^E \right>_I -\left<\mathcal S_{n_1}^E\big| t  \mathcal S_{n_2}^E  \right>_I   \left<\mathcal C_{n_1}^E\big| t  \mathcal C_{n_4}^E \right>_I \right) \nonumber \\   
&+B_{n_1}^2 B_{n_2} B_{n_4} \left(\left<\mathcal C_{n_1}^E\big| t  \mathcal S_{n_2}^E \right>_I  \left<\mathcal C_{n_1}^E\big| t  \mathcal S_{n_4}^E \right>_I +\left<\mathcal S_{n_1}^E\big| t  \mathcal S_{n_2}^E  \right>_I   \left<\mathcal S_{n_1}^E\big| t  \mathcal S_{n_4}^E \right>_I \right) \nonumber \\      
&-B_{n_1}^2 B_{n_2}  A_{n_4} \left(\left<\mathcal C_{n_1}^E\big| t  \mathcal S_{n_2}^E \right>_I  \left<\mathcal C_{n_1}^E\big| t  \mathcal C_{n_4}^E \right>_I +\left<\mathcal S_{n_1}^E\big| t  \mathcal S_{n_2}^E  \right>_I   \left<\mathcal S_{n_1}^E\big| t  \mathcal C_{n_4}^E \right>_I \right) \nonumber \\   
&-B_{n_1} A_{n_2} A_{n_1} B_{n_4}\left(\left<\mathcal C_{n_1}^E\big| t  \mathcal C_{n_2}^E \right>_I  \left<\mathcal S_{n_1}^E\big| t  \mathcal S_{n_4}^E \right>_I -\left<\mathcal S_{n_1}^E\big| t  \mathcal C_{n_2}^E  \right>_I   \left<\mathcal C_{n_1}^E\big| t  \mathcal S_{n_4}^E \right>_I \right)  \nonumber \\  
&+B_{n_1} A_{n_2} A_{n_1} A_{n_4} \left(\left<\mathcal C_{n_1}^E\big| t  \mathcal C_{n_2}^E \right>_I  \left<\mathcal S_{n_1}^E\big| t  \mathcal C_{n_4}^E \right>_I -\left<\mathcal S_{n_1}^E\big| t  \mathcal C_{n_2}^E  \right>_I   \left<\mathcal C_{n_1}^E\big| t  \mathcal C_{n_4}^E \right>_I \right) \nonumber \\  
&-B_{n_1}^2 A_{n_2} B_{n_4} \left(\left<\mathcal C_{n_1}^E\big| t  \mathcal C_{n_2}^E \right>_I  \left<\mathcal C_{n_1}^E\big| t  \mathcal S_{n_4}^E \right>_I +\left<\mathcal S_{n_1}^E\big| t  \mathcal C_{n_2}^E  \right>_I   \left<\mathcal S_{n_1}^E\big| t  \mathcal S_{n_4}^E \right>_I \right)  \nonumber \\          
&+B_{n_1}^2 A_{n_2} A_{n_4} \left(\left<\mathcal C_{n_1}^E\big| t  \mathcal C_{n_2}^E \right>_I  \left<\mathcal C_{n_1}^E\big| t  \mathcal C_{n_4}^E \right>_I +\left<\mathcal S_{n_1}^E\big| t  \mathcal C_{n_2}^E  \right>_I   \left<\mathcal S_{n_1}^E\big| t  \mathcal C_{n_4}^E \right>_I \right)   \bigg\} ,\nonumber             
\end{align}

\begin{align}
\label{eq:xi_1-xi_a-eccentric-(2)}
\left< \xi_1 \xi_a \right> =\left< \xi_a \xi_1 \right> = \frac{\pi }{T^2} {\mathop \sum \limits_{I=1}^{N_{\rm P}}}  & {\mathop \sum \limits_{n_1=1}^{N_I}}{\mathop \sum \limits_{n_2=1}^{N_I}}{\mathop \sum \limits_{n_4=1}^{N_I}} n_2  \bigg\{
 \ A_{n_1}^2 B_{n_2} D_{a, n_4}  \left(\left<\mathcal S_{n_1}^E\big| t  \mathcal S_{n_2}^E \right>_I  \left<\mathcal S_{n_1}^E\big|   \mathcal S_{n_4}^E \right>_I+ \left<\mathcal C_{n_1}^E\big| t  \mathcal S_{n_2}^E  \right>_I   \left<\mathcal C_{n_1}^E\big|   \mathcal S_{n_4}^E \right>_I \right)          \\  
&+A_{n_1}^2 B_{n_2} E_{a, n_4} \left(\left<\mathcal S_{n_1}^E\big| t  \mathcal S_{n_2}^E \right>_I  \left<\mathcal S_{n_1}^E\big|   \mathcal C_{n_4}^E \right>_I+ \left<\mathcal C_{n_1}^E\big| t  \mathcal S_{n_2}^E  \right>_I   \left<\mathcal C_{n_1}^E\big|   \mathcal C_{n_4}^E \right>_I \right) \nonumber \\  
&+A_{n_1} B_{n_2} B_{n_1} D_{a, n_4}\left(\left<\mathcal S_{n_1}^E\big| t  \mathcal S_{n_2}^E \right>_I  \left<\mathcal C_{n_1}^E\big|   \mathcal S_{n_4}^E \right>_I -\left<\mathcal C_{n_1}^E\big| t  \mathcal S_{n_2}^E  \right>_I   \left<\mathcal S_{n_1}^E\big|   \mathcal S_{n_4}^E \right>_I \right)  \nonumber \\  
&+A_{n_1} B_{n_2} B_{n_1} E_{a, n_4} \left(\left<\mathcal S_{n_1}^E\big| t  \mathcal S_{n_2}^E \right>_I  \left<\mathcal C_{n_1}^E\big|   \mathcal C_{n_4}^E \right>_I -\left<\mathcal C_{n_1}^E\big| t  \mathcal S_{n_2}^E  \right>_I   \left<\mathcal S_{n_1}^E\big|   \mathcal C_{n_4}^E \right>_I \right)  \nonumber \\  
&-A_{n_1}^2 A_{n_2} D_{a, n_4} \left(\left<\mathcal S_{n_1}^E\big| t  \mathcal C_{n_2}^E \right>_I  \left<\mathcal S_{n_1}^E\big|   \mathcal S_{n_4}^E \right>_I +\left<\mathcal C_{n_1}^E\big| t  \mathcal C_{n_2}^E  \right>_I   \left<\mathcal C_{n_1}^E\big|   \mathcal S_{n_4}^E \right>_I \right) \nonumber \\             
&-A_{n_1}^2 A_{n_2}  E_{a, n_4} \left(\left<\mathcal S_{n_1}^E\big| t  \mathcal C_{n_2}^E \right>_I  \left<\mathcal S_{n_1}^E\big|   \mathcal C_{n_4}^E \right>_I +\left<\mathcal C_{n_1}^E\big| t  \mathcal C_{n_2}^E  \right>_I   \left<\mathcal C_{n_1}^E\big|   \mathcal C_{n_4}^E \right>_I \right) \nonumber \\    
&-A_{n_1} A_{n_2} B_{n_1}D_{a, n_4} \left(\left<\mathcal S_{n_1}^E\big| t  \mathcal C_{n_2}^E \right>_I  \left<\mathcal C_{n_1}^E\big|   \mathcal S_{n_4}^E \right>_I -\left<\mathcal C_{n_1}^E\big| t  \mathcal C_{n_2}^E  \right>_I   \left<\mathcal S_{n_1}^E\big|   \mathcal S_{n_4}^E \right>_I \right)  \nonumber \\   
&-A_{n_1} A_{n_2} B_{n_1} E_{a, n_4} \left(\left<\mathcal S_{n_1}^E\big| t  \mathcal C_{n_2}^E \right>_I  \left<\mathcal C_{n_1}^E\big|   \mathcal C_{n_4}^E \right>_I -\left<\mathcal C_{n_1}^E\big| t  \mathcal C_{n_2}^E  \right>_I   \left<\mathcal S_{n_1}^E\big|   \mathcal C_{n_4}^E \right>_I \right)  \nonumber \\   
&+B_{n_1} B_{n_2} A_{n_1} D_{a, n_4} \left(\left<\mathcal C_{n_1}^E\big| t  \mathcal S_{n_2}^E \right>_I  \left<\mathcal S_{n_1}^E\big|   \mathcal S_{n_4}^E \right>_I -\left<\mathcal S_{n_1}^E\big| t  \mathcal S_{n_2}^E  \right>_I   \left<\mathcal C_{n_1}^E\big|   \mathcal S_{n_4}^E \right>_I \right) \nonumber \\    
&+B_{n_1} B_{n_2} A_{n_1} E_{a, n_4} \left(\left<\mathcal C_{n_1}^E\big| t  \mathcal S_{n_2}^E \right>_I  \left<\mathcal S_{n_1}^E\big|   \mathcal C_{n_4}^E \right>_I -\left<\mathcal S_{n_1}^E\big| t  \mathcal S_{n_2}^E  \right>_I   \left<\mathcal C_{n_1}^E\big|   \mathcal C_{n_4}^E \right>_I \right) \nonumber \\   
&+B_{n_1}^2 B_{n_2} D_{a, n_4} \left(\left<\mathcal C_{n_1}^E\big| t  \mathcal S_{n_2}^E \right>_I  \left<\mathcal C_{n_1}^E\big|   \mathcal S_{n_4}^E \right>_I +\left<\mathcal S_{n_1}^E\big| t  \mathcal S_{n_2}^E  \right>_I   \left<\mathcal S_{n_1}^E\big|   \mathcal S_{n_4}^E \right>_I \right) \nonumber \\      
&+B_{n_1}^2 B_{n_2}  E_{a, n_4} \left(\left<\mathcal C_{n_1}^E\big| t  \mathcal S_{n_2}^E \right>_I  \left<\mathcal C_{n_1}^E\big|   \mathcal C_{n_4}^E \right>_I +\left<\mathcal S_{n_1}^E\big| t  \mathcal S_{n_2}^E  \right>_I   \left<\mathcal S_{n_1}^E\big|   \mathcal C_{n_4}^E \right>_I \right) \nonumber \\   
&-B_{n_1} A_{n_2} A_{n_1} D_{a, n_4}\left(\left<\mathcal C_{n_1}^E\big| t  \mathcal C_{n_2}^E \right>_I  \left<\mathcal S_{n_1}^E\big|   \mathcal S_{n_4}^E \right>_I -\left<\mathcal S_{n_1}^E\big| t  \mathcal C_{n_2}^E  \right>_I   \left<\mathcal C_{n_1}^E\big|   \mathcal S_{n_4}^E \right>_I \right)  \nonumber \\  
&-B_{n_1} A_{n_2} A_{n_1} E_{a, n_4} \left(\left<\mathcal C_{n_1}^E\big| t  \mathcal C_{n_2}^E \right>_I  \left<\mathcal S_{n_1}^E\big|   \mathcal C_{n_4}^E \right>_I -\left<\mathcal S_{n_1}^E\big| t  \mathcal C_{n_2}^E  \right>_I   \left<\mathcal C_{n_1}^E\big|   \mathcal C_{n_4}^E \right>_I \right) \nonumber \\  
&-B_{n_1}^2 A_{n_2} D_{a, n_4} \left(\left<\mathcal C_{n_1}^E\big| t  \mathcal C_{n_2}^E \right>_I  \left<\mathcal C_{n_1}^E\big|   \mathcal S_{n_4}^E \right>_I +\left<\mathcal S_{n_1}^E\big| t  \mathcal C_{n_2}^E  \right>_I   \left<\mathcal S_{n_1}^E\big|   \mathcal S_{n_4}^E \right>_I \right)  \nonumber \\          
&-B_{n_1}^2 A_{n_2} E_{a, n_4} \left(\left<\mathcal C_{n_1}^E\big| t  \mathcal C_{n_2}^E \right>_I  \left<\mathcal C_{n_1}^E\big|   \mathcal C_{n_4}^E \right>_I +\left<\mathcal S_{n_1}^E\big| t  \mathcal C_{n_2}^E  \right>_I   \left<\mathcal S_{n_1}^E\big|   \mathcal C_{n_4}^E \right>_I \right)   \bigg\} ,\nonumber             
\end{align}
for $a \neq 1$, and
\begin{align}
\label{eq:xi_a-xi_b-eccentric-(2)}
\left< \xi_a \xi_b \right> = \frac{1 }{2} {\mathop \sum \limits_{I=1}^{N_{\rm P}}}  & {\mathop \sum \limits_{n_1=1}^{N_I}}{\mathop \sum \limits_{n_2=1}^{N_I}}{\mathop \sum \limits_{n_4=1}^{N_I}}  \bigg\{
 \ A_{n_1}^2 D_{a, n_2} D_{b, n_4}  \left(\left<\mathcal S_{n_1}^E\big|  \mathcal S_{n_2}^E \right>_I  \left<\mathcal S_{n_1}^E\big|   \mathcal S_{n_4}^E \right>_I+ \left<\mathcal C_{n_1}^E\big|  \mathcal S_{n_2}^E  \right>_I   \left<\mathcal C_{n_1}^E\big|   \mathcal S_{n_4}^E \right>_I \right)          \\  
&+A_{n_1}^2 D_{a, n_2} E_{b, n_4} \left(\left<\mathcal S_{n_1}^E\big|  \mathcal S_{n_2}^E \right>_I  \left<\mathcal S_{n_1}^E\big|   \mathcal C_{n_4}^E \right>_I+ \left<\mathcal C_{n_1}^E\big|  \mathcal S_{n_2}^E  \right>_I   \left<\mathcal C_{n_1}^E\big|   \mathcal C_{n_4}^E \right>_I \right) \nonumber \\  
&+A_{n_1} D_{a, n_2} B_{n_1} D_{b, n_4}\left(\left<\mathcal S_{n_1}^E\big|  \mathcal S_{n_2}^E \right>_I  \left<\mathcal C_{n_1}^E\big|   \mathcal S_{n_4}^E \right>_I -\left<\mathcal C_{n_1}^E\big|  \mathcal S_{n_2}^E  \right>_I   \left<\mathcal S_{n_1}^E\big|   \mathcal S_{n_4}^E \right>_I \right)  \nonumber \\  
&+A_{n_1} D_{a, n_2} B_{n_1} E_{b, n_4} \left(\left<\mathcal S_{n_1}^E\big|  \mathcal S_{n_2}^E \right>_I  \left<\mathcal C_{n_1}^E\big|   \mathcal C_{n_4}^E \right>_I -\left<\mathcal C_{n_1}^E\big|  \mathcal S_{n_2}^E  \right>_I   \left<\mathcal S_{n_1}^E\big|   \mathcal C_{n_4}^E \right>_I \right)  \nonumber \\  
&+A_{n_1}^2 E_{a, n_2} D_{b, n_4} \left(\left<\mathcal S_{n_1}^E\big|  \mathcal C_{n_2}^E \right>_I  \left<\mathcal S_{n_1}^E\big|   \mathcal S_{n_4}^E \right>_I +\left<\mathcal C_{n_1}^E\big|  \mathcal C_{n_2}^E  \right>_I   \left<\mathcal C_{n_1}^E\big|   \mathcal S_{n_4}^E \right>_I \right) \nonumber \\             
&+A_{n_1}^2 E_{a, n_2}  E_{b, n_4} \left(\left<\mathcal S_{n_1}^E\big|  \mathcal C_{n_2}^E \right>_I  \left<\mathcal S_{n_1}^E\big|   \mathcal C_{n_4}^E \right>_I +\left<\mathcal C_{n_1}^E\big|  \mathcal C_{n_2}^E  \right>_I   \left<\mathcal C_{n_1}^E\big|   \mathcal C_{n_4}^E \right>_I \right) \nonumber \\    
&+A_{n_1} E_{a, n_2} B_{n_1}D_{b, n_4} \left(\left<\mathcal S_{n_1}^E\big|  \mathcal C_{n_2}^E \right>_I  \left<\mathcal C_{n_1}^E\big|   \mathcal S_{n_4}^E \right>_I -\left<\mathcal C_{n_1}^E\big|  \mathcal C_{n_2}^E  \right>_I   \left<\mathcal S_{n_1}^E\big|   \mathcal S_{n_4}^E \right>_I \right)  \nonumber \\   
&+A_{n_1} E_{a, n_2} B_{n_1} E_{b, n_4} \left(\left<\mathcal S_{n_1}^E\big|  \mathcal C_{n_2}^E \right>_I  \left<\mathcal C_{n_1}^E\big|   \mathcal C_{n_4}^E \right>_I -\left<\mathcal C_{n_1}^E\big|  \mathcal C_{n_2}^E  \right>_I   \left<\mathcal S_{n_1}^E\big|   \mathcal C_{n_4}^E \right>_I \right)  \nonumber \\   
&+B_{n_1} D_{a, n_2} A_{n_1} D_{b, n_4} \left(\left<\mathcal C_{n_1}^E\big|  \mathcal S_{n_2}^E \right>_I  \left<\mathcal S_{n_1}^E\big|   \mathcal S_{n_4}^E \right>_I -\left<\mathcal S_{n_1}^E\big|  \mathcal S_{n_2}^E  \right>_I   \left<\mathcal C_{n_1}^E\big|   \mathcal S_{n_4}^E \right>_I \right) \nonumber \\    
&+B_{n_1} D_{a, n_2} A_{n_1} E_{b, n_4} \left(\left<\mathcal C_{n_1}^E\big|  \mathcal S_{n_2}^E \right>_I  \left<\mathcal S_{n_1}^E\big|   \mathcal C_{n_4}^E \right>_I -\left<\mathcal S_{n_1}^E\big|  \mathcal S_{n_2}^E  \right>_I   \left<\mathcal C_{n_1}^E\big|   \mathcal C_{n_4}^E \right>_I \right) \nonumber \\   
&+B_{n_1}^2 D_{a, n_2} D_{b, n_4} \left(\left<\mathcal C_{n_1}^E\big|  \mathcal S_{n_2}^E \right>_I  \left<\mathcal C_{n_1}^E\big|   \mathcal S_{n_4}^E \right>_I +\left<\mathcal S_{n_1}^E\big|  \mathcal S_{n_2}^E  \right>_I   \left<\mathcal S_{n_1}^E\big|   \mathcal S_{n_4}^E \right>_I \right) \nonumber \\      
&+B_{n_1}^2 D_{a, n_2}  E_{b, n_4} \left(\left<\mathcal C_{n_1}^E\big|  \mathcal S_{n_2}^E \right>_I  \left<\mathcal C_{n_1}^E\big|   \mathcal C_{n_4}^E \right>_I +\left<\mathcal S_{n_1}^E\big|  \mathcal S_{n_2}^E  \right>_I   \left<\mathcal S_{n_1}^E\big|   \mathcal C_{n_4}^E \right>_I \right) \nonumber \\   
&+B_{n_1} E_{a, n_2} A_{n_1} D_{b, n_4}\left(\left<\mathcal C_{n_1}^E\big|  \mathcal C_{n_2}^E \right>_I  \left<\mathcal S_{n_1}^E\big|   \mathcal S_{n_4}^E \right>_I -\left<\mathcal S_{n_1}^E\big|  \mathcal C_{n_2}^E  \right>_I   \left<\mathcal C_{n_1}^E\big|   \mathcal S_{n_4}^E \right>_I \right)  \nonumber \\  
&+B_{n_1} E_{a, n_2} A_{n_1} E_{b, n_4} \left(\left<\mathcal C_{n_1}^E\big|  \mathcal C_{n_2}^E \right>_I  \left<\mathcal S_{n_1}^E\big|   \mathcal C_{n_4}^E \right>_I -\left<\mathcal S_{n_1}^E\big|  \mathcal C_{n_2}^E  \right>_I   \left<\mathcal C_{n_1}^E\big|   \mathcal C_{n_4}^E \right>_I \right) \nonumber \\  
&+B_{n_1}^2 E_{a, n_2} D_{b, n_4} \left(\left<\mathcal C_{n_1}^E\big|  \mathcal C_{n_2}^E \right>_I  \left<\mathcal C_{n_1}^E\big|   \mathcal S_{n_4}^E \right>_I +\left<\mathcal S_{n_1}^E\big|  \mathcal C_{n_2}^E  \right>_I   \left<\mathcal S_{n_1}^E\big|   \mathcal S_{n_4}^E \right>_I \right)  \nonumber \\          
&+B_{n_1}^2 E_{a, n_2} E_{b, n_4} \left(\left<\mathcal C_{n_1}^E\big|  \mathcal C_{n_2}^E \right>_I  \left<\mathcal C_{n_1}^E\big|   \mathcal C_{n_4}^E \right>_I +\left<\mathcal S_{n_1}^E\big|  \mathcal C_{n_2}^E  \right>_I   \left<\mathcal S_{n_1}^E\big|   \mathcal C_{n_4}^E \right>_I \right)   \bigg\} ,\nonumber             
\end{align}
for both $a \neq 1$ and $b \neq 1$.

\section{Coefficients $A$, $B$, $D_a$ and $E_a$ for the Circular Case} \label{Sec: D_a-E_a}

The coefficients $A$ and $B$ in the template \eqref{eq:template-rewritten-circular} are respectively
\begin{align}
\label{eq:A-B}
A=\mathcal A_m (1+c_i^2) (\cos 2\psi F_{I +}-\sin 2\psi F_{I \times}), ~~~~~~~~~~~~~~~
B= 2 \mathcal A_m  c_i  (\sin 2\psi F_{I +}+\cos 2\psi F_{I \times}).
\end{align}

The coefficients $D_{a}$ and $E_{a}$ over $a=1, 2, ...,7$ in Eq. \eqref{eq:s_I^E-coefficient-circular} are respectively
\begin{align}
\label{eq:D1-E1}
D_{1}=\frac{4\pi B}{T^2} t,~~~~~~~~~~~~~~~~~~~~~~
E_{1}=-\frac{4\pi A}{T^2} t,
\end{align}
\begin{align}
D_{2}=\frac{A}{\mathcal A_m},~~~~~~~~~~~~~~~~~~~~~~~~~~~~~~
E_{2}=\frac{B}{\mathcal A_m},
\end{align}
\begin{align}
D_{3}=\mathcal A_m (1+c_i^2) (\cos 2\psi F_{I +}^{(3)}-\sin 2\psi F_{I \times}^{(3)}), ~~~~~~~~~~~~
E_{3}=2 \mathcal A_m  c_i  (\sin 2\psi F_{I +}^{(3)}+\cos 2\psi F_{I \times}^{(3)}),
\end{align}
with $F_{I+}^{(3)}$ and $F_{I\times}^{(3)}$ given by Eq. \eqref{eq:F^(3)},
\begin{align}
D_{4}=\mathcal A_m (1+c_i^2) (\cos 2\psi F_{I +}^{(4)}-\sin 2\psi F_{I \times}^{(4)}), ~~~~~~~~~~~~
E_{4}=2 \mathcal A_m   c_i  (\sin 2\psi F_{I +}^{(4)}+\cos 2\psi F_{I \times}^{(4)}),
\end{align}
with $F_{I+}^{(4)}$ and $F_{I\times}^{(4)}$ given by Eq. \eqref{eq:F^(4)},
\begin{align}
D_5=-2 \mathcal A_m (1+c_i^2) (\sin 2\psi F_{I +}+\cos 2\psi F_{I \times}),  ~~~~~~~~~~~~
E_5=4 \mathcal A_m  c_i  (\cos 2\psi F_{I +}-\sin 2\psi F_{I \times}),
\end{align}
\begin{align}
D_6=2 \mathcal A_m  \cdot c_i (\cos 2\psi F_{I +}-\sin 2\psi F_{I \times}),  ~~~~~~~~~~~~~
E_6=2 \mathcal A_m  (\sin 2\psi F_{I +}+\cos 2\psi F_{I \times}),
\end{align}
and
\begin{align}
\label{eq:D7-E7}
D_7= 2 B,~~~~~~~~~~~~~~~~  E_7= - 2 A.
\end{align}

\section{Analytical Results of the Inner-Product Terms for the circular case}\label{Sec: circular-time-evolving}

From Eq. \eqref{eq:inner-product-weak}, we can solve the following inner-product terms in Eqs. \eqref{eq:Gamma_11-circular}-\eqref{eq:Gamma_ab-circular} and \eqref{eq:xi_1-xi_1-circular-(2)}-\eqref{eq:xi_a-xi_b-circular-(2)}:
\begin{align}
\label{eq:inner-product-circular}
\left<\mathcal S^E\big| \mathcal S^E \right>_I &=\frac{1}{\sigma_I^2 \delta t_I} \left[\frac{t}{2}-\frac{T \sin \left( \frac{8 \pi t}{T}+4 \varphi^E \right)}{16 \pi}+\frac{T \sin 4 \varphi^E }{16 \pi} \right],
\end{align}
\begin{align}
\left<\mathcal C^E\big| \mathcal C^E \right>_I &=\frac{1}{\sigma_I^2 \delta t_I} \left[\frac{t}{2}+\frac{T \sin \left( \frac{8 \pi t}{T}+4 \varphi^E \right)}{16 \pi}-\frac{T \sin 4 \varphi^E }{16 \pi} \right],
\end{align}
\begin{align}
\left<\mathcal S^E\big| \mathcal C^E \right>_I &=\left<\mathcal C^E\big| \mathcal S^E \right>_I=\frac{1}{\sigma_I^2 \delta t_I} \left[ \frac{T \cos \left( \frac{8 \pi t}{T}+4 \varphi^E \right)}{16 \pi}-\frac{T \cos 4 \varphi^E}{16 \pi} \right],
\end{align}
\begin{align}
\left<\mathcal S^E\big| t \mathcal S^E \right>_I =\frac{1}{\sigma_I^2 \delta t_I}\left[\frac{t^2}{4}-\frac{t T \sin \left( \frac{8 \pi t}{T}+4 \varphi^E \right)}{16 \pi}-\frac{T^2 \cos \left( \frac{8 \pi t}{T} +4 \varphi^E \right)}{128 \pi^2}+\frac{T^2 \cos  4 \varphi^E}{128 \pi^2} \right],
\end{align}
\begin{align}
\left<\mathcal C^E\big| t \mathcal C^E \right>_I &=\frac{1}{\sigma_I^2 \delta t_I}\left[\frac{t^2}{4}+\frac{t T \sin \left( \frac{8 \pi t}{T}+4 \varphi^E \right)}{16 \pi}+\frac{T^2 \cos \left( \frac{8 \pi t}{T}+4 \varphi^E \right)}{128 \pi^2}-\frac{T^2 \cos  4 \varphi^E}{128 \pi^2} \right],
\end{align}
\begin{align}
\label{eq:inner-product-circular-(2)}
\left<\mathcal S^E\big| t \mathcal C^E \right>_I =\left<\mathcal C^E\big| t \mathcal S^E \right>_I=  \frac{1}{\sigma_I^2 \delta t_I}\bigg[  -\frac{t T\cos \left( \frac{8 \pi t}{T}+4 \varphi^E \right)}{16 \pi}
 +\frac{T^2 \sin \left( \frac{8 \pi t}{T}+4 \varphi^E \right)}{128 \pi^2}-\frac{T^2 \sin 4 \varphi^E }{128 \pi^2} \bigg].
\end{align}
\begin{align}
\label{eq:inner-product-circular-(3)}
\left<t \mathcal S^E\big| t \mathcal S^E \right>_I =\frac{1}{\sigma_I^2 \delta t_I}\bigg[ \frac{t^3}{6}-\frac{t^2 T \sin \left( \frac{8 \pi t}{T}+4 \varphi^E \right)}{16 \pi} -\frac{t T^2 \cos \left( \frac{8 \pi t}{T}+4 \varphi^E \right)}{64 \pi^2}
 +\frac{T^3 \sin \left( \frac{8 \pi t}{T}+4 \varphi^E \right)}{512 \pi^3}-\frac{T^3 \sin 4 \varphi^E}{512 \pi^3} \bigg],
\end{align}
\begin{align}
\left<t \mathcal C^E\big| t \mathcal C^E \right>_I =\frac{1}{\sigma_I^2 \delta t_I}\bigg[  \frac{t^3}{6}+\frac{t^2 T \sin \left(
\frac{8 \pi t}{T}+4 \varphi^E \right)}{16 \pi} +\frac{t T^2 \cos \left( \frac{8 \pi t}{T}+4 \varphi^E \right)}{64 \pi^2}
 -\frac{T^3 \sin \left( \frac{8 \pi t}{T}+4 \varphi^E \right)}{512 \pi^3}+\frac{T^3 \sin 4 \varphi^E}{512 \pi^3} \bigg], \
\end{align}
and
\begin{align}
\label{eq:inner-product-circular-(4)}
\left<t \mathcal S^E\big| t \mathcal C^E \right>_I &=\left<t \mathcal C^E\big| t \mathcal S^E \right>_I    \\
&=\frac{1}{\sigma_I^2 \delta t_I}\bigg[  -\frac{t^2 T \cos \left( \frac{8 \pi t}{T}+4 \varphi^E \right)}{16 \pi}  +\frac{t T^2 \sin \left( \frac{8 \pi t}{T}+4 \varphi^E \right)}{64 \pi^2}
 +\frac{T^3 \cos \left( \frac{8 \pi t}{T}+4 \varphi^E \right)}{512\pi^3} -\frac{T^3 \cos 4 \varphi^E}{512\pi^3} \bigg].  \nonumber
\end{align}

\section{Matrix $\left< \xi_a \xi_b  \right>$ for the circular case} \label{Sec: results-circular}~

In the circular case $e=0$,
the mean-squared noise-projection matrix $\left< \xi_a \xi_b \right>$ given by \eqref{eq:xi_1-xi_1-eccentric-(2)}-\eqref{eq:xi_a-xi_b-eccentric-(2)}
reduces to
\begin{align}
\label{eq:xi_1-xi_1-circular-(2)}
& \left< \xi_1 \xi_1 \right>  = \frac{8 \pi^2 }{T^4} {\mathop \sum \limits_{I=1}^{N_{\rm P}}} \bigg[  A^2  B^2  \left(\left<\mathcal S^E\big| t  \mathcal S^E \right>_I  \left<\mathcal S^E\big| t  \mathcal S^E \right>_I+ \left<\mathcal C^E\big| t  \mathcal S^E  \right>_I   \left<\mathcal C^E\big| t  \mathcal S^E \right>_I \right)          \\  
&- A^3  B     \left(\left<\mathcal S^E\big| t  \mathcal S^E \right>_I  \left<\mathcal S^E\big| t  \mathcal C^E \right>_I+ \left<\mathcal C^E\big| t  \mathcal S^E  \right>_I   \left<\mathcal C^E\big| t  \mathcal C^E \right>_I \right)   
- A^2 B^2 \left(\left<\mathcal S^E\big| t  \mathcal S^E \right>_I  \left<\mathcal C^E\big| t  \mathcal C^E \right>_I -\left<\mathcal C^E\big| t  \mathcal S^E  \right>_I   \left<\mathcal S^E\big| t  \mathcal C^E \right>_I \right)  \nonumber \\  
&- A^3   B  \left(\left<\mathcal S^E\big| t  \mathcal C^E \right>_I  \left<\mathcal S^E\big| t  \mathcal S^E \right>_I +\left<\mathcal C^E\big| t  \mathcal C^E  \right>_I   \left<\mathcal C^E\big| t  \mathcal S^E \right>_I \right)              
+ A^4  \left(\left<\mathcal S^E\big| t  \mathcal C^E \right>_I  \left<\mathcal S^E\big| t  \mathcal C^E \right>_I +\left<\mathcal C^E\big| t  \mathcal C^E  \right>_I   \left<\mathcal C^E\big| t  \mathcal C^E \right>_I \right) \nonumber \\    
&- A^2 B^2  \left(\left<\mathcal S^E\big| t  \mathcal C^E \right>_I  \left<\mathcal C^E\big| t  \mathcal S^E \right>_I -\left<\mathcal C^E\big| t  \mathcal C^E  \right>_I   \left<\mathcal S^E\big| t  \mathcal S^E \right>_I \right)    
- A^2 B^2  \left(\left<\mathcal C^E\big| t  \mathcal S^E \right>_I  \left<\mathcal S^E\big| t  \mathcal C^E \right>_I -\left<\mathcal S^E\big| t  \mathcal S^E  \right>_I   \left<\mathcal C^E\big| t  \mathcal C^E \right>_I \right) \nonumber \\   
&+ B^4  \left(\left<\mathcal C^E\big| t  \mathcal S^E \right>_I  \left<\mathcal C^E\big| t  \mathcal S^E \right>_I +\left<\mathcal S^E\big| t  \mathcal S^E  \right>_I   \left<\mathcal S^E\big| t  \mathcal S^E \right>_I \right)     
- A B^3 \left(\left<\mathcal C^E\big| t  \mathcal S^E \right>_I  \left<\mathcal C^E\big| t  \mathcal C^E \right>_I +\left<\mathcal S^E\big| t  \mathcal S^E  \right>_I   \left<\mathcal S^E\big| t  \mathcal C^E \right>_I \right) \nonumber \\   
&- A^2 B^2 \left(\left<\mathcal C^E\big| t  \mathcal C^E \right>_I  \left<\mathcal S^E\big| t  \mathcal S^E \right>_I -\left<\mathcal S^E\big| t  \mathcal C^E  \right>_I   \left<\mathcal C^E\big| t  \mathcal S^E \right>_I \right)    
- A B^3   \left(\left<\mathcal C^E\big| t  \mathcal C^E \right>_I  \left<\mathcal C^E\big| t  \mathcal S^E \right>_I +\left<\mathcal S^E\big| t  \mathcal C^E  \right>_I   \left<\mathcal S^E\big| t  \mathcal S^E \right>_I \right)     \nonumber \\       
&+A^2 B^2  \left(\left<\mathcal C^E\big| t  \mathcal C^E \right>_I  \left<\mathcal C^E\big| t  \mathcal C^E \right>_I +\left<\mathcal S^E\big| t  \mathcal C^E  \right>_I   \left<\mathcal S^E\big| t  \mathcal C^E \right>_I \right)   \bigg],    \nonumber          
\end{align}
\begin{align}
\label{eq:xi_1-xi_a-circular-(2)}
&       \left< \xi_1 \xi_a \right>=\left< \xi_a \xi_1 \right>
=\frac{2 \pi }{T^2}{\mathop \sum \limits_{I=1}^{N_{\rm P}}} \bigg[    A^2 B D_a \left(\left<\mathcal S^E\big| t  \mathcal S^E \right>_I \left<\mathcal S^E\big|   \mathcal S^E \right>_I+ \left<\mathcal C^E\big| t  \mathcal S^E  \right>_I  \left<\mathcal C^E\big|   \mathcal S^E \right>_I \right)      \\  
&+A^2 B E_a \left(\left<\mathcal S^E\big| t  \mathcal S^E \right>_I \left<\mathcal S^E\big|   \mathcal C^E \right>_I+ \left<\mathcal C^E\big| t  \mathcal S^E  \right>_I   \left<\mathcal C^E\big|   \mathcal C^E \right>_I \right)    
+A B^2 D_a \left(\left<\mathcal S^E\big| t  \mathcal S^E \right>_I  \left<\mathcal C^E\big|   \mathcal S^E \right>_I -\left<\mathcal C^E\big| t  \mathcal S^E  \right>_I   \left<\mathcal S^E\big|  \mathcal S^E \right>_I \right)    \nonumber \\    
&+A B^2 E_a \left(\left<\mathcal S^E\big| t  \mathcal S^E \right>_I  \left<\mathcal C^E\big|   \mathcal C^E \right>_I -\left<\mathcal C^E\big| t  \mathcal S^E  \right>_I   \left<\mathcal S^E\big|   \mathcal C^E \right>_I \right)    
-A^3 D_a \left(\left<\mathcal S^E\big| t  \mathcal C^E \right>_I  \left<\mathcal S^E\big|   \mathcal S^E \right>_I +\left<\mathcal C^E\big| t  \mathcal C^E  \right>_I   \left<\mathcal C^E\big|   \mathcal S^E \right>_I \right) \nonumber \\          
&-A^3 E_a \left(\left<\mathcal S^E\big| t  \mathcal C^E \right>_I  \left<\mathcal S^E\big|   \mathcal C^E \right>_I +\left<\mathcal C^E\big| t  \mathcal C^E  \right>_I   \left<\mathcal C^E\big|   \mathcal C^E \right>_I \right)    
-A^2 B D_a \left(\left<\mathcal S^E\big| t  \mathcal C^E \right>_I  \left<\mathcal C^E\big|   \mathcal S^E \right>_I -\left<\mathcal C^E\big| t  \mathcal C^E  \right>_I   \left<\mathcal S^E\big|   \mathcal S^E \right>_I \right)   \nonumber \\           
&-A^2 B E_a \left(\left<\mathcal S^E\big| t  \mathcal C^E \right>_I  \left<\mathcal C^E\big|   \mathcal C^E \right>_I -\left<\mathcal C^E\big| t  \mathcal C^E  \right>_I   \left<\mathcal S^E\big|   \mathcal C^E \right>_I \right)    
+A B^2 D_a \left(\left<\mathcal C^E\big| t  \mathcal S^E \right>_I  \left<\mathcal S^E\big|   \mathcal S^E \right>_I -\left<\mathcal S^E\big| t  \mathcal S^E  \right>_I   \left<\mathcal C^E\big|   \mathcal S^E \right>_I \right) \nonumber \\               
&+A B^2 E_a \left(\left<\mathcal C^E\big| t  \mathcal S^E \right>_I  \left<\mathcal S^E\big|   \mathcal C^E \right>_I -\left<\mathcal S^E\big| t  \mathcal S^E  \right>_I   \left<\mathcal C^E\big|   \mathcal C^E \right>_I \right)    
+B^3 D_a   \left(\left<\mathcal C^E\big| t  \mathcal S^E \right>_I  \left<\mathcal C^E\big|   \mathcal S^E \right>_I +\left<\mathcal S^E\big| t  \mathcal S^E  \right>_I   \left<\mathcal S^E\big|   \mathcal S^E \right>_I \right) \nonumber \\                  
&+ B^3 E_a  \left(\left<\mathcal C^E\big| t  \mathcal S^E \right>_I  \left<\mathcal C^E\big|   \mathcal C^E \right>_I +\left<\mathcal S^E\big| t  \mathcal S^E  \right>_I   \left<\mathcal S^E\big|   \mathcal C^E \right>_I \right)      
-A^2 B D_a \left(\left<\mathcal C^E\big| t  \mathcal C^E \right>_I  \left<\mathcal S^E\big|   \mathcal S^E \right>_I -\left<\mathcal S^E\big| t  \mathcal C^E  \right>_I   \left<\mathcal C^E\big|   \mathcal S^E \right>_I \right)  \nonumber \\                  
&-A^2 B E_a \left(\left<\mathcal C^E\big| t  \mathcal C^E \right>_I  \left<\mathcal S^E\big|   \mathcal C^E \right>_I -\left<\mathcal S^E\big| t  \mathcal C^E  \right>_I   \left<\mathcal C^E\big|   \mathcal C^E \right>_I \right)      
-A B^2 D_a \left(\left<\mathcal C^E\big| t  \mathcal C^E \right>_I  \left<\mathcal C^E\big|   \mathcal S^E \right>_I +\left<\mathcal S^E\big| t  \mathcal C^E  \right>_I   \left<\mathcal S^E\big|   \mathcal S^E \right>_I \right)  \nonumber \\                 
&-A B^2 E_a \left(\left<\mathcal C^E\big| t  \mathcal C^E \right>_I  \left<\mathcal C^E\big|  \mathcal C^E \right>_I +\left<\mathcal S^E\big| t  \mathcal C^E  \right>_I   \left<\mathcal S^E\big|   \mathcal C^E \right>_I \right) \bigg],     \nonumber     
\end{align}
for $a \neq 1$, and
\begin{align}
\label{eq:xi_a-xi_b-circular-(2)}
& \left< \xi_a \xi_b \right> =\frac{1}{2}{\mathop \sum \limits_{I=1}^{N_{\rm P}}} \bigg[    A^2 D_a D_b \left(\left<\mathcal S^E\big|  \mathcal S^E \right>_I \left<\mathcal S^E\big|   \mathcal S^E \right>_I+ \left<\mathcal C^E\big|  \mathcal S^E  \right>_I  \left<\mathcal C^E\big|  \mathcal S^E \right>_I \right)  \\ 
&+A^2 D_a E_b \left(\left<\mathcal S^E\big|   \mathcal S^E \right>_I \left<\mathcal S^E\big|   \mathcal C^E \right>_I+ \left<\mathcal C^E\big|  \mathcal S^E  \right>_I   \left<\mathcal C^E\big|   \mathcal C^E \right>_I \right)   
+A B D_a E_b \left(\left<\mathcal S^E\big|  \mathcal S^E \right>_I  \left<\mathcal C^E\big|   \mathcal C^E \right>_I -\left<\mathcal C^E\big|  \mathcal S^E  \right>_I   \left<\mathcal S^E\big|   \mathcal C^E \right>_I \right)   \nonumber \\     
&+A^2 E_a D_b \left(\left<\mathcal S^E\big|  \mathcal C^E \right>_I  \left<\mathcal S^E\big|   \mathcal S^E \right>_I +\left<\mathcal C^E\big|  \mathcal C^E  \right>_I   \left<\mathcal C^E\big|   \mathcal S^E \right>_I \right)            
+A^2 E_a E_b \left(\left<\mathcal S^E\big|   \mathcal C^E \right>_I  \left<\mathcal S^E\big|   \mathcal C^E \right>_I +\left<\mathcal C^E\big|  \mathcal C^E  \right>_I   \left<\mathcal C^E\big|   \mathcal C^E \right>_I \right) \nonumber \\   
&+A B E_a D_b \left(\left<\mathcal S^E\big|   \mathcal C^E \right>_I  \left<\mathcal C^E\big|   \mathcal S^E \right>_I -\left<\mathcal C^E\big|  \mathcal C^E  \right>_I   \left<\mathcal S^E\big|   \mathcal S^E \right>_I \right)           
+A B D_a E_b \left(\left<\mathcal C^E\big|   \mathcal S^E \right>_I  \left<\mathcal S^E\big|   \mathcal C^E \right>_I -\left<\mathcal S^E\big|   \mathcal S^E  \right>_I   \left<\mathcal C^E\big|   \mathcal C^E \right>_I \right) \nonumber \\  
&+B^2 D_a D_b \left(\left<\mathcal C^E\big|   \mathcal S^E \right>_I  \left<\mathcal C^E\big|   \mathcal S^E \right>_I +\left<\mathcal S^E\big|   \mathcal S^E  \right>_I   \left<\mathcal S^E\big|   \mathcal S^E \right>_I \right)        
+B^2 D_a E_b \left(\left<\mathcal C^E\big|   \mathcal S^E \right>_I  \left<\mathcal C^E\big|   \mathcal C^E \right>_I +\left<\mathcal S^E\big|   \mathcal S^E  \right>_I   \left<\mathcal S^E\big|   \mathcal C^E \right>_I \right)  \nonumber \\ 
&+A B E_a D_b \left(\left<\mathcal C^E\big|   \mathcal C^E \right>_I  \left<\mathcal S^E\big|   \mathcal S^E \right>_I -\left<\mathcal S^E\big|   \mathcal C^E  \right>_I   \left<\mathcal C^E\big|   \mathcal S^E \right>_I \right)       
+B^2 E_a D_b \left(\left<\mathcal C^E\big|   \mathcal C^E \right>_I  \left<\mathcal C^E\big|   \mathcal S^E \right>_I +\left<\mathcal S^E\big|  \mathcal C^E  \right>_I   \left<\mathcal S^E\big|   \mathcal S^E \right>_I \right)  \nonumber \\         
&+ B^2 E_a E_b \left(\left<\mathcal C^E\big|   \mathcal C^E \right>_I  \left<\mathcal C^E\big|  \mathcal C^E \right>_I +\left<\mathcal S^E\big|  \mathcal C^E  \right>_I   \left<\mathcal S^E\big|   \mathcal C^E \right>_I \right) \bigg],  \nonumber  
\end{align}
for both $a \neq 1$ and $b \neq 1$.

\bibliography{reference}{}
\bibliographystyle{aasjournal}



\end{document}